\documentclass[aps,prd,11pt,superscriptaddress,notitlepage,longbibliography,nofootinbib,tightenlines,groupedaddress]{revtex4-1}

\usepackage{amsmath,amssymb,amsfonts}
\usepackage{graphicx}
\usepackage{subfigure}
\usepackage{color}
\usepackage{hyperref}
\usepackage{dsfont}
\usepackage{slashed}
\definecolor{darkred}{rgb}{0.8,0.1,0.1}
\hypersetup{colorlinks=true, linkcolor=darkred, citecolor=blue, linktoc=page}

\newcommand{\N}[1]{\ensuremath{\mathcal N=#1}}

\DeclareMathOperator{\sech}{sech}

\newcommand{\GammaAdS}{\ensuremath{\Gamma_{\mathrm{AdS}}}}
\newcommand{\GammaS}{\ensuremath{\Gamma_{\mathrm{S}^5}}}
\newcommand{\GammaChi}{\ensuremath{\Gamma_{\vec{\chi}}}}
\newcommand{\RAdS}{\ensuremath{R_\mathrm{AdS}}}
\newcommand{\RS}{\ensuremath{R_{\mathrm{S}^5}}}

\newcommand{\varGammaAdS}{\ensuremath{\varGamma_{\mathrm{AdS}}}}
\newcommand{\varGammaS}{\ensuremath{\varGamma_{\mathrm{S}^5}}}
\newcommand{\varGammaChi}{\ensuremath{\varGamma_{\underline{\vec{\chi}}}}}

\makeatletter\def\l@subsubsection#1#2{}%
\makeatother

\begin{document}

\title{Supersymmetric D3/D7 for holographic flavors on curved space}

  \author{Andreas Karch}
  \email{akarch@uw.edu}
  \author{Brandon Robinson}
  \email{robinb22@uw.edu}
  \author{Christoph F.~Uhlemann}
  \email{uhlemann@uw.edu}

 \affiliation{Department of Physics, University of Washington, Seattle, WA 98195-1560, USA}

\begin{abstract}
We derive a new class of supersymmetric D3/D7 brane configurations,
which allow to holographically describe \N{4} SYM coupled to massive \N{2} flavor degrees of freedom
on spaces of constant curvature.
We systematically solve the $\kappa$-symmetry condition for D7-brane embeddings into AdS$_4$-sliced AdS$_5\times$S$^5$,
and find supersymmetric embeddings in a simple closed form.
Up to a critical mass, these embeddings come in surprisingly diverse families,
and we present a first study of their (holographic) phenomenology.
We carry out the holographic renormalization, compute the one-point functions and attempt a field-theoretic interpretation
of the different families.
To complete the catalog of supersymmetric D3/D7 configurations, we construct analogous
embeddings for flavored \N{4} SYM on S$^4$ and dS$_4$.
\end{abstract}

\maketitle
\tableofcontents

\section{Introduction}
The study of quantum field theory (QFT) on curved space has a long history, and has revealed numerous interesting insights.
From Hawking radiation to cosmological particle production, many interesting phenomena are tied to curved backgrounds,
which makes Minkowski space appear as only a very special case.
Minkowski space is also artificial from a more conceptual point of view,
as in our universe it only arises as an approximate description on length scales which are short compared to those associated with curvature.
From that perspective, QFT should in general be understood in curved backgrounds, and only be specialized to Minkowski space where appropriate,
as opposed to the other way around.
Conceptually attractive as that approach may be, it is technically challenging, at best, for conventional methods.
Nicely enough, though, many interesting phenomena in curved space arise already in free field theory,
which often is as far as one can get with direct methods.

The non-trivial nature of even free field theory in curved space makes one suspect interesting things to happen
if curved backgrounds are combined with strong interactions,
as already on flat space physics at strong coupling can likewise be qualitatively different from the more easily accessible physics at weak coupling.
For traditional QFT methods, combining strong coupling with curved backgrounds certainly is challenging.
But from the AdS/CFT perspective, which has become one of the few established tools to quantitatively access
strongly-coupled QFTs, the increase in difficulty is not that dramatic after all.
Going from Minkowski space to curved space QFT in the simplest cases just corresponds to choosing a
different conformal compactification of AdS.
That certainly makes it interesting and worthwhile to study strongly-coupled QFTs in curved spacetimes.
Some early holographic investigations of QFT on (A)dS$_4$ were already initiated in \cite{Karch:2000ct,Karch:2000gx,Buchel:2002wf},
and more recent and comprehensive work can be found e.g.\ in  \cite{Marolf:2013ioa,Rangamani:2015qga}.
Spacetimes of constant curvature are certainly the natural starting point for departure from Minkowski space,
and we will focus on AdS$_4$ in the main part of this work.

When it comes to the choice of theory, \N{4} SYM is a natural starting point for holographic investigations,
and detailed studies on AdS$_4$ and dS$_4$ were initiated in \cite{Marolf:2010tg,Aharony:2010ay}.
But by itself, \N{4} SYM also is a rather special theory, with its conformal (super)symmetry and all fields in the adjoint representation.
As in flat space, it is desirable to bridge the gap to the theories we actually find realized in nature.
One aspect of that is adding fundamental matter, which can be done holographically by adding D7-branes \cite{Karch:2002sh}.
The resulting theory is \N{2} supersymmetric and has a non-trivial UV fixed point in the quenched approximation,
where the rank of the gauge group is large compared to the number of ``quarks''.
For massless quenched flavors, the theory is actually conformal, and the AdS$_4$ discussion could just as well be carried out on flat space.
The primary focus of this work will be to add massive \N{2} flavors and thereby explicitly break conformal symmetry.
We will also employ the quenched or probe approximation throughout, such that the D7-branes on the holographic side can be
described by a classical action, with Dirac-Born-Infeld (DBI) and Wess-Zumino (WZ) terms, and backreaction effects are small.

A holographic study of \N{4} SYM coupled to massive flavor hypermultiplets on AdS$_4$ can be found already in \cite{Karch:2009ph,Clark:2013mfa}.
In those works, interesting embeddings were found by solving the non-linear field equation for the brane embedding numerically,
which yields embeddings that generically break all supersymmetries.
It is desirable, however, to preserve the \N{2} supersymmetry that the theory has on flat space.
Besides the general argument that a larger amount of symmetry makes the theory more tractable,
supersymmetry actually offers a chance to make quantitative statements on the field theory side,
using, for example, localization \cite{Pestun:2007rz}.
Formulating supersymmetric field theories in curved space needs some care, however, as just minimally coupling a flat-space supersymmetric theory
to a curved metric does in general not result in a supersymmetric theory.
Non-minimal curvature couplings may be needed, and can be understood systematically by consistently coupling to supergravity
and then restricting to a fixed background \cite{Pestun:2007rz,Festuccia:2011ws}.
The desire to include supersymmetry makes AdS the preferred curved space to look at in Lorentzian signature,
as the formulation of supersymmetric QFTs on dS faces additional issues with unitarity unless the theory is conformal \cite{Anous:2014lia}.

On the holographic side, preserving supersymmetry for flavored \N{4} SYM on AdS$_4$ also adds a new aspect to the discussion.
Instead of just straightforwardly solving the field equations resulting from the D7-brane action,
we now have to deal with $\kappa$-symmetry \cite{Bergshoeff:1996tu,Cederwall:1996pv,Cederwall:1996ri}.
This extra fermionic gauge symmetry projects out part of the fermionic modes, to obtain matching numbers of bosonic
and fermionic brane degrees of freedom, as required by supersymmetry.
Demanding some amount of supersymmetry to be preserved then amounts to preserving $\kappa$-symmetry,
which can be further translated into a set of necessary conditions for the embedding and worldvolume fluxes.
Extracting these conditions, however, is technically challenging.
The extra terms needed on the field theory side to preserve supersymmetry suggest that varying the slipping mode alone will not be enough to
get massive supersymmetric embeddings.
So we will also have to include worldvolume flux, which additionally complicates the discussion.
Once the step of extracting necessary conditions for the embedding and worldvolume gauge field is carried out,
however, the $\kappa$-symmetry condition promises 1$^\mathrm{st}$-order BPS equations, as opposed to the 2$^\mathrm{nd}$-order field equations.
This will allow us to find analytic solutions, and so is well worth the trouble.

We systematically analyze the constraint imposed by $\kappa$-symmetry on the embeddings
and extract necessary conditions for the slipping mode and worldvolume flux in Sec.~\ref{sec:susy-d3d7}.
From the resulting conditions we will be able to extract analytic supersymmetric D7-brane embeddings in a nice closed form,
which are given in Sec.~\ref{sec:fully-massive-embedding-solution}, with the conventions laid out in \ref{sec:GeometryKillingAction}.
The solutions we find allow to realize a surprisingly rich set of supersymmetric embeddings,
which we categorize into short, long and connected embeddings.
We study those in more detail in Sec.~\ref{sec:embedding-classes}.
In Sec.~\ref{sec:N4SYM} we focus on implications for flavored \N{4} SYM.
We carry out the holographic renormalization, compute the chiral and scalar condensates,
and attempt an interpretation of the various embeddings found in Sec.~\ref{sec:embedding-classes} from the QFT perspective.
This raises some interesting questions, and we close with a more detailed summary and discussion
in Sec.~\ref{sec:discussion}.

In App.~\ref{app:global-Euclidean-kappa} we similarly construct supersymmetric D7-brane embeddings
into S$^4$-sliced and dS$_4$-sliced AdS$_5\times$S$^5$,
so we end up with a comprehensive catalog of D7-brane embeddings to holographically describe massive
\N{2} supersymmetric flavors on spaces of constant curvature.
These will be used in a companion paper to compare the free energy obtained from the holographic
calculation for S$^4$ to a QFT calculation using supersymmetric localization.

\section{Supersymmetric D7 branes in \texorpdfstring{A\lowercase{d}S$_4$}{AdS4}-sliced \texorpdfstring{A\lowercase{d}S$_5\times$S$^5$}{AdS5xS5}}\label{sec:susy-d3d7}
In this section we evaluate the constraint imposed by $\kappa$ symmetry to find supersymmetric D7-brane embeddings into AdS$_4$-sliced AdS$_5\times$S$^5$.
The $\kappa$-symmetry constraint for embeddings with non-trivial fluxes has a fairly non-trivial Clifford-algebra structure,
and the explicit expressions for AdS$_5\times$S$^5$ Killing spinors are themselves not exactly simple.
That makes it challenging to extract the set of necessary equations for the embedding and flux from it,
and this task will occupy most of the next section.
On the other hand, the non-trivial Clifford-algebra structure will allow us to separate the equations for flux and embedding.
Once the $\kappa$-symmetry analysis is done, the pay-off is remarkable. Instead of heaving to solve the square-root
non-linear coupled differential equations resulting from
variation of the DBI action with Wess-Zumino term, we will be able to explicitly solve for the worldvolume gauge field in terms of the slipping mode.
The remaining equation then is a non-linear but reasonably simple differential equation for the slipping mode alone.
As we verified explicitly to validate our derivation, these simple equations indeed imply the full non-linear DBI equations of motion.
We set up the background, establish conventions
and motivate our choices for the embedding ansatz and worldvolume flux in Sec.~\ref{sec:GeometryKillingAction}.
Generalities on $\kappa$-symmetry are set up in Sec.~\ref{sec:kappa-generalities},
and infinitesimally massive embeddings are discussed in Sec.~\ref{sec:small-mass-kappa}.
The finite mass embeddings are in Sec.~\ref{sec:fully-massive}.
To find the solutions, we take a systematic approach to the $\kappa$-symmetry analysis,
which is also necessary to show that the solutions we find are indeed supersymmetric.
Readers interested mainly in the results can directly proceed from Sec.~\ref{sec:GeometryKillingAction} to the
the embeddings given in Sec.~\ref{sec:fully-massive-embedding-solution}.

\subsection{Geometry and embedding ansatz}\label{sec:GeometryKillingAction}

Our starting point will be Lorentzian signature and the AdS$_4$ slices in Poincar\'{e} coordinates.
For the global structure, it does make a difference whether we choose global AdS$_4$
or the Poincar\'e patch as slices, and the explicit expressions for the metric,
Killing spinors etc.\ are also different.
However, the field equations and the $\kappa$-symmetry constraint are local conditions,
and our final solutions will thus be valid for both choices.

We choose coordinates such that the AdS$_5\times$S$^5$ background geometry has a metric
\begin{align}\label{eqn:AdS5S5-metric}
 g_{\mathrm{AdS}_5}^{}&=d\rho^2+\cosh^2\rho\left[dr^2+e^{2r}(-dt^2+d\vec{x}^2)\right]~,
 &
 g_{\mathrm{S}^5}&=d\theta^2+\cos^2\theta\:d\psi^2+\sin^2\theta\:d\Omega_3^2~,
\end{align}
where $d\Omega_3^2=d\chi_1^2+\sin^2\chi_1 (d\chi_2^2+\sin^2\chi_2 d\chi_3^2)$.
We use the AdS$_5\times$S$^5$ Killing spinor equation in the conventions of \cite{Grisaru:2000zn}
\begin{align}\label{eqn:Killing-spinor-eq}
 D_\mu\epsilon&=\frac{i}{2}\,\GammaAdS\Gamma_\mu\epsilon~,\quad \mu=0\dots 4~,&
 D_\mu\epsilon&=\frac{i}{2}\,\GammaS\Gamma_\mu\epsilon~,\quad \mu=5\dots 9~,&
\end{align}
and we have $\GammaAdS:=\Gamma^{\underline{01234}}=-\Gamma_{\underline{01234}}$
along with $\GammaS:=\Gamma^{\underline{56789}}$.
Generally, we follow the usual convention and denote coordinate indices by Greek letters from the middle of the alphabet and
local Lorentz indices by latin letters from the beginning of the alphabet.
We will use an underline to distinguish Lorentz indices from coordinate indices whenever explicit values appear.
The ten-dimensional chirality matrix is $\Gamma_{11}=\GammaAdS\GammaS$.

For the $\kappa$-symmetry analysis we will need the explicit expressions for the Killing spinors solving (\ref{eqn:Killing-spinor-eq}).
They can be constructed from a constant chiral spinor $\epsilon_0$ with $\Gamma_{11}\epsilon_0=\epsilon_0$ as
\begin{align}\label{eqn:Killing-spinors-general}
 \epsilon=R_{\mathrm{S}^5}\times R_{\mathrm{AdS}}\times\epsilon_0~.
\end{align}
The matrices $R_{\mathrm{AdS}}$, $R_{\mathrm{S}^5}$
denote products of exponentials of even numbers of $\Gamma$-matrices with indices in AdS$_5$ and S$^5$, respectively.
For the S$^5$ part we find\footnote{%
For $\psi=0$ our (\ref{eqn:Killing-spinors}) agrees with the S$^4$ Killing spinors constructed in \cite{Lu:1998nu}.
But this is different from (86) of \cite{Grisaru:2000zn} by factors of $i$ in the S$^3$ part.
}
\begin{align}\label{eqn:Killing-spinors}
  R_{\mathrm{S}^5}&=
   e^{\frac{\theta}{2} i\Gamma_{}^{\underline{\psi}}\GammaChi}
   \:e^{\frac{\psi}{2} i\GammaChi\Gamma^{\underline{\theta}}}
   \:e^{\frac{1}{2}\chi_1\Gamma^{\underline{\theta\chi_1}}}
   \:e^{\frac{1}{2}\chi_2\Gamma^{\underline{\chi_1\chi_2}}}
   \:e^{\frac{1}{2}\chi_3\Gamma^{\underline{\chi_2\chi_3}}}~,
\end{align}
where we have defined $\GammaChi:=\Gamma^{\underline{\chi_1}}\Gamma^{\underline{\chi_2}}\Gamma^{\underline{\chi_3}}$.
The exponent in all the exponentials is the product of a real function $f$ and a matrix $A$ which squares to~$-\mathds{1}$.
We will also encounter the product of a real function and a matrix $B$ which squares to $+\mathds{1}$ in the exponential.
The explicit expansions are
\begin{align}
 e^{f A}&=\cos f\cdot\mathds{1} + \sin f\cdot A~,&
 e^{f B}&=\cosh f\cdot\mathds{1} + \sinh f\cdot B~.
\end{align}
The corresponding $R$-matrix for AdS$_4$-sliced AdS$_5$ can be constructed easily, starting from the AdS Killing spinors given in \cite{Lu:1996rhb,Lu:1998nu}.
With the projectors $P_{r\pm}=\frac{1}{2}(\mathds{1}\pm i\Gamma_{\underline{r}}\GammaAdS)$,
the AdS$_5$ part reads
\begin{align}\label{eqn:AdS5S5-Killing-from-AdS4-L}
 R_\mathrm{AdS}&=e_{}^{\frac{\rho}{2}i\Gamma_{\underline{\rho}}\GammaAdS}R_{\mathrm{AdS}_4}~,&
 R_{\mathrm{AdS}_4}&=e^{\frac{r}{2}i\Gamma_{\underline{r}}\GammaAdS}
 +ie^{r/2}x^\mu\Gamma_{\underline{x_\mu}}\GammaAdS P_{r-}~.
\end{align}

For the D7 branes we explicitly spell out the DBI action and WZ term to fix conventions.
For the $\kappa$-symmetry analysis we will not actually need it,
but as a consistency check we want to verify that our final solutions solve the equations of motion derived from it.
We take
\begin{align}\label{eqn:DBI-action}
 S_\mathrm{D7}&=-T_7 \int_{\Sigma_8}d^8\xi\sqrt{-\det\left(g+2\pi\alpha^\prime F\right)}
 +2(2\pi\alpha^\prime)^2T_7\int_{\Sigma_8}C_4\wedge F\wedge F
 ~,
\end{align}
with $g$ denoting the pullback of the background metric and the pullback on the four-form gauge field $C_4$ is understood.
We absorb $2\pi\alpha^\prime$ by a rescaling of the gauge field, so it is implicit from now on.
To fix conventions on the five-form field strength we use \cite{Kim:1985ez}:
to get $R_{\mu\nu}=4 L^{-2}g_{\mu\nu}$, we need $F_5=L^{-1}(1+\star)\operatorname{vol}(\mathrm{AdS}_5)$.
So we take
\begin{align}\label{eqn:RR-gauge-field}
 C_4&=L^{-1}\zeta(\rho)\operatorname{vol}(\mathrm{AdS}_4)+\dots~,&\zeta^\prime(\rho)=\cosh^4\rho~.
\end{align}
The dots in the expression for $C_4$ denote the part producing the volume form on S$^5$ in $F_5$,
which will not be relevant in what follows.
As usual, $C_4$ is determined by $F_5$ only up to gauge transformations, and we in particular have an undetermined constant in $\zeta$,
which will not play any role in the following.

\subsubsection{Embedding ansatz}\label{sec:embedding-ansatz}
We will be looking for D7-brane embeddings to holographically describe \N{4} SYM coupled to massive \N{2} flavors on AdS$_4$.
So we in particular want to preserve the AdS$_4$ isometries.
The ansatz for the embedding will be such that the D7-branes wrap entire AdS$_4\times$S$^3$ slices in AdS$_5\times$S$^5$,
starting at the conformal boundary and reaching into the bulk possibly only up to a finite value of the radial coordinate $\rho$.
The S$^3$ is parametrized as usual by the ``slipping mode'' $\theta$ as function of the radial coordinate $\rho$ only.
We choose static gauge such that the entire embedding is characterized by $\theta$.

To gain some intuition for these embeddings, we recall the Poincar\'e AdS analysis of \cite{Karch:2002sh}.
From that work we already know the $\theta\equiv \pi/2$ embedding, which is
a solution regardless of the choice of coordinates on AdS$_5$.
So we certainly expect to find that again, also with our ansatz.
This particular D3/D7 configuration preserves half of the background supersymmetries, corresponding to the
breaking from \N{4} to \N{2} superconformal symmetry in the boundary theory.\footnote{%
The preserved conformal symmetry is a feature of the quenched approximation with $N_f/N_c\ll 1$ only.}
For Poincar\'e AdS$_5$ with radial coordinate $z$, turning on a non-trivial slipping mode $\theta=\arcsin m z$
breaks additional, but not all supersymmetries.
The configuration is still 1/4 BPS \cite{Karch:2002sh}, corresponding to the breaking of
\N{2} superconformal symmetry to just \N{2} supersymmetry in the boundary theory on Minkowski space.

Our embedding ansatz, on the other hand, is chosen such that it preserves AdS$_4$ isometries,
and the slipping mode depends non-trivially on a different radial coordinate.
These are, therefore, geometrically different embeddings.
As we will see explicitly below, supersymmetric embeddings can not be found in that case by just turning on a non-trivial slipping mode.
From the field-theory analyses in \cite{Pestun:2007rz,Festuccia:2011ws}, we know that in addition to the
mass term for the flavor hypermultiplets we will have to add another purely scalar mass term to preserve some supersymmetry
on curved backgrounds.
This term holographically corresponds to a certain mode of the worldvolume gauge field on the S$^3\subset$ S$^5$,
an $\ell=1,-$ mode in the language of \cite{Kruczenski:2003be}.
Including such worldvolume flux breaks the SO(4) isometries of the S$^3$ to SU(2)$\times$U(1).
The same indeed applies to the extra scalar mass term on the field theory side: it breaks the R-symmetry from SU(2) to U(1).
The SU(2) acting on the \N{2} adjoint hypermultiplet coming from the \N{4} vector multiplet
is not altered by the flavor mass term (see e.g.\ \cite{Hong:2003jm}).
The bottom line for our analysis is that we should not expect to get away with a non-trivial slipping mode only.

For the analysis below we will not use the details of these arguments as input.
Our ansatz is a non-trivial slippling mode $\theta(\rho)$ and a worldvolume gauge field $A=f(\rho)\omega$,
where $\omega$ is a generic one-form on S$^3$.
This ansatz can be motivated just by the desire to preserve the AdS$_4$ isometries.\footnote{%
A generalization which we will not study here is to also allow for non-trivial $\rho$-dependence in~$\psi$.}
Whether the supersymmetric embeddings we will find reflect the field-theory analysis will then be a nice consistency check,
rather than input.
As we will see,  the $\kappa$-symmetry constraint is enough to determine $\omega$ completely,
and the result is indeed consistent with the field-theory analysis.

\subsection{\texorpdfstring{$\kappa$}{kappa}-symmetry generalities}\label{sec:kappa-generalities}
The $\kappa$-symmetry condition projecting on those Killing spinors which are preserved by a given brane embedding was
derived in \cite{Bergshoeff:1996tu,Cederwall:1996pv,Cederwall:1996ri}.
We follow the conventions of \cite{Bergshoeff:1996tu}.
The pullback of the ten-dimensional vielbein $E^a$ to the D7 worldvolume is denoted by $e^a=E_\mu^a(\partial_i X^\mu)dx^i$,
and the Clifford algebra generators pulled back to the worldvolume are denoted by $\gamma_i=e_i^a\Gamma_a$.
We follow \cite{Bergshoeff:1996tu} and define $X^{i}_{\hphantom{i}j}:=g^{ik}F_{kj}$.
The $\kappa$-symmetry condition then is $\Gamma_\kappa\epsilon=\epsilon$, where
\begin{subequations}\label{eqn:kappa}
\begin{align}\label{eqn:kappa-1}
  \Gamma_\kappa&=\frac{1}{\sqrt{\det(1+X)}}\sum_{n=0}^\infty\frac{1}{2^n n!}\gamma^{j_1k_1\dots j_nk_n}X_{j_1k_1}\dots X_{j_nk_n}J_{(p)}^{(n)}~,\\
  J_{(p)}^{(n)}&=(-1)^n\left(\sigma_3\right)^{n+(p-3)/2}i\sigma_2\otimes\Gamma_{(0)}~,\\
  \Gamma_{(0)}&=\frac{1}{(p+1)!\sqrt{-\det g}}\,\varepsilon^{i_1\dots i_{p+1}}\gamma^{}_{i_1\dots i_{p+1}}~.
\end{align}
\end{subequations}
For embeddings characterized by a non-trivial slipping mode as described above, the induced metric on the D7-branes reads
\begin{align}\label{eqn:induced-metric}
 g&=\big(1+{\theta^\prime}^2\big)d\rho^2+\cosh^2\!\rho\, ds_{\mathrm{AdS}_4}^2+\sin^2\!\theta\,d\Omega_3^2~.
\end{align}
The pullback of the ten-bein to the D7 worldvolume is given by
 \begin{align}\label{eqn:D7-vielbein}
e^{a}&=E^{a}~,\quad a=\underline{0}\dots \underline{7}~,
&
e^{\underline{8}}&=\theta^\prime d\rho~,
& e^{\underline{9}}&=0~.
 \end{align}
The $\kappa$-symmetry condition (\ref{eqn:kappa}) for type IIB supergravity is formulated for a pair of Majorana-Weyl spinors.
We will find it easier to change to complex notation, such that we deal with a single Weyl Killing spinor without the Majorana condition.

\subsubsection{Complex notation}
Eq.~(\ref{eqn:kappa}) is formulated for a pair of Majorana-Weyl Killing spinors $(\epsilon_1,\epsilon_2)$,
and it is the index labeling the two spinors on which the Pauli matrices act.
To switch to complex notation we define a single Weyl spinor $\epsilon=\epsilon_1+i\epsilon_2$.
With the Pauli matrices
\begin{align}
 \sigma_2&=\begin{pmatrix}0 & -i\\i &0\end{pmatrix}~,&
 \sigma_3&=\begin{pmatrix} 1 &0\\0 &-1\end{pmatrix}~,
\end{align}
we then find that $i\sigma_2 (\epsilon_1,\epsilon_2)$ translates to $-i\epsilon$ and
$\sigma_3(\epsilon_1,\epsilon_2)$ to $C\epsilon^\star$.
With these replacements the action of $\sigma_{2/3}$ commutes with multiplication by $\Gamma$-matrices, as it should
(the $\Gamma$-matrices in (\ref{eqn:kappa}) should be understood as $\mathds{1}_2\otimes\Gamma$).
We thus find
\begin{align}\label{eqn:switch-to-complex}
 J_{(7)}^{(n)}\begin{pmatrix}\epsilon_1\\ \epsilon_2\end{pmatrix}&\rightarrow -i \Gamma_{(0)}\epsilon \quad  \text{for $n$ even}~,&
 J_{(7)}^{(n)}\begin{pmatrix}\epsilon_1\\ \epsilon_2\end{pmatrix}&\rightarrow -i C\left(\Gamma_{(0)}\epsilon\right)^\star \quad  \text{for $n$ odd}~.
\end{align}
Note that $\Gamma_\kappa$ contains an involution and does not act as a $\mathbb{C}$-linear operator.
To fix conventions, we choose the matrix $B_1$ defined in the appendix of \cite{Polchinski:1998rr}, and set $C=B_1$.
$C$ then is the product of four Hermitian $\Gamma$-matrices that square to $\mathds{1}$,
so we immediately get $C^\dagger=C$ and $C^2=\mathds{1}$.
Furthermore, we have
\begin{align}
 C\Gamma^\mu&=(\Gamma^\mu)^\star C~,&
 C^\star C&=\mathds{1}~.
\end{align}
With (\ref{eqn:switch-to-complex}) it is straightforward now to switch to complex notation in (\ref{eqn:kappa}).

\subsubsection{Projection condition for our embedding ansatz}

We now set up the $\kappa$-symmetry condition in complex notation for our specific ansatz for embedding and worldvolume flux.
As explained above, for our analysis we do not make an a priori restriction on the S$^3$ gauge field to be turned on.
So we set $A=f(\rho)\omega$, with $\omega$ a generic one-form on the S$^3$.
The field strength is $F=df\wedge \omega+fd\omega$, and we find the components $F_{\rho \alpha}=f^\prime\omega_\alpha$
and $F_{\alpha\beta}=\partial_\alpha\omega_\beta-\partial_\beta\omega_\alpha$.
We only have $4$ non-vanishing components of $F$, which means that the
sum in (\ref{eqn:kappa-1}) terminates at $n=2$.
We thus find
\begin{align}\label{eqn:kappa-with-our-ansatz}
 \Gamma_\kappa\epsilon&=\frac{-i}{\sqrt{\det(1+X)}}\left[
 \Big(\mathds{1}+\frac{1}{8}\gamma^{ijkl}F_{ij}F_{kl}\Big)\Gamma_{(0)}\epsilon
 +\frac{1}{2}\gamma^{ij}F_{ij}C\left(\Gamma_{(0)}\epsilon\right)^\star\right]~.
\end{align}
The pullback of the vielbein to the D7 branes has been given in (\ref{eqn:D7-vielbein}) above,
and we have
\begin{align}\label{eqn:Gamma0-Gammahat}
 \Gamma_{(0)}&= \frac{-1}{\sqrt{1+{\theta^\prime}^2}}\hat\Gamma~,
 &
 \hat\Gamma&=\left[\mathds{1}+\theta^\prime\Gamma_{\underline{\theta}}\Gamma_{\underline{\rho}}\right]\GammaAdS\GammaChi~.
\end{align}
The equations (\ref{eqn:kappa-with-our-ansatz}) and (\ref{eqn:Gamma0-Gammahat}) are the starting point for our analysis in the next subsections.

\subsection{Infinitesimally massive embeddings}\label{sec:small-mass-kappa}

Our construction of supersymmetric embeddings will proceed in two steps.
We first want to know what exactly the preserved supersymmetries are and what the general form of the S$^3$ gauge field is.
These questions can be answered from a linearized analysis, which we carry out  in this section.
With that information in hand, the full non-linear analysis will be easier to carry out, and we come to that in the next section.

So, for now, want to solve the $\kappa$-symmetry condition in a small-mass expansion,
starting from the $\theta\equiv\pi/2$, $F\equiv 0$ massless configuration which we know as solution from the flat slicing.
We expand $\theta=\frac{\pi}{2}+\delta\theta+\dots$ and analogously for $f$.
We use $f$ without explicit $\delta$, though, as it is zero for the massless embedding and there should be no confusion.
The $\kappa$-symmetry condition $\Gamma_\kappa\epsilon=\epsilon$ can then be expanded up to linear order in $\delta m$.
The leading-order equation reads
\begin{align}\label{eqn:kappa-condition-massless}
 \Gamma^{(0)}_\kappa\epsilon^{(0)}&=\epsilon^{(0)}~,&
 \Gamma_\kappa^{(0)}&=i\GammaAdS\GammaChi~,&
 \epsilon^{(0)}&=\epsilon\vert_{\theta=\pi/2}~,
\end{align}
where we use the superscript to indicate the order in the expansion in $\delta m$.
For the next-to-leading order we need to take into account that not only the projector changes, but also the location where the Killing spinor is
evaluated -- the $\kappa$-symmetry condition is evaluated on the D7s. This way we get
\begin{align}\label{eqn:kappa-perturbatively}
 \Gamma_\kappa^{(0)}\epsilon^{(1)}+\Gamma_\kappa^{(1)}\epsilon^{(0)}&=\epsilon^{(1)}~,&
 \epsilon^{(1)}&=\frac{i}{2}\delta\theta\,\Gamma_{}^{\underline{\psi}}\GammaChi\,\epsilon^{(0)}~.
\end{align}

To see which supersymmetries can be preserved, if any, we need to find out under which circumstances the projection
conditions (\ref{eqn:kappa-condition-massless}), (\ref{eqn:kappa-perturbatively}) can be satisfied.
To work this out, we note that we can only impose {\it constant} projection conditions on the constant spinor $\epsilon_0$
that was used to construct the Killing spinors in (\ref{eqn:Killing-spinors-general}):
any projector with non-trivial position dependence would only allow for trivial solutions when imposed on a constant spinor.
For the massless embedding we can straightforwardly find that projector on $\epsilon_0$,
by acting on the projection condition in (\ref{eqn:kappa-condition-massless}) with inverse $R$-matrices.
This gives
\begin{align}\label{eqn:kappa-massless}
 \epsilon_0&=R_{\mathrm{AdS}}^{-1}R_{\mathrm{S}^5}^{-1} \Gamma_\kappa^{(0)}R_{\mathrm{S}^5} R_{\mathrm{AdS}}\epsilon_0
 =-\GammaAdS\Gamma^{\underline{\psi}}\epsilon_0~.
\end{align}
We have used the fact that $\GammaAdS$ commutes with all the $\Gamma$-matrices in the AdS$_5$ part, and also with $R_{\mathrm{S}^5}$.
The last equality holds only when the left hand side is evaluated at $\theta=\pi/2$.
Using that $\Gamma_{11}\epsilon_0=\epsilon_0$, this can be written as a projector involving S$^5$ $\Gamma$-matrices only
\begin{align}\label{eqn:kappa-massless-projector}
 P_{0}\epsilon_0&=\epsilon_0~,& P_{0}&=\frac{1}{2}\big(\mathds{1}+\Gamma^{\underline{\theta}}\GammaChi\big)~.
\end{align}
This is the desired projection condition on the constant spinor: those AdS$_5\times$S$^5$ Killing spinors constructed from (\ref{eqn:Killing-spinors})
with $\epsilon_0$ satisfying (\ref{eqn:kappa-massless-projector}) generate supersymmetries that are preserved by the D3/D7 configuration.
We are left with half the supersymmetries of the AdS$_5\times$S$^5$ background.

\subsubsection{Projection condition at next-to-leading order}

For the small-mass embeddings we expect that additional supersymmetries will be broken, namely those corresponding to
the special conformal supersymmetries in the boundary theory.
We can use the massless condition, (\ref{eqn:kappa-condition-massless}), to simplify the projection
condition (\ref{eqn:kappa-perturbatively}) before evaluating it.
With $\lbrace \Gamma_\kappa^{(0)},\Gamma^{\underline{\psi}}\GammaChi\rbrace=0$
and $\Gamma_\kappa^{(0)}\epsilon^{(0)}=\epsilon^{(0)}$,
we immediately see that $\Gamma_\kappa^{(0)}\epsilon^{(1)}=-\epsilon^{(1)}$.
The next-to-leading-order condition given in (\ref{eqn:kappa-perturbatively}) therefore simply becomes
\begin{align}\label{eqn:kappa-NLO}
 \Gamma_\kappa^{(1)}\epsilon^{(0)}&=2\epsilon^{(1)}
\end{align}
The determinants entering $\Gamma_\kappa$ in (\ref{eqn:kappa-with-our-ansatz}) contribute only at quadratic order, so we find
\begin{align}\label{eqn:Gamma-kappa-1}
 \Gamma_\kappa^{(1)}&=\theta^\prime\Gamma_{\underline{\theta}}\Gamma_{\underline{\rho}}\Gamma_{\kappa}^{(0)}
 -\frac{1}{2}\gamma^{ij}F_{ij}C\big(\Gamma_\kappa^{(0)}\,\cdot\,\big)^\star~.
\end{align}
We use that in (\ref{eqn:kappa-NLO}) and multiply both sides by  $\Gamma^{\underline{\psi}}\GammaChi$.
With $\Gamma_\kappa^{(0)}\epsilon^{(0)}=\epsilon^{(0)}$ and $\GammaS\epsilon^{(0)}=-\GammaAdS\epsilon^{(0)}$,
we find the explicit projection condition
\begin{align}\label{eqn:kappa-NLO-2}
 \left[\delta\theta^\prime\Gamma_{\underline{\rho}}\GammaAdS-i\delta\theta \mathds{1}\right]\epsilon^{(0)}&=
 \frac{1}{2}\Gamma^{\underline{\psi}}\GammaChi\gamma^{ij}F_{ij}C\epsilon^{(0)\star}~.
\end{align}
The left hand side has no $\Gamma$-structures on S$^5$, except for those implicit in the Killing spinor.
We turn to evaluating the right hand side further, and note that
\begin{align}\label{eqn:gammaF}
 \frac{1}{2}\gamma^{ij}F_{ij}&=f^\prime\omega_\alpha \gamma^{\rho\alpha}+f\partial_\alpha\omega_\beta\gamma^{\alpha\beta}~.
\end{align}
For the perturbative analysis, the pullback to the D7 brane for the $\gamma$-matrices is to be evaluated with the zeroth-order
embedding, i.e.\ for the massless $\theta\equiv\pi/2$ one.
Then (\ref{eqn:kappa-NLO-2}) becomes
\begin{align}\label{eqn:kappa-NLO-3}
 \left[\delta\theta^\prime\Gamma_{\underline{\rho}}\GammaAdS-i\delta\theta \mathds{1}\right]\epsilon^{(0)}&=
 \Gamma^{\underline{\psi}}\GammaChi\left[f^\prime\omega_\alpha \Gamma^{\underline{\rho}}\Gamma^{\chi_\alpha}+f\partial_\alpha\omega_\beta\Gamma^{\chi_\alpha\chi_\beta}\right]C\epsilon^{(0)\star}~.
\end{align}
Note that some of the S$^3$ $\Gamma$-matrices on the right hand side 
include non-trivial dependence on the S$^3$ coordinates through the vielbein.

\subsubsection{Next-to-leading order solutions: projector and \texorpdfstring{S$^3$}{S3} harmonic}
We now come to evaluating (\ref{eqn:kappa-NLO-3}) more explicitly, starting with the complex conjugation on $\epsilon^{(0)}$.
Commuting $R_\mathrm{AdS}$ and $R_{\mathrm{S}^5}$ through $C$ acts as just complex conjugation on the coefficients
in (\ref{eqn:Killing-spinors}) and (\ref{eqn:AdS5S5-Killing-from-AdS4-L}).
We define $R$-matrices with a tilde such that $\tilde R_\mathrm{AdS}C=C\RAdS^\star$ and analogously for $\tilde R_{\mathrm{S}^5}$.
Acting on (\ref{eqn:kappa-NLO-3}) with $\RS^{-1}$, we then find
\begin{align}\label{eqn:kappa-NLO-4}
 \left[\delta\theta^\prime\Gamma_{\underline{\rho}}\GammaAdS-i\delta\theta \mathds{1}\right]\RAdS\epsilon_0&=
 \RS^{-1}
 \Gamma^{\underline{\psi}}\GammaChi\left[f^\prime\omega_\alpha \Gamma^{\underline{\rho}}\Gamma^{\chi_\alpha}+f\partial_\alpha\omega_\beta\Gamma^{\chi_\alpha\chi_\beta}\right]\tilde R_{\mathrm{S}^5}\tilde R_\mathrm{AdS} C\epsilon_0^\star~,
\end{align}
The noteworthy feature of this equation is that the left hand side has no more dependence on S$^3$ directions.
To have a chance at all to satisfy this equation, we therefore have to ensure that any S$^3$ dependence drops out on the right hand side as well.
Since $f$ and $f^\prime$ are expected to be independent as functions of $\rho$, this has to happen for each of the two terms individually.
We start with the first one, proportional to $f^\prime$,
and solve for an $\omega$ s.t.\ the dependence on S$^3$ coordinates implicit in the $\Gamma^{\chi_\alpha}$ matrices drops out.
The Clifford-algebra structure on S$^5$ is dictated by the terms we get from evaluating
$\RS^{-1} \Gamma^{\underline{\psi}}\GammaChi \Gamma^{\chi_\alpha}\tilde R_{\mathrm{S}^5}$,
but we want to solve for the coefficients to be constants.
That is, with three constants $c_i$ we solve for
\begin{align}\label{eqn:omega-projection}
\omega_\alpha \RS^{-1} \Gamma^{\underline{\psi}}\GammaChi \Gamma^{\chi_\alpha}\tilde R_{\mathrm{S}^5}P_0
&=c_i\Gamma^{\underline{\chi_i}}\Gamma^{\underline{\psi}}\Gamma^{\underline{\theta}}P_0~.
\end{align}
We only need this equation to hold when acting on $\epsilon^{(0)}$, i.e.\ only when projected on $P_0$.
This fixes $\omega$.
We can find a solution for arbitrary $c_i$, and the generic solution satisfies $\star_{\mathrm{S}^3} d\omega=-(\ell+1)\omega$ with $\ell=1$.
The S$^3$ one-form $\omega$ thus is precisely the $\ell=1,-$ mode we had speculated to find in Sec.~\ref{sec:embedding-ansatz} when we
set up the ansatz, and the bulk analysis indeed reproduces the field-theory results.
This is a result solely about matching symmetries and may not be overly surprising, but it is a nice consistency check anyway.
The solutions parametrized by $c_i$ are equivalent for our purposes, and we choose a simple one with $c_1\,{=}\,1$, $c_2\,{=}\,c_3\,{=}\,0$.
This yields
\begin{align}\label{eqn:omega}
 \omega&=-\cos \chi_2d\chi_1+\sin\chi_1\cos\chi_1 \sin\chi_2d\chi_2+\sin^2\!\chi_1\sin^2\!\chi_2d\chi_3~.
\end{align}
The second term on the right hand side of (\ref{eqn:kappa-NLO-4}) can then easily be evaluated
using that $(\partial_\alpha\omega_\beta)\gamma^{\alpha\beta}= -2\csc\theta\,\omega_\alpha \GammaChi\gamma^{\alpha}$,
for $\omega$ as given in (\ref{eqn:omega}).
The $\kappa$-symmetry condition becomes
\begin{align}\label{eqn:kappa-NLO-5}
 \left[\delta\theta^\prime\Gamma_{\underline{\rho}}\GammaAdS-i\delta\theta \mathds{1}\right]\RAdS\epsilon_0&=
 \left[f^\prime\Gamma_{\underline{\rho}}\GammaAdS+2if\mathds{1}\right]\Gamma^{\underline{\theta}}\Gamma^{\underline{\chi_1}}\tilde R_\mathrm{AdS} C\epsilon_0^\star~.
\end{align}

There are no more S$^5$ $\Gamma$-matrices on the left hand side, so to have solutions those on the right hand side have to drop out as well.
We expected to find at most one fourth of the background supersymmetries preserved,
and now indeed see that we can not get away without demanding an additional projection condition on $\epsilon_0$.
We will demand that
\begin{align}\label{eqn:massive-projector}
  \tilde\Gamma C\epsilon_0^\star&=\lambda\epsilon_0~,&
  \tilde\Gamma&=\Gamma_{\underline{\rho}}\GammaAdS\Gamma^{\underline{\chi_1}}\Gamma^{\underline{\theta}}~,
\end{align}
where $\lambda^\star\lambda=1$.
We can achieve that by setting $\epsilon_0=\eta+\lambda^\star\tilde\Gamma C\eta^\star$, noting that $\tilde\Gamma^2=\mathds{1}$
and $C\tilde\Gamma^\star=\tilde\Gamma C$.
We also see that, due to $[\GammaAdS\Gamma^{\underline{\psi}},\tilde\Gamma]=0$, $\epsilon_0$ satisfies (\ref{eqn:kappa-massless}) if $\eta$ does.
So the two conditions are compatible.
Note that in Majorana-Weyl notation (\ref{eqn:massive-projector}) relates the two spinors to each other,
rather than acting as projection condition on each one of them individually.
This is different from the flat slicing.
With (\ref{eqn:massive-projector}), eq.~(\ref{eqn:kappa-NLO-5}) then becomes
\begin{align}\label{eqn:kappa-NLO-6}
 \left[\delta\theta^\prime\Gamma_{\underline{\rho}}\GammaAdS-i\delta\theta \mathds{1}\right]\RAdS\epsilon_0&=
 \lambda\left[2i f \GammaAdS-f^\prime\Gamma_{\underline{\rho}}\right]\tilde R_\mathrm{AdS} \Gamma_{\underline{\rho}}\epsilon_0~.
\end{align}
There is no dependence on the S$^5$ $\Gamma$-matrices anymore.
As a final step we just act with $\RAdS^{-1}$ on both sides.
To evaluate the result we use the following relation between $\tilde R_\mathrm{AdS}$ and $\RAdS$,
and define a short hand $\tilde\Gamma_{\rho\mathrm{A}}$ as
\begin{align}\label{eqn:Rtilde-R-relation}
 \Gamma_{\underline{\rho}}\tilde R_\mathrm{AdS}\Gamma_{\underline{\rho}}&=e_{}^{-\rho\, i\Gamma_{\underline{\rho}} \GammaAdS}\RAdS~,&
 \tilde\Gamma_{\rho\mathrm{A}}&:=\RAdS^{-1}\Gamma_{\underline{\rho}}\GammaAdS\RAdS~.
\end{align}
We then find that acting with $\RAdS^{-1}$ on (\ref{eqn:kappa-NLO-6}) yields
\begin{align}\label{eqn:kappa-NLO-7}
 \left[\delta\theta^\prime-i\lambda\left(2f\cosh\rho+f^\prime\sinh\rho\right)\right]\tilde\Gamma_{\rho\mathrm{A}}\epsilon_0&=
 \left(i\delta\theta-\lambda f^\prime\cosh\rho-2f\sinh\rho\right)\epsilon_0~.
\end{align}
These are independent $\Gamma$-matrix structures on the left and on the right hand side, so the coefficients have to vanish separately.

The main results for this section are the massive projector (\ref{eqn:massive-projector})
and the one-form on the S$^3$ given in (\ref{eqn:omega}).
They will be the input for the full analysis with finite masses in the next section.
To validate our results so far, we still want to verify that the $\kappa$-symmetry condition (\ref{eqn:kappa-NLO-7})
for small masses can indeed be satisfied with the linearized solutions for $\theta$ and $f$.
The solutions to the linearized equations of motion resulting from (\ref{eqn:DBI-action})
with (\ref{eqn:omega}) (or simply (\ref{eqn:D7-action-theta-f}) below) read
\begin{align}\label{eqn:linear-solutions}
 f&=\mu\sech^2\rho\left(1-\rho\tanh\rho\right)~,&
 \theta&=m\sech\rho\left(\rho\sech^2\rho+\tanh\rho\right)~.
\end{align}
We find that both conditions encoded in (\ref{eqn:kappa-NLO-7}) are indeed satisfied exactly if $i\lambda\mu=m$.
To get a real gauge field, $\lambda$ should be chosen imaginary, which is compatible with consistency of (\ref{eqn:massive-projector}).

Before coming to the finite mass embeddings, we want to better understand the projector (\ref{eqn:massive-projector}).
The massive embedding is expected to break what acts on the conformal boundary of AdS$_5$ as
special conformal supersymmetries, leaving only the usual supersymmetries intact.
Now, what exactly the usual supersymmetries are depends on the boundary geometry.
To explain this point better, we view the \N{4} SYM theory on the boundary as naturally being in
a (fixed) background of \N{4} conformal supergravity.
How the conformal supergravity multiplet and its transformations arise from the AdS supergravity fields
has been studied in detail
for \N{1,2} subsectors in \cite{Balasubramanian:2000pq,Ohl:2010au}.
The Q- and S-supersymmetry transformations of the gravitino in the \N{4} conformal supergravity multiplet schematically take the form
\begin{align}\label{eqn:N4confsugra-trafos}
 \delta_Q \psi_\mu&=D_\mu\epsilon_Q+\dots~,&
 \delta_S \psi_\mu&=i\gamma_\mu\epsilon_S+\dots~,
\end{align}
where the dots denote the contribution from other fields in the multiplet.
Holographically, these transformations arise as follows:
for a local bulk supersymmetry transformation parametrized by a bulk spinor $\epsilon$,
the two classes of transformations arise from the two chiral components of $\epsilon$
with respect to the operator we called $\Gamma_{\underline{\rho}}\GammaAdS$ above \cite{Balasubramanian:2000pq,Ohl:2010au}.
A quick way to make our point is to compare this to the transformation for four-dimensional (non-conformal)
Poincar\'e and AdS supergravities.
They take the form $\delta\psi_\mu=D_\mu\varepsilon$ for Poincar\'e and $\delta\psi_\mu=D_\mu\varepsilon-i\gamma_\mu\varepsilon$
for AdS supergravities.
If we now break conformal symmetry on Minkowski space, we expect to preserve those conformal supergravity transformations
which correspond to the former,
for AdS$_4$ those corresponding to the latter.
From (\ref{eqn:N4confsugra-trafos}) we see that the Poincar\'e supergravity transformations arise purely as Q-supersymmetries.
So holographically we expect a simple chirality projection on the bulk Killing spinor,
of the form $\Gamma_{\underline{\rho}}\GammaAdS\epsilon=\epsilon$,
to give the supersymmetries preserved by a massive D7-brane embedding, and this is indeed the case.
For an AdS$_4$ background, on the other hand, the transformations arise as a particular combination of Q- and S-supersymmetries of the
background \N{4} conformal supergravity multiplet. That means we need both chiral components of the bulk spinor, with specific relations between them.
This is indeed reflected in our projector (\ref{eqn:massive-projector}).\footnote{%
The global fermionic symmetries of \N{4} SYM actually arise from the conformal supergravity transformations as
those combinations of Q- and S-supersymmetries which leave the background invariant.
A more careful discussion should thus be phrased in terms of the resulting (conformal) Killing spinor equations
along similar lines.
For a nice discussion of (conformal) Killing spinor equations on curved space we refer to \cite{Klare:2012gn}.}

\subsection{Finite mass embeddings}\label{sec:fully-massive}
We now turn to the full non-linear $\kappa$-symmetry condition,
i.e.\ with the full non-linear slipping mode and gauge field dependence.
From the linearized analysis we will take the precise form of $\omega$ given in (\ref{eqn:omega}),
and the projection conditions on the constant spinors $\epsilon_0$, (\ref{eqn:kappa-massless-projector}), (\ref{eqn:massive-projector}).
We will assume that $\psi$ is constant, and then set $\psi=0$ w.o.l.g.\ whenever explicit expressions are given.

We have two overall factors in the definition of $\Gamma^{(0)}$ and $\Gamma_\kappa$,
and we pull those out by defining
\begin{align}\label{eqn:h-def}
 h(\rho):=\sqrt{1+{\theta^\prime}^2}\sqrt{\det(1+X)}=
 \sqrt{1+4 f^2 \csc^4\theta}
 \sqrt{1+{\theta^\prime}^2+{f^\prime}^2 \csc^2\theta}~.
\end{align}
For the explicit evaluation we used (\ref{eqn:omega}).
Note that there is no dependence on the S$^3$ coordinates in $h$.
We can then write the $\kappa$-symmetry condition (\ref{eqn:kappa-with-our-ansatz}) as
\begin{align}\label{eqn:kappa-nonlinear-2}
 \Big(\mathds{1}+\frac{1}{8}\gamma^{ijkl}F_{ij}F_{kl}\Big)\hat\Gamma\epsilon
 +\frac{1}{2}\gamma^{ij}F_{ij}\hat\Gamma C\epsilon^\star
 &=-i h\epsilon~,
\end{align}
where we have used $C\hat\Gamma^\star=\hat\Gamma C$ since $\theta$ is supposed to be real.
This compact enough expression will be our starting point,
and we now evaluate the individual terms more explicitly.
With the expression for $\omega$ in (\ref{eqn:omega}), the $F^2$-term evaluates to
\begin{align}\label{eqn:F2-term}
 \frac{1}{8}\gamma^{ijkl}F_{ij}F_{kl}\hat\Gamma&=
 \frac{1}{2}\gamma_{\rho}\gamma_{\chi_1\chi_2\chi_3}\hat\Gamma\epsilon_{ijk}F^{\rho \chi_i}F^{\chi_j\chi_k}
 =
 -2f^\prime f\csc^3\!\theta\Gamma_{\underline{\rho}}\GammaAdS~.
\end{align}
For the last equality we used $\gamma^\rho\GammaChi\hat\Gamma=\Gamma_{\underline{\rho}}\GammaAdS$.
To evaluate $C\epsilon^\star$ in (\ref{eqn:kappa-nonlinear-2}), we recall the definition of $\tilde R_\mathrm{AdS}$
by $C\RAdS=\tilde R_\mathrm{AdS}C$ and analogously for {\RS} (see above (\ref{eqn:kappa-NLO-4})), and use  (\ref{eqn:massive-projector}).
With (\ref{eqn:F2-term}) we then find
\begin{align}\label{eqn:kappa-nonlinear-4}
 \hat\Gamma\epsilon
 +\frac{\lambda}{2}\gamma^{ij}F_{ij}\hat\Gamma \tilde R_{\mathrm{S}^5}\tilde R_\mathrm{AdS}\tilde\Gamma\epsilon_0
 &=2f^\prime f \csc^3\!\theta\,\Gamma_{\underline{\rho}}\GammaAdS\epsilon-ih\epsilon~.
\end{align}
There are no more S$^5$ $\Gamma$-matrices except for those implicit in $\epsilon$ due to {\RS}
on the right hand side, and also no explicit dependence on the S$^3$ coordinates.
So the remaining task is to find out whether we can dispose of all the non-trivial S$^3$ dependence and S$^5$ $\Gamma$-matrices on the
left hand side with just the projectors we already have derived in Sec.~\ref{sec:small-mass-kappa}
-- the amount of preserved supersymmetry and the form of the Killing spinors
are not expected to change when going from infinitesimally small to finite masses.

\subsubsection{Explicit \texorpdfstring{S$^3$}{S3} dependence}
To evaluate the left hand side of (\ref{eqn:kappa-nonlinear-4}) further, we have to work out the term linear in $F$.
With the specific form of $\omega$ given in (\ref{eqn:omega}), we find $(\partial_\alpha\omega_\beta)\gamma^{\alpha\beta}= -2\csc\theta\,\omega_\alpha \GammaChi\gamma^{\alpha}$.
From (\ref{eqn:gammaF}) we then get
\begin{align}
 \frac{1}{2}\gamma^{ij}F_{ij}\hat\Gamma&=
 \left[f^\prime\gamma^\rho-2f\csc\theta \GammaChi\right]\omega_\alpha\gamma^\alpha\hat\Gamma
 =
 -\left[f^\prime\Gamma_{\underline{\rho}}\GammaAdS+2f\csc\theta\,\hat\Gamma\right]\GammaChi\omega_\alpha\gamma^\alpha~.
\end{align}
For the last equality we have used $\big[\GammaChi\Gamma_{\underline{\chi_i}},\hat\Gamma\big]=0$
and $\gamma^\rho\gamma^\alpha\hat\Gamma=-\Gamma_{\underline{\rho}}\GammaAdS\GammaChi\gamma^\alpha$.
With (\ref{eqn:omega}) we easily find the generalization of (\ref{eqn:omega-projection}) to generic $\theta$, and this allows us
to eliminate all explicit S$^3$ dependence. We have
\begin{align}\label{eqn:omega-projection-general-theta}
 \omega_\alpha\gamma^\alpha\tilde R_{\mathrm{S}^5}P_0
 &=
 -\csc\theta\,\Gamma^{\underline{\psi}}\GammaChi \RS \Gamma^{\underline{\theta}}\Gamma^{\underline{\psi}}\Gamma^{\underline{\chi_1}}P_0~.
\end{align}
Since $P_0$ commutes with {\RS} and $\tilde\Gamma$, we can pull it out of $\epsilon_0$ in (\ref{eqn:kappa-nonlinear-4})
and use it when applying (\ref{eqn:omega-projection-general-theta}).
When acting on $\epsilon_0$ as in (\ref{eqn:kappa-nonlinear-4}), we thus find
\begin{align}
 \frac{1}{2}\gamma^{ij}F_{ij}\hat\Gamma \tilde R_{\mathrm{S}^5}&=
 \csc\theta\,\left[f^\prime\Gamma_{\underline{\rho}}\GammaAdS+2f\csc\theta\, \hat\Gamma \right]\Gamma^{\underline{\psi}} \RS\Gamma^{\underline{\theta}}\Gamma^{\underline{\psi}}\Gamma^{\underline{\chi_1}}~.
\end{align}
As desired, the right hand side does not depend on the S$^3$ coordinates anymore.
Using the explicit expression for $\tilde\Gamma$ and the massless projector (\ref{eqn:kappa-massless}), we find
$\Gamma^{\underline{\theta}}\Gamma^{\underline{\psi}}\Gamma^{\underline{\chi_1}}\tilde R_\mathrm{AdS}\tilde\Gamma\epsilon_0=\tilde R_\mathrm{AdS}\Gamma_{\underline{\rho}}\epsilon_0$.
So we get
\begin{align}
 \frac{1}{2}\gamma^{ij}F_{ij}\hat\Gamma \tilde R_{\mathrm{S}^5}\tilde R_\mathrm{AdS}\tilde\Gamma\epsilon_0
 &=
 \csc\theta\,\left[f^\prime\Gamma_{\underline{\rho}}\GammaAdS+2f\csc\theta\, \hat\Gamma \right]\Gamma^{\underline{\psi}} \RS\tilde R_\mathrm{AdS}\Gamma_{\underline{\rho}}\epsilon_0
 \\
 &=
\csc\theta\,\left[f^\prime\Gamma_{\underline{\rho}}\GammaAdS\Gamma^{\underline{\psi}}
 -2f\csc\theta\, \big(\Gamma^{\underline{\theta}}-\theta^\prime\Gamma_{\underline{\rho}}\big)\right] \RS\tilde R_\mathrm{AdS}\Gamma_{\underline{\rho}}\epsilon_0~.
 \label{eqn:gamma-F-term}
\end{align}
For the second equality we have used $\hat\Gamma\Gamma^{\underline{\psi}}=\big(\Gamma^{\underline{\theta}}-\theta^\prime\Gamma_{\underline{\rho}}\big)\Gamma_{11}$.
The second term in round brackets does not have any S$^5$ $\Gamma$-matrices, and can go to the right hand side of (\ref{eqn:kappa-nonlinear-4}).
So combining (\ref{eqn:kappa-nonlinear-4}) with (\ref{eqn:gamma-F-term}), we find
\begin{align}\label{eqn:kappa-nonlinear-5}
\begin{split}
 \mathrm{LHS}:=\hat\Gamma\epsilon
 +&\lambda\csc\theta\,\left[f^\prime\Gamma_{\underline{\rho}}\GammaAdS\Gamma^{\underline{\psi}}
 -2f\csc\theta\, \Gamma^{\underline{\theta}}\right] \RS\tilde R_\mathrm{AdS}\Gamma_{\underline{\rho}}\epsilon_0\\
 &=
 -ih\epsilon+2f^\prime f \csc^3\!\theta\,\Gamma_{\underline{\rho}}\GammaAdS\epsilon
 -2\lambda f\theta^\prime \csc^2\!\theta\,\Gamma_{\underline{\rho}}\RS\tilde R_\mathrm{AdS}\Gamma_{\underline{\rho}}\epsilon_0=:\mathrm{RHS}~.
\end{split}
\end{align}
Nicely enough, the left hand side is linear in the gauge field and its derivative -- it appears non-linearly only on the right hand side.

\subsubsection{Solving the \texorpdfstring{$\kappa$}{kappa}-symmetry condition}
The $\kappa$-symmetry condition (\ref{eqn:kappa-nonlinear-5}) still has coordinate dependences implicit in $\epsilon$, through $\RAdS$ and $\RS$.
To eliminate those, we want to act with $\RS^{-1}\RAdS^{-1}$ on both sides, and evaluate the result.
That is cumbersome, and we derive the required identities in App.~\ref{sec:s5-gamma-identities}.
For notational convenience, we define the operator $\mathcal R[\Gamma]:=\RS^{-1}\Gamma\RS$.
With (\ref{eqn:LHS-aux-1}), (\ref{eqn:LHS-aux-2}) and (\ref{eqn:LHS-aux-3}),
we can then evaluate (\ref{eqn:kappa-nonlinear-5}) explicitly.
For the left hand side we find
\begin{align}\label{eqn:kappa-nonlinear-5-LHS}
\begin{split}
 \RAdS^{-1}\RS^{-1}\mathrm{LHS}=&
 \left(i\cot\theta+\lambda\csc\theta f^\prime\cosh\rho+2f\lambda\csc^3\theta\sinh\rho\right)\mathcal R[\Gamma^{\underline{\theta}}\GammaChi]\epsilon_0\\
 &+\left(\theta^\prime-i\lambda\csc\theta f^\prime \sinh\rho-2i f \lambda \csc^3\theta\cosh\rho\right)\mathcal R[\Gamma^{\underline{\theta}}\GammaChi]\tilde\Gamma_{\rho\mathrm{A}}\epsilon_0\\
 &-i\csc\theta\epsilon_0+2f\lambda\csc^2\theta\cot\theta\left(i\cosh\rho\,\tilde\Gamma_{\rho\mathrm{A}}-\sinh\rho\,\mathds{1}\right)\epsilon_0~,
 \end{split}
\end{align}
where $\tilde\Gamma_{\rho\mathrm{A}}$ was defined in (\ref{eqn:Rtilde-R-relation}).
Note that the RHS in (\ref{eqn:kappa-nonlinear-5}) has no dependence on the S$^3$-directions,
but $\mathcal R[\Gamma^{\underline{\theta}}\GammaChi]$ in (\ref{eqn:kappa-nonlinear-5-LHS}) does.
So the coefficients of the two terms involving $\mathcal R[\Gamma^{\underline{\theta}}\GammaChi]$ in (\ref{eqn:kappa-nonlinear-5-LHS}) have to vanish.
Moreover, they have to vanish separately, since they multiply different AdS$_5$ $\Gamma$-matrix structures.
So we find the two conditions
\begin{subequations}\label{eqn:f-theta-nonlinear-1}
\begin{align}
 i\cot\theta+\lambda\csc\theta f^\prime\cosh\rho+2f\lambda\csc^3\theta\sinh\rho&=0~,\\
 \theta^\prime-i\lambda\csc\theta f^\prime \sinh\rho-2i f \lambda \csc^3\theta\cosh\rho&=0~.
\end{align}
\end{subequations}
The non-trivial Clifford-algebra structure of the $\kappa$-symmetry condition has thus given us two independent $1^\mathrm{st}$-order
differential equations.
Moreover,
since $f$ and $f^\prime$ only appear linearly, we can actually solve (\ref{eqn:f-theta-nonlinear-1}) for $f$ and $f^\prime$.
This yields
\begin{align}\label{eqn:f-fprime-solutions}
 f&=\frac{i}{2\lambda}\sin^3\theta\left(\sinh\rho\,\cot\theta-\theta^\prime\cosh\rho\right)~,
 &
 f^\prime&=\frac{i}{\lambda}\left(\theta^\prime\sin\theta\sinh\rho-\cosh\rho\cos\theta\right)~.
\end{align}
Note that the expression for $f^\prime$ does not contain second-order derivatives of $\theta$,
which we would get if we just took the expression for $f$ and differentiate.
Comparing the expressions for $f$ and $f^\prime$, we can thus derive a second-order ODE for $\theta$ alone.
It reads
\begin{align}\label{eqn:theta-second-order-ode}
 \theta^{\prime\prime}+3 {\theta^\prime}^2 \cot\theta+4 \tanh\rho\, \theta^\prime-\cot\theta \left(1+2 \csc ^2\theta\right)=0~.
\end{align}

With the solutions for $f$ and $f^\prime$ s.t.\ the first two lines of (\ref{eqn:kappa-nonlinear-5-LHS}) vanish, the
$\kappa$-symmetry condition (\ref{eqn:kappa-nonlinear-5}) simplifies quite a bit.
Collecting the remaining terms according to their $\Gamma$-matrix structure gives
\begin{align}\label{eqn:kappa-remaining}
\begin{split}
 0=&\left[i h-i\csc\theta+2f\lambda\csc^2\theta\left(\theta^\prime\cosh\rho-\cot\theta\sinh\rho\right)\right]\epsilon_0\\
 &-
 2f\csc^2\theta
 \left[f^\prime\csc\theta+i\lambda\left(\theta^\prime\sinh\rho-\cot\theta\cosh\rho\right)\right]
 \tilde\Gamma_{\rho\mathrm{A}}\epsilon_0~.
\end{split}
\end{align}
With the solution for $f^\prime$ in terms of $\theta$ given in (\ref{eqn:f-fprime-solutions}),
we see that the term in square brackets in the second line vanishes
exactly if $\lambda$ is purely imaginary, s.t.\ $\lambda^{-1}=-\lambda$.
So we are left with the first line only.
This once again vanishes when plugging in the explicit expressions of (\ref{eqn:h-def}) and (\ref{eqn:f-fprime-solutions}),
and using imaginary $\lambda$.
So any solution for the slipping mode satisfying (\ref{eqn:theta-second-order-ode}), which is accompanied
by the gauge field (\ref{eqn:f-fprime-solutions}), gives a supersymmetric D7-brane embedding into AdS$_4$-sliced AdS$_5$.
These equations are our first main result.

As a consistency check, one wants to verify that each such combination
of slipping mode satisfying (\ref{eqn:theta-second-order-ode}) with gauge field (\ref{eqn:f-fprime-solutions})
indeed satisfies the highly non-linear and coupled
equations of motion resulting from the D7-brane action.
To derive those, we first express (\ref{eqn:DBI-action}) explicitly in terms of $\theta$ and $f$.
That is, we use $A=f\omega$ with $\omega$ given in (\ref{eqn:omega}), but not any of the other $\kappa$-symmetry relations.
Also, for $\omega$ we only use that our $\omega$ satisfies  $\star_{\mathrm{S}^3} d\omega=-2\omega$,
i.e.\ that we found an $\ell=1,-$ mode in the language of \cite{Kruczenski:2003be}.
The combination of DBI action and WZ term then becomes
\begin{align}\label{eqn:D7-action-theta-f}
 S_\mathrm{D7}&=-T_7 V_{\mathrm{S}^3}\int d^5\xi\sqrt{g_{\mathrm{AdS}_4}} \left[
 \zeta^\prime\sqrt{\sin^4\theta+4f^2}\sqrt{{f^\prime}^2+(1+{\theta^\prime}^2)\sin^2\theta}
 +8 \zeta f^\prime f\right]~,
\end{align}
where $\zeta^\prime=\cosh^4\rho$ as defined in (\ref{eqn:RR-gauge-field}), and we have integrated over the S$^3$.
Working out the resulting equations of motion gives two coupled second-order non-linear equations.
In the equation for the slipping mode one can at least dispose of the square root, by a suitable rescaling of the equation.
But for the gauge field even that is not possible, due to the WZ term.
The resulting equations are bulky, and we will not spell them out explicitly.
Finding an analytic solution to these equations right away certainly seems hopeless.
But we do find that using (\ref{eqn:f-fprime-solutions}) to replace $f$, along with
replacing $\theta^{\prime\prime}$ using (\ref{eqn:theta-second-order-ode}), actually
solves both of the equations of motion resulting from (\ref{eqn:D7-action-theta-f}).

\subsection{Solutions}\label{sec:fully-massive-embedding-solution}

We now have a decoupled equation for the slipping mode alone in (\ref{eqn:theta-second-order-ode}), and an immediate solution for the accompanying $f$
in (\ref{eqn:f-fprime-solutions}).
So it does not seem impossible to find an explicit solution for the embedding in closed form.
To simplify (\ref{eqn:theta-second-order-ode}), we reparametrize the slipping mode as
$\cos\theta(\rho)=2\cos\left(\frac{1}{3}\cos^{-1}\tau(\rho)\right)$,
which turns it into a simple linear equation for $\tau$. Namely,
\begin{align}
\tau^{\prime\prime}+4 \tanh\rho\, \tau^\prime+3 \tau&=0~.
\end{align}
This can be solved in closed form, and as a result we get three two-parameter families of solutions for $\theta$,
corresponding to the choice of branch for the $\cos^{-1}$.
Restricting $\cos^{-1}$ to the principle branch, where it takes values in $[0,\pi]$, we can write them as
\begin{align}\label{eqn:soltheta}
 \theta&=\cos^{-1}\left(2\cos\frac{2\pi k+\cos^{-1}\tau}{3}\right)~,&
 \tau&=\frac{6(m \rho-c) +3 m \sinh (2 \rho )}{4\cosh^3\rho}~,
\end{align}
with $k\in\lbrace 0,1,2\rbrace$.
Only $k=2$ gives real $\theta$, though:
To get real $\theta$, we need
$|\cos\frac{2\pi k+\cos^{-1}\tau}{3}|\leq \frac{1}{2}$.
This translates to $\cos^{-1}\tau\in [\pi,2\pi]+(3n-2k)\pi$.
Since we have chosen the branch with $\cos^{-1}\tau\in[0,\pi]$ in (\ref{eqn:soltheta}), this only happens for $k=2$.
For $\rho\rightarrow\infty$ we then have $\tau\rightarrow 0$ and $\theta\rightarrow \frac{\pi}{2}$, so the branes wrap an equatorial S$^3$ in the S$^5$.
As $\rho$ is decreased, $\tau$ increases and the branes potentially cap off -- we need $|\tau|\leq 1$ to have real $\theta$.
The remaining constant $c$ may then be fixed from regularity constraints, and we will look at this in more detail below.
These are finally the supersymmetric embeddings we were looking for:
the slipping mode $\theta$ given in (\ref{eqn:soltheta}) with $k=2$, accompanied by the gauge field $A=f\omega$,
with $f$ given (\ref{eqn:f-fprime-solutions}) and $\omega$ in (\ref{eqn:omega}).
The naming of the constants is anticipating our results for the one-point functions in  (\ref{eqn:theta-near-boundary}) below:
$m$ will be the flavor mass in the boundary theory and $c$ will appear in the chiral condensate.

\section{Topologically distinct classes of embeddings}\label{sec:embedding-classes}
In the previous section we have obtained the general solution to the $\kappa$-symmetry condition,
giving the two-parameter family of embeddings in (\ref{eqn:soltheta}) with the accompanying gauge field (\ref{eqn:f-fprime-solutions}).
In this section we will study the parameter space $(m,c)$, and whether and where the branes cap off depending on these parameters.
A crucial part in that discussion will be demanding regularity of the configurations,
e.g.\ that the worldvolume of the branes does not have a conical singularity
and a similar condition for the worldvolume gauge field.

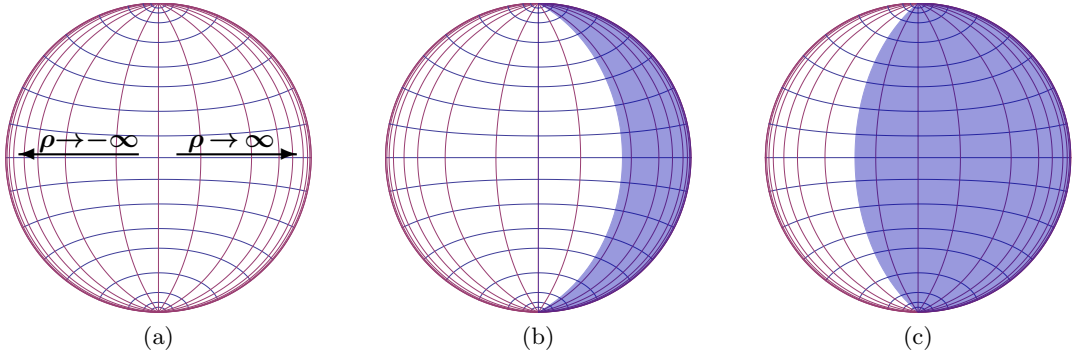
\begin{figure}[htb]
\center
\subfigure[][]{ \label{fig:ads-sliced-ads}
  \includegraphics[width=0.25\linewidth]{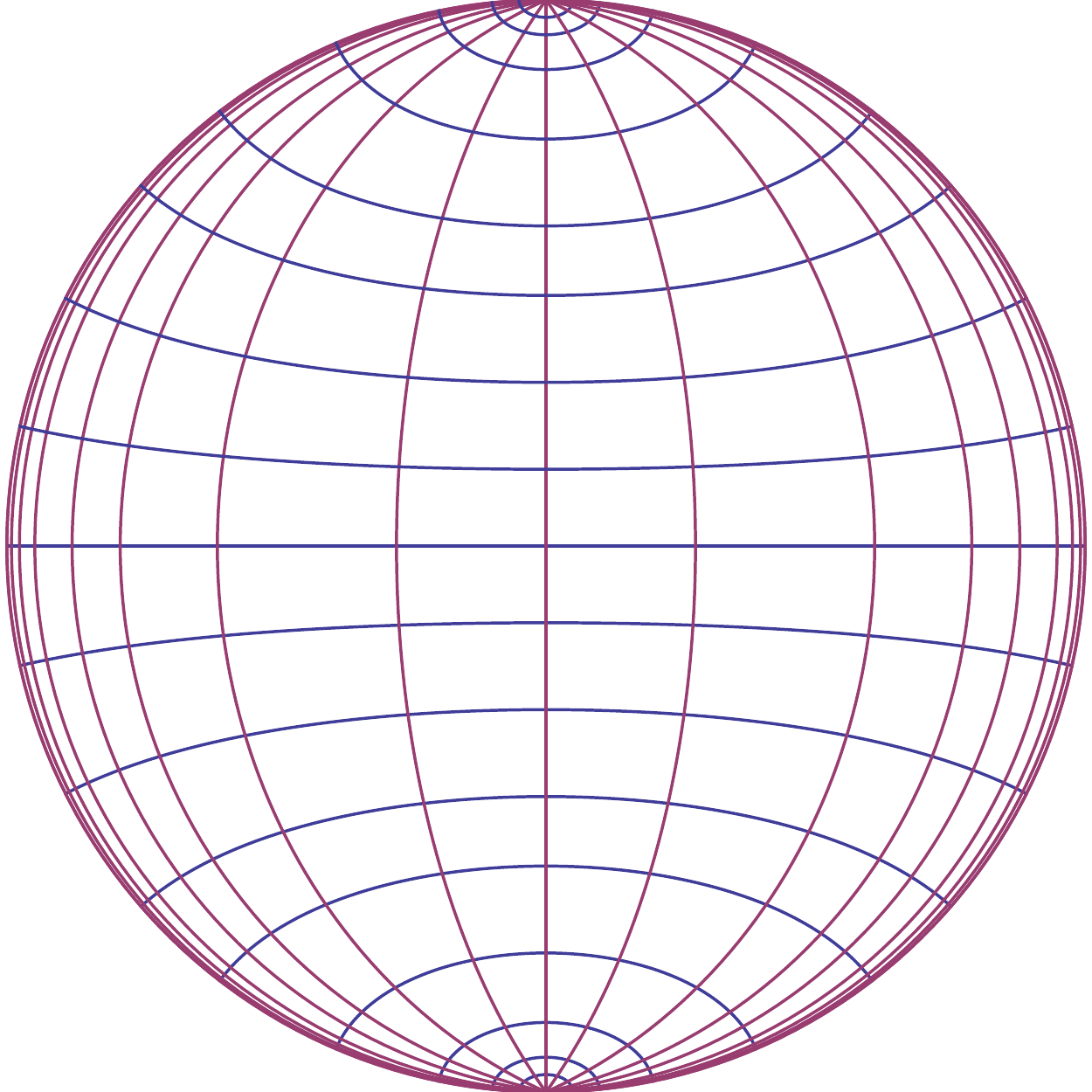}
}\qquad
\subfigure[][]{ \label{fig:D7-short-embedding}
  \includegraphics[width=0.25\linewidth]{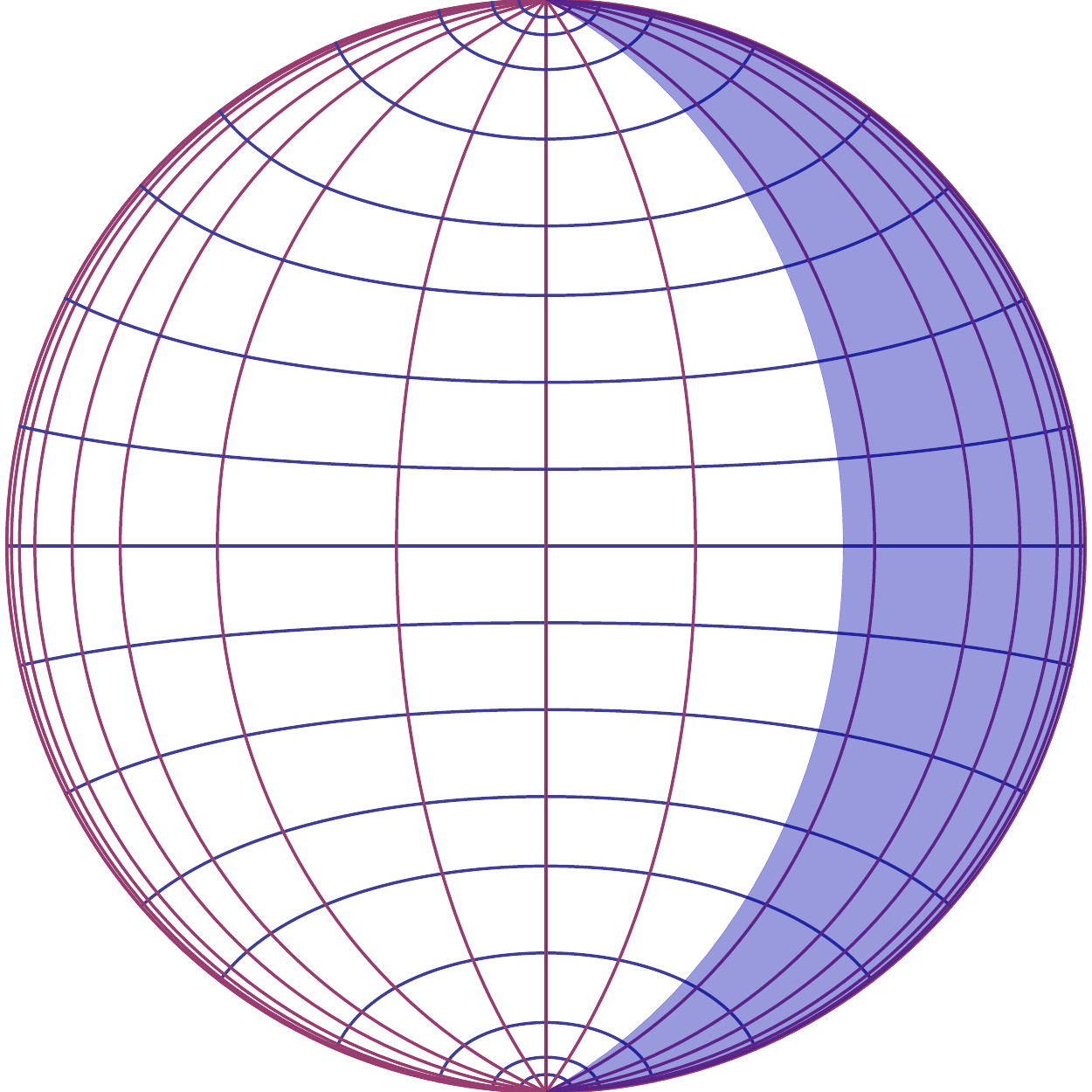}
}\qquad
\subfigure[][]{ \label{fig:D7-long-embedding}
  \includegraphics[width=0.25\linewidth]{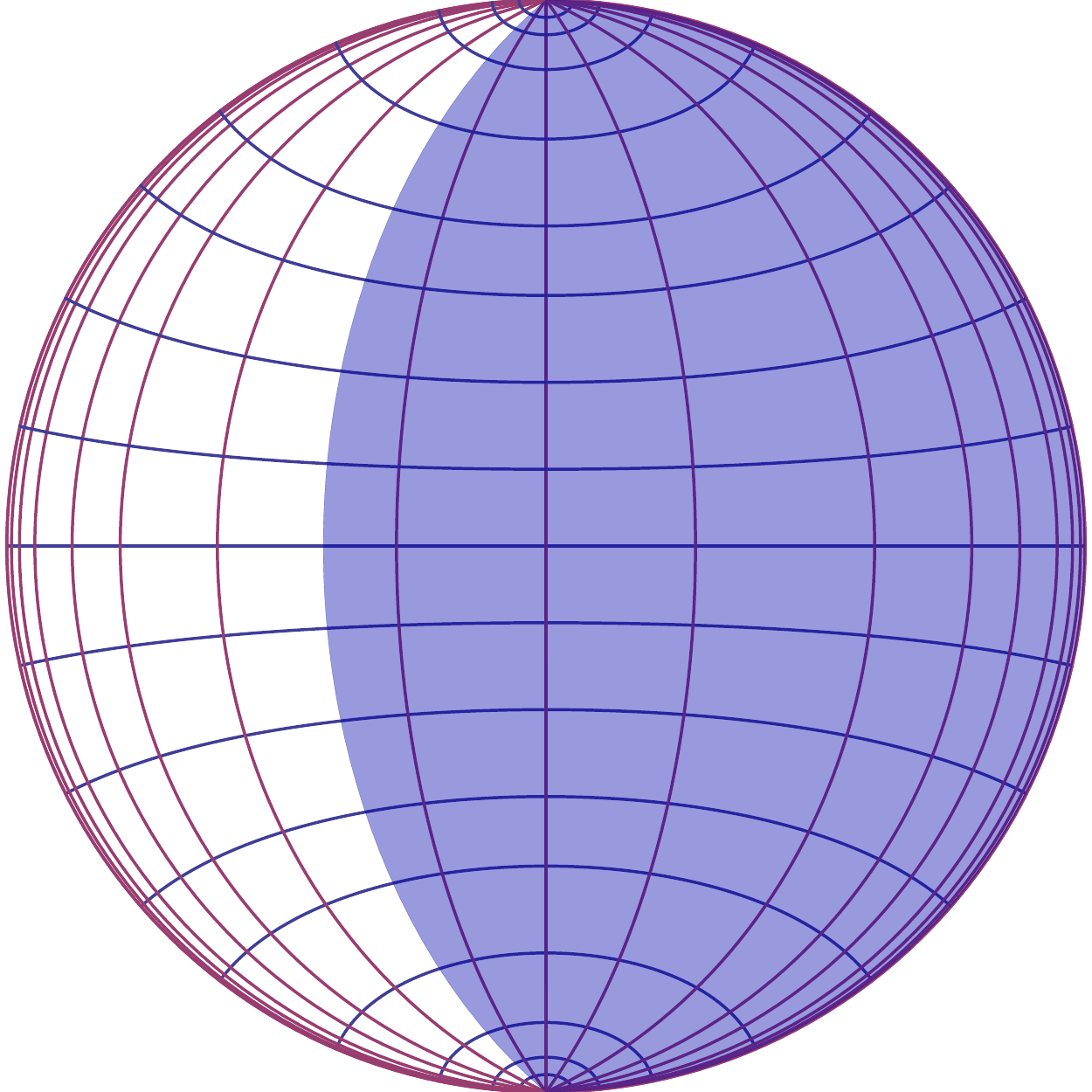}
}
\begin{picture}(0,0)
\setlength{\unitlength}{1cm}
\thicklines
\put(-12.00,2.2){\boldmath{$\rho\,{\to}\, \infty$}}
\put(-12.15,2.1){\vector(1,0){1.6}}
\put(-13.98,2.2){\boldmath{$\rho {\to} {-}\infty$}}
\put(-12.65,2.1){\vector(-1,0){1.6}}
\end{picture}
\caption{
The left hand side illustrates how global AdS$_5$ is sliced by AdS$_4$ slices.
The vertical lines are slices of constant $\rho$, the horizontal ones correspond to constant AdS$_4$ radial coordinate.
Connected embeddings cover the entire AdS$_5$ and stretch out to the conformal boundary in both patches.
The figure in the middle illustrates what we call a short embedding, where the branes, wrapping the shaded region,
stretch out to the conformal boundary in one patch and then cap off at a finite value of the radial coordinate in the same patch.
On the right hand side is what we call a long embedding, where the branes stretch out to one half of the conformal boundary,
cover the entire patch and cap off in the other one.
\label{fig:D7-embeddings}}
\end{figure}

To cover either of global or Poincar\'{e} AdS$_5$ with AdS$_4$ slices,
we need two coordinate patches with the corresponding choice of global or Poincar\'{e} AdS$_4$ slices, as illustrated in Fig.~\ref{fig:D7-embeddings}.
They can be realized by just letting $\rho$ run through the entire $\mathbb{R}$.
The figure illustrates global AdS, but we do not need to commit to one choice at this point.
For the massless embeddings in Poincar\'{e} AdS, where $\theta\equiv \pi/2$ is the known supersymmetric solution,
the D7 branes wrap all of the AdS$_5$ part.
For the massive case, again from Poincar\'{e}-AdS$_5$ intuition, we naively expect this to be different.
However, that discussion will turn out to be more nuanced for AdS$_4$-sliced AdS$_5$.

The two options for the branes to cap off are the (arbitrarily assigned) north and south poles of the S$^5$,
which we take as $\theta=0$ or $\tau=-1$ and $\theta=\pi$ or $\tau=1$, respectively.
With (\ref{eqn:soltheta}), the condition for the branes to cap off at the north/south pole at $\rho=\rho_\star$ then becomes
\begin{align}\label{eqn:disconnected-c-from-rho-s}
 c&=m\rho_\star +\frac{1}{2}m\sinh(2\rho_\star) \pm \frac{2}{3}\cosh^3\rho_\star=:c_\mathrm{n/s}(\rho_\star)~.
\end{align}
There is no a priori relation between the masses we choose for the two patches, so we start the discussion from one patch, say the one with $\rho\geq0$.
For $\rho\rightarrow\infty$ we have $\theta\rightarrow\pi/2$, and what happens as we move into the bulk depends on whether and what sort of solutions $\rho_\star$
there are to (\ref{eqn:disconnected-c-from-rho-s}). Depending on $m$ and $c$, we can distinguish 3 scenarios:
\begin{itemize}
 \item[(i)] There is a $\rho_\star\geq0$ such that either $c=c_\mathrm{n}(\rho_\star)$ or $c=c_\mathrm{s}(\rho_\star)$.
            In that case, the branes cap off in the patch in which they started, and we call this a short embedding.
 \item[(ii)] There is no $\rho_\star\geq 0$ as above, but there is a $\rho_\star<0$ such that $c=c_\mathrm{n}(\rho_\star)$ or $c=c_\mathrm{s}(\rho_\star)$.
             In that case, the branes cover the entire $\rho>0$ patch and part of the $\rho<0$ patch. We call this a long embedding.
 \item[(iii)] We have $c\neq c_\mathrm{n}(\rho_\star)$ and $c\neq c_\mathrm{s}(\rho_\star)$ for all $\rho_\star\in\mathbb{R}$.
              In that case the branes never reach either of the poles and do not cap off at all.
              They cover all of AdS$_5$, connecting both parts of the conformal boundary. We call these connected embeddings.
\end{itemize}
The types of embeddings are illustrated in Fig.~\ref{fig:D7-embeddings}, and we will study them in more detail below.
If the branes do cap off, demanding regularity at the cap-off point imposes an additional constraint, and we find one-parameter families.
However, the masses in the two patches can then be chosen independently, and one 
can also combine e.g.\ a long embedding in one patch with a short one in the other.
For the connected embeddings there is no such freedom, and the flavor masses on the two copies of AdS$_4$ are related.

\subsection{Cap-off points}
If a cap-off point exists at all, the branes should cap off smoothly.
The relevant piece of the induced metric to check for a conical singularity is $g=(1+{\theta^\prime}^2)d\rho^2+\sin^2\theta d\Omega_3^2+\dots$.
Expanding around $\rho=\rho_\star$ with (\ref{eqn:soltheta}) and (\ref{eqn:disconnected-c-from-rho-s}) gives
\begin{align}\label{eqn:tau-expansion}
 \sin\theta\big\vert_{c=c^\star_\mathrm{n/s}(\rho_\star)}&=\alpha_\mathrm{n/s}(\rho-\rho_\star)^{1/4}+\mathcal O(\sqrt{\rho-\rho_\star})~,&
 \alpha_\mathrm{n/s}&=\sqrt[4]{8\sech\rho_\star(\sinh\rho_\star\pm m)}~.
\end{align}
The induced metric with that scaling is smooth without a conical singularity.
To examine the regularity of the gauge field $A=f\omega$,
we fix $\chi_1=\chi_2=\pi/2$, s.t.\ we look at a plane around $\rho=\rho_\star$.
The pullback of the gauge field to the plane is $A=f d\chi_1$,
and regularity at the origin, $\rho=\rho_\star$, demands $f(\rho_\star)=0$.
For small $\sin\theta$, we see from (\ref{eqn:f-fprime-solutions}) that
\begin{align}
 f=-\frac{i}{2\lambda}\cosh\rho\,\theta^\prime\sin^3\theta+\mathcal O(\sin^2\theta)~.
\end{align}
From the expansion (\ref{eqn:tau-expansion}),
we then find that $f(\rho_\star)=0$ for branes capping off at the north/south pole translates to
\begin{align}\label{eqn:disconnected-m-rhos}
 \theta^\prime\sin^3\theta\big\vert_{\rho=\rho_\star}&=\pm \frac{1}{4}\alpha_\mathrm{n/s}^4\stackrel{!}{=}0
 &\Longleftrightarrow&
 &\rho_\star&=\rho_\mathrm{n/s}~,&\rho_\mathrm{n/s}&=\mp \sinh^{-1}\!m~.
\end{align}
We thus find that for any given $m$ there are in principle two options for the branes to cap off smoothly.
For positive $m$ they can potentially cap off smoothly at the south pole in the $\rho>0$ patch
and at the north pole in the $\rho<0$ patch, and for negative $m$ the other way around.
With $\alpha_\mathrm{n/s}$ fixed like that, the slipping mode shows the usual square root behavior as it approaches the north/south pole.

The constraint (\ref{eqn:disconnected-m-rhos}) can also be obtained from the on-shell action.
From the $\kappa$-symmetry discussion we know that 
the combination in square brackets in the first line of (\ref{eqn:kappa-remaining}) vanishes,
which implies that on shell
\begin{align}\label{eqn:h-from-kappa}
 h&=\csc\theta\left(1-4\lambda^2f^2\csc^4\!\theta\right)~.
\end{align}
This allows us to eliminate the square root in the on-shell DBI Lagrangian, which will also be useful for the discussions below.
The DBI Lagrangian of (\ref{eqn:DBI-action}) expressed in terms of $h$ reads
\begin{align}\label{eqn:DBI-Lagrangian-h}
 L_\mathrm{DBI}&=-T_7\cosh^4\!\rho\,\sin^3\!\theta\,\cdot h\cdot \sqrt{g^{}_{\mathrm{AdS}_4}}\sqrt{g^{}_{\mathrm{S}^3}}~,
\end{align}
where $\sqrt{g^{}_{\mathrm{AdS}_4}}$ and $\sqrt{g^{}_{\mathrm{S}^3}}$ are the standard volume elements on AdS$_4$ and S$^3$ of unit curvature radius, respectively.
For the full D7-brane action (\ref{eqn:DBI-action}), we then find
\begin{align}\label{eqn:D7-action-simplified}
 S_\mathrm{D7}&=-T_7\int d^7\xi d\rho\,\sqrt{g^{}_{\mathrm{AdS}_4}}\sqrt{g^{}_{\mathrm{S}^3}} \left[h\cosh^4\!\rho\,\sin^3\!\theta\,+8 \zeta\, f^\prime f\right]~.
\end{align}
To have the first term in square brackets finite at $\rho=\rho_\star$ we once again need $f(\rho_\star)=0$,
leading to (\ref{eqn:disconnected-m-rhos}).
We will look at the two options for the branes to cap off smoothly in more detail now.

\subsection{Short embeddings}\label{sec:embeddings-short}
The first option we want to discuss are the short embeddings
illustrated in Fig.~\ref{fig:D7-short-embedding}, 
where the branes cap off in the same patch in which they reach out to the conformal boundary.
This kind of embedding can be realized for arbitrary $m$:
For the $\rho>0$ patch,
we simply take $\rho_\mathrm{s/n}$ from (\ref{eqn:disconnected-m-rhos})
and fix
\begin{align}\label{eqn:short-embedding}
 c&=c_\mathrm{s}(\rho_\mathrm{s})\quad \text{for $m>0$}~,
 &c&=c_\mathrm{n}(\rho_\mathrm{n})\quad \text{for $m<0$}~,
\end{align}
with $c_\mathrm{n/s}$ defined in (\ref{eqn:disconnected-c-from-rho-s}).
This gives a smooth cap-off point at $\rho=\rho_\mathrm{s/n}\geq 0$ -- in the same patch where we assumed the D7 branes to extend to the conformal boundary.
For the other patch the choices are simply reversed.

There is a slight subtlety with that, though, which gives us some useful insight into the curves $c_\mathrm{s/n}(\rho_\star)$.
For the embeddings to actually be smooth, there must be no additional cap-off points between the conformal boundary and 
the smooth cap-off point at $\rho_\mathrm{s/n}$.
This is indeed not the case, which can be seen  as follows.
For given $(m,c)$, the cap-off points are determined as solutions $\rho_\star$ to (\ref{eqn:disconnected-c-from-rho-s}),
so we want to look at $c_\mathrm{n/s}$ as functions of $\rho_\star$.
The specific values $\rho_\mathrm{n/s}$, found from regularity considerations above, are also the only extrema of the curves $c_\mathrm{n/s}$.
\begin{figure}[htb]
\center
\subfigure[][]{
  \includegraphics[width=0.46\linewidth]{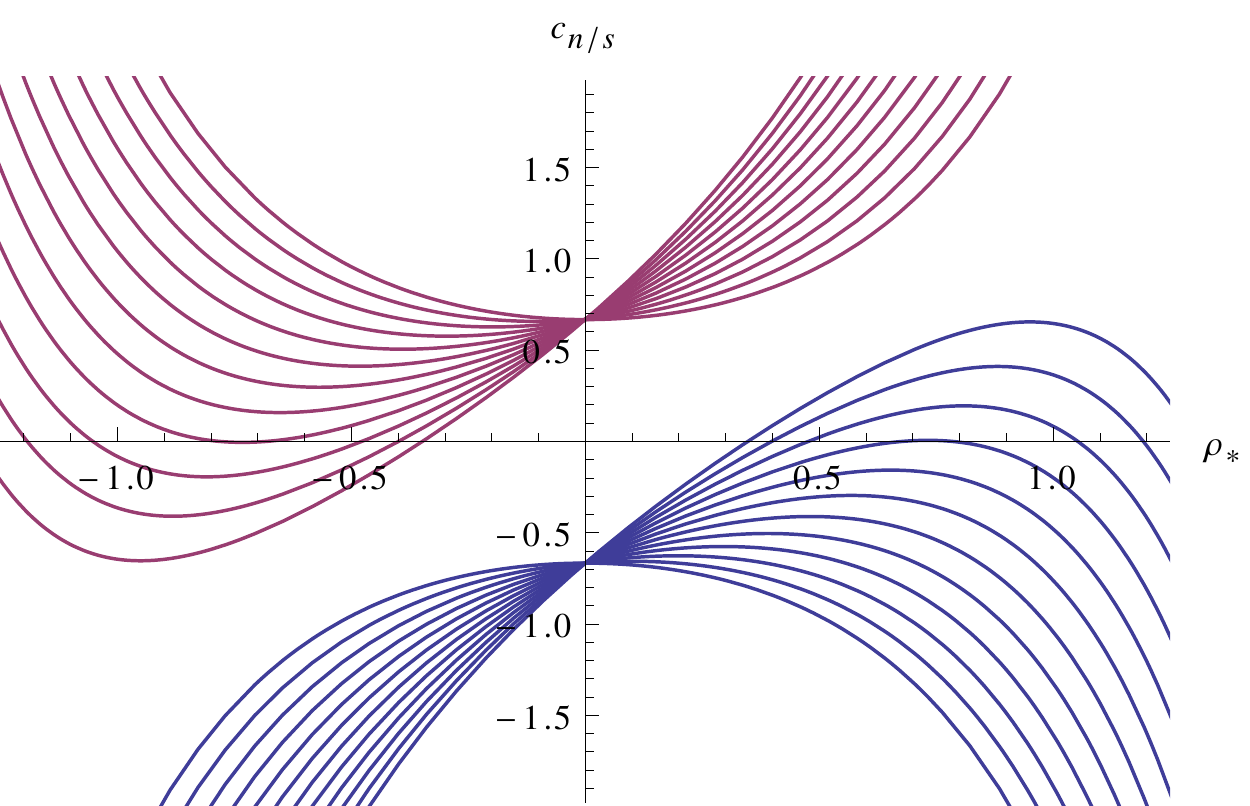} \label{fig:cap-off-cNS-branches}
}\qquad
\subfigure[][]{
  \includegraphics[width=0.46\linewidth]{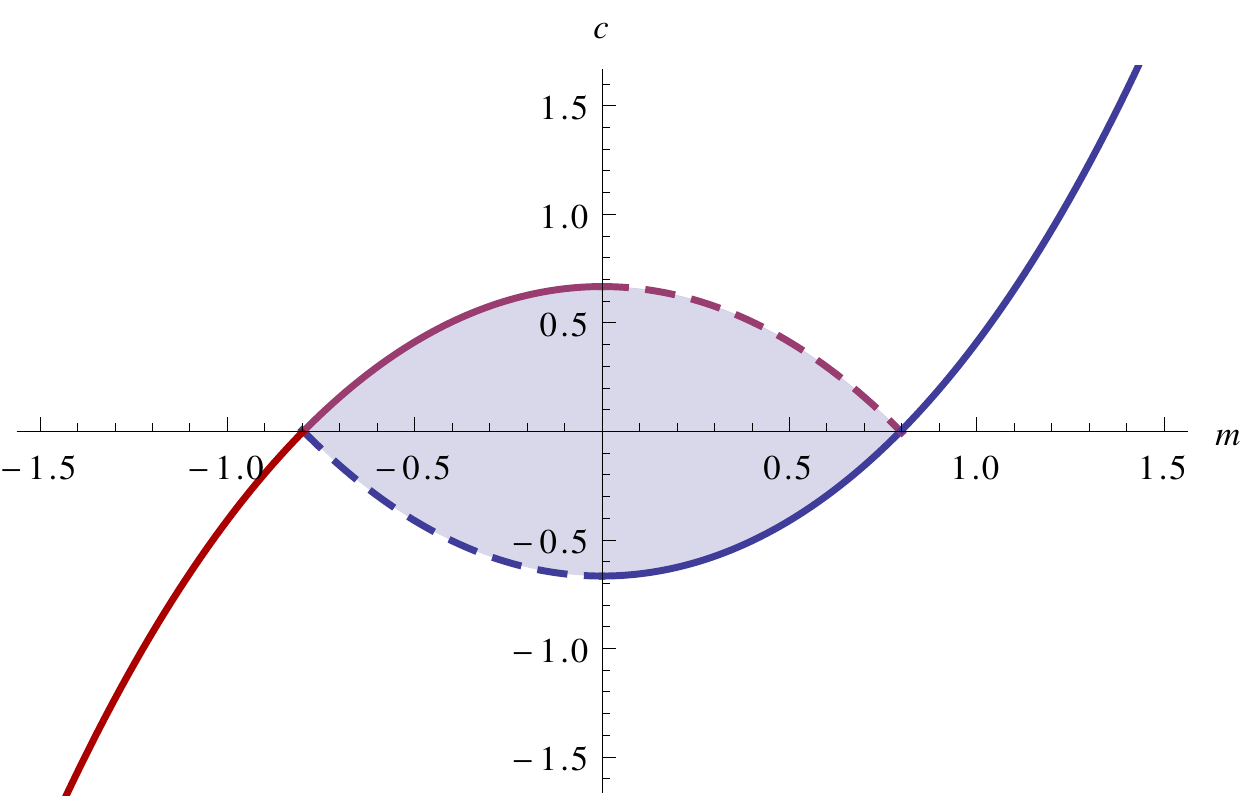} \label{fig:m-c-diagram}
}
\caption{
The plot on the left hand side shows $c_\mathrm{n}$ and $c_\mathrm{s}$, as defined
in (\ref{eqn:disconnected-c-from-rho-s}), as the family of red upper and blue lower curves, respectively.
The symmetric curves are $m=0$ for both, and as $m$ is increased the curves tilt.
For $m=0$ and $|c|\leq \frac{2}{3}$,
there is no solution $\rho_\star$ to (\ref{eqn:disconnected-c-from-rho-s}), and the branes do not cap off.
As $m$ is increased, the maximum value taken on the lower curve increases and the minimum taken on the upper curve decreases.
So the window for $c$ to get continuous embeddings shrinks.
The plot on the right hand side shows the smooth brane embeddings in the $(m,c)$ plane.
For large $m$ these are given by the thick blue curve only, which corresponds to the disconnected embeddings.
For lower $m$ the blue-shaded region is possible and corresponds to connected embeddings.
On the dashed lines the embedding covers all of one patch and caps off smoothly in the other one.
The $\theta\equiv\frac{\pi}{2}$ embedding corresponds to $c=m=0$.
$\mathbb{Z}_2$-symmetric connected embeddings correspond to the axes, i.e.\ $c=0$ or $m=0$ in this plot.
\label{fig:m-c-plots}
}
\end{figure}
That is, $c_\mathrm{n}$ has a minimum at $\rho_\mathrm{n}$ and $c_\mathrm{s}$ has a maximum at $\rho_\mathrm{s}$.
That means we only get that one smooth cap-off point from the curve we used to set $c$ in (\ref{eqn:short-embedding}).
Moreover, in the patch where $c_\mathrm{n/s}$ take their minimum/maximum, $c_\mathrm{s/n}$ is always
strictly smaller/greater than $c_\mathrm{n/s}$.
This ensures that there are no cap-off points in between coming from the other curve either.
See Fig.~\ref{fig:cap-off-cNS-branches} for an illustration.

\begin{figure}[htb]
\center
\subfigure[][]{
  \includegraphics[width=0.40\linewidth]{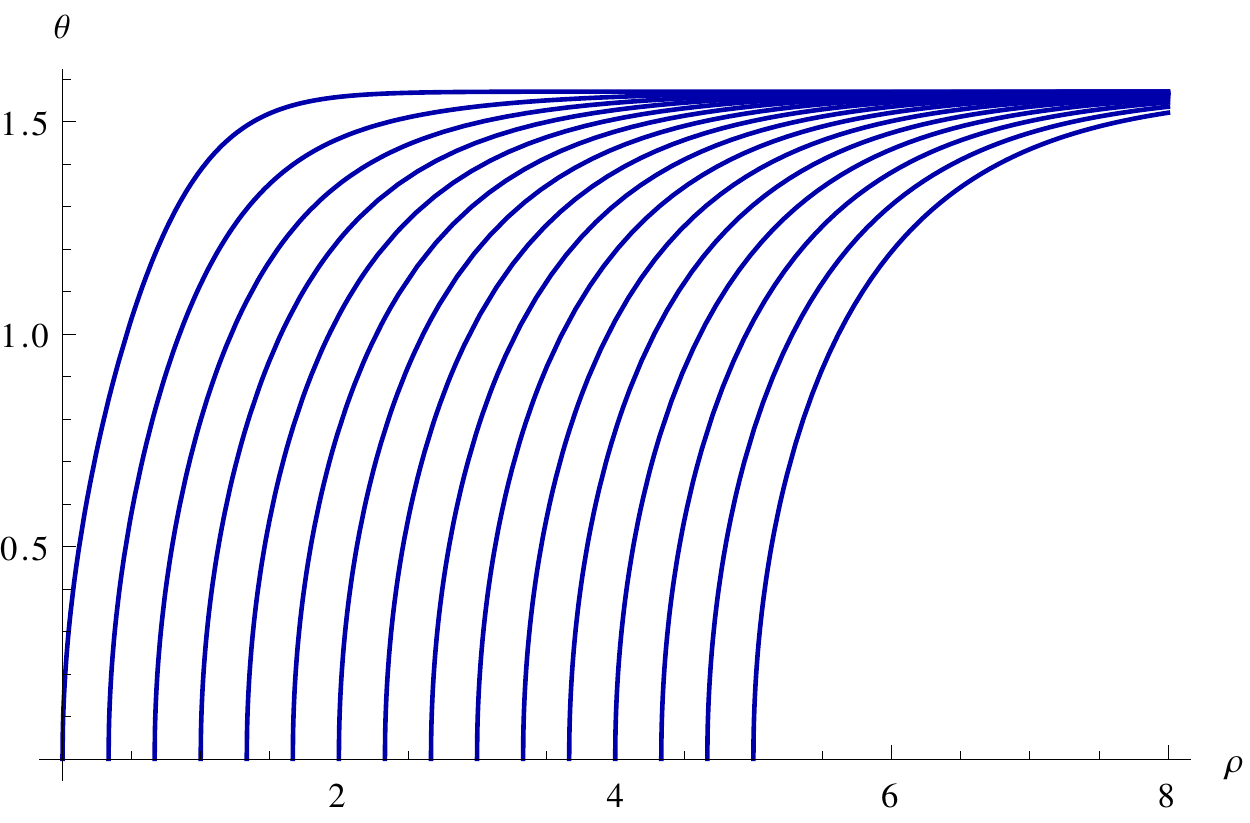}
}\qquad
\subfigure[][]{
    \includegraphics[width=0.40\linewidth]{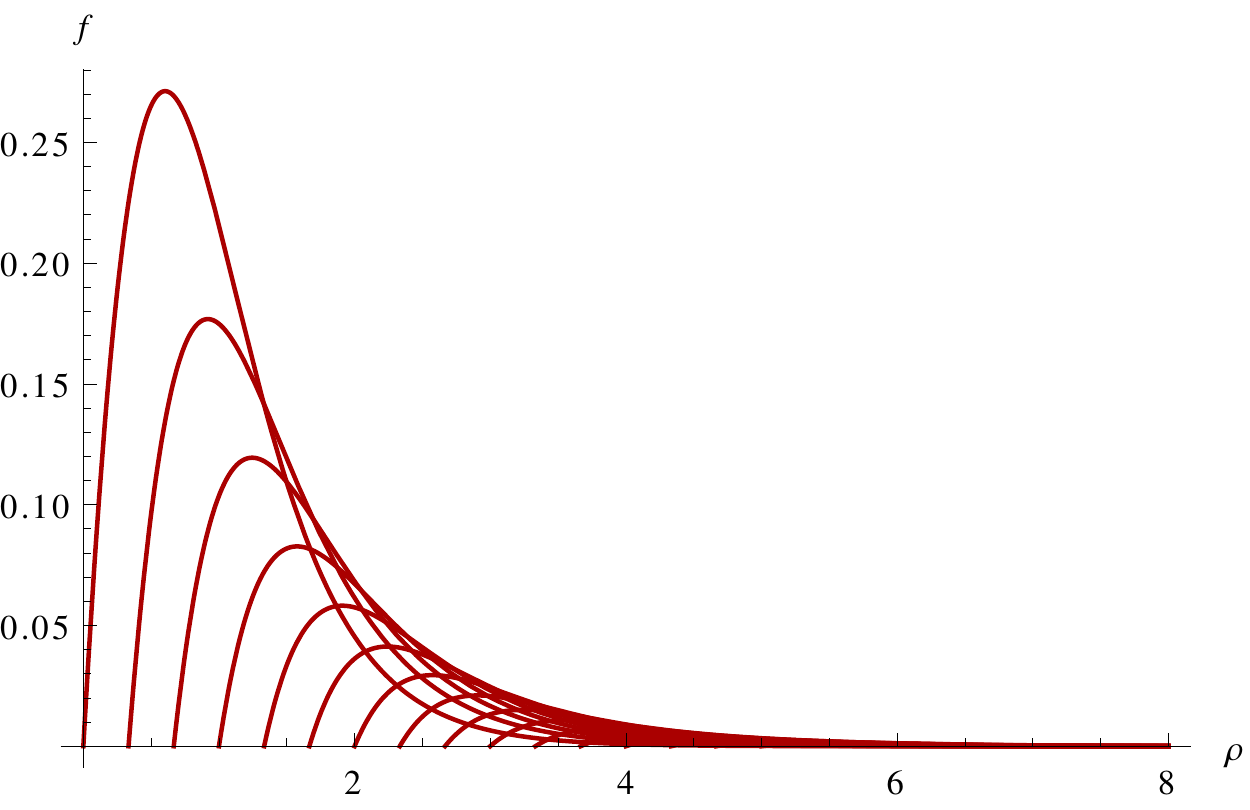}
}
\caption{
Slipping mode on the left hand side and the accompanying gauge field on the right hand side for short embeddings
with masses $m$ in $[0,5]$, as function of the radial coordinate $\rho$.
The smaller the mass, the deeper the branes extend into the bulk.
For $m=0$ we find an embedding which caps off at $\rho=0$, which is the boundary to the other patch.
So we already see that, with $\theta\equiv\pi/2$, there are at least two massless embeddings.
\label{fig:slipping-gauge-field-short}
}
\end{figure}

In the $(m,c)$-plane, these short embeddings correspond to the thick solid lines in Fig.~\ref{fig:m-c-diagram}.
Let us see what happens when we depart from the choice (\ref{eqn:short-embedding}) for large masses.
Already from Fig.~\ref{fig:cap-off-cNS-branches} we see that for large enough masses 
$\operatorname{Im} c_\mathrm{n}\cup\operatorname{Im} c_\mathrm{s}=\mathbb{R}$.
So there will be solutions $\rho_\star$ to (\ref{eqn:disconnected-c-from-rho-s}) for any real $c$.
But these cap-off points with $\rho_\star$ different from $\rho_\mathrm{n/s}$ will not be regular in the sense discussed around (\ref{eqn:disconnected-m-rhos}).
So for large masses the short embeddings with (\ref{eqn:short-embedding}) are also the only regular ones.
They are the only generic embeddings, in the sense that they exist for any $m$,
and sample plots of the slipping mode and gauge field can be found in Fig.~\ref{fig:slipping-gauge-field-short}.

\subsection{Long embeddings}
As seen in the previous section, there is at least one smooth D7-brane embedding for any $m$.
In this section we start to look at less generic configurations, which are possible for small enough masses only.
Fig.~\ref{fig:cap-off-cNS-branches} already indicates that small $m$ is special, and we study this in more detail now.

\begin{figure}[htb]
\center
\subfigure[][]{
  \includegraphics[width=0.40\linewidth]{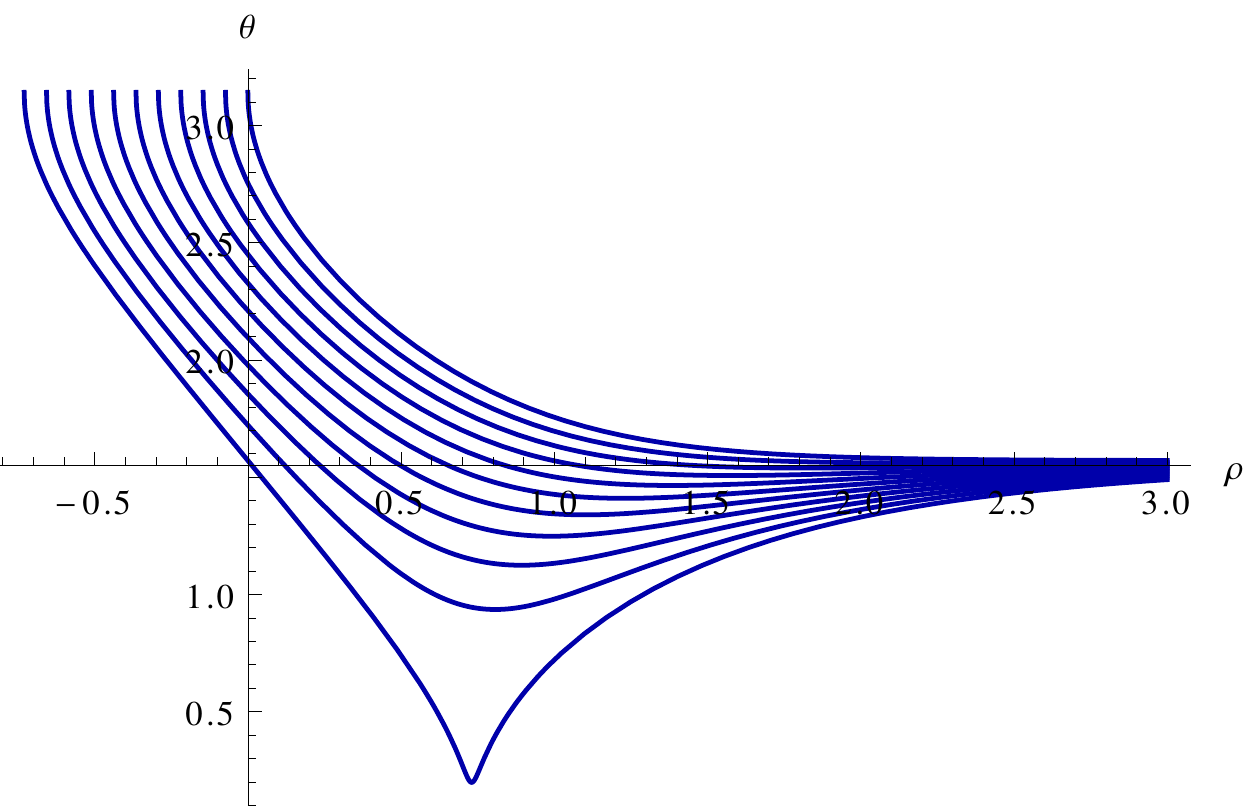}
}\qquad
\subfigure[][]{
    \includegraphics[width=0.40\linewidth]{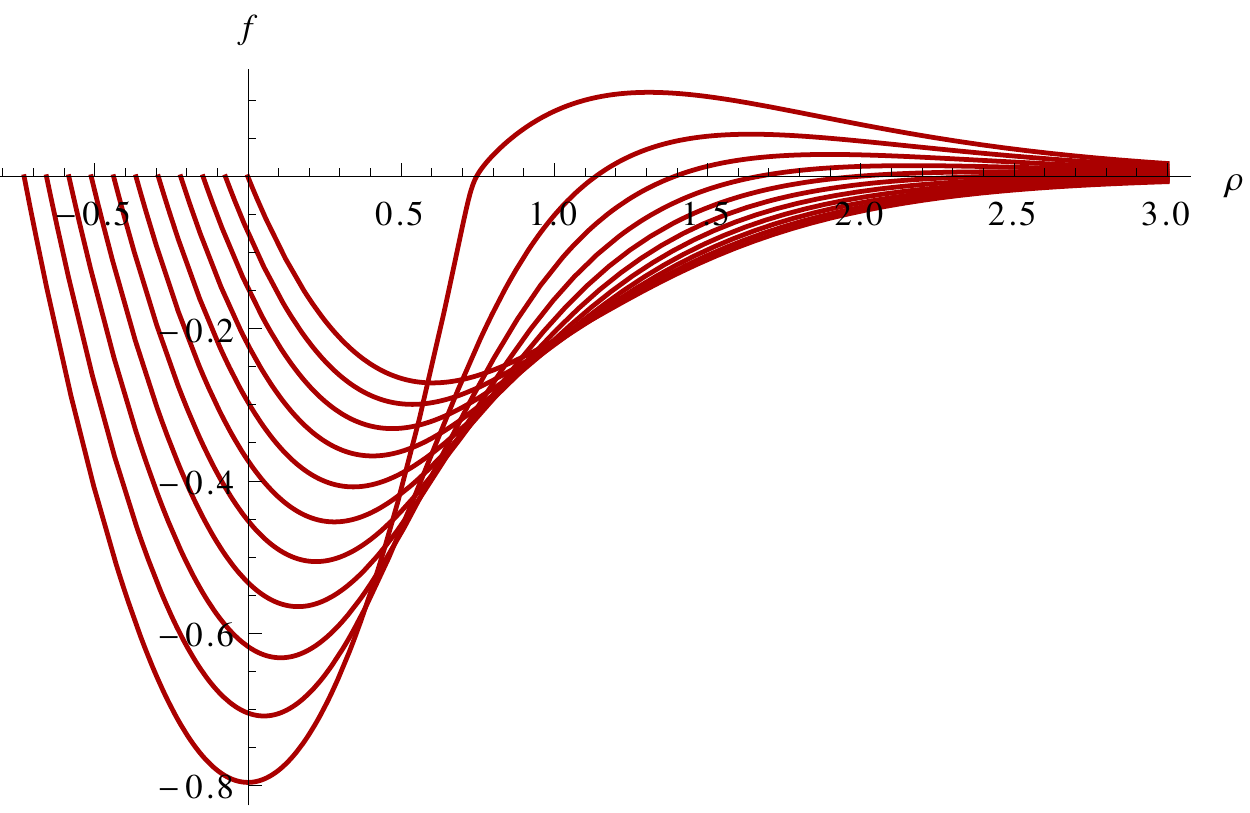}
}
\caption{
The left hand side shows the slipping mode for long embeddings,
from top to bottom corresponding to the mass $m$ increasing from $0$ to $m_\ell$,
as function of the radial coordinate $\rho$.
The right hand side shows the accompanying gauge field.
As compared to the short embeddings with the same mass, the long ones cap off at the other pole,
and there is a maximal depth up to which the long embeddings can extend into the second patch.
The sharp feature developing for the bottom curve on the left hand side turns into a cap-off for a short
embedding as $m\rightarrow m_\ell$, as given in (\ref{eqn:critical-mass}).
This way the plot connects to Fig.~\ref{fig:D7-short-embedding}.
In the $(m,c)$ plane of Fig.~\ref{fig:m-c-diagram},
this corresponds to following one of the thick solid lines coming from large $|m|$, and then at $m_\ell$
switching to the dashed line instead of further following the solid one.
\label{fig:slipping-gauge-field-long}
}
\end{figure}

For small enough mass, the maximum of the lower curve in Fig.~\ref{fig:cap-off-cNS-branches}, which is $c_\mathrm{s}$,
is strictly smaller than the minimum of the upper curve, which is $c_\mathrm{n}$.
We denote the critical value of the mass, below which this happens, by $m_\mathrm{\ell}$.
It can be characterized as the mass for which the maximum of $c_\mathrm{s}$ is equal to the minimum of $c_\mathrm{n}$,
which translates to
\begin{align}\label{eqn:critical-mass}
c_\mathrm{s}(\rho_\mathrm{s})&=c_\mathrm{n}(\rho_\mathrm{n}) &
&\Longleftrightarrow&
\sqrt{m_\ell^2+1} \left(m_\ell^2-2\right)+3 m_\ell \sinh ^{-1}m_\ell&=0~,
\end{align}
or $m_\ell\approx 0.7968$.
For $m<m_\ell$ we can make the opposite choice for $c$ as compared to (\ref{eqn:short-embedding}),
and still get a smooth cap-off point.
As discussed above, if we were to reverse the choice of $c$ for larger mass, the branes would hit a non-smooth cap-off point before reaching the other patch.
But for $m<m_\ell$ we can fix
\begin{align}\label{eqn:long-embedding}
 c&=c_\mathrm{n}(\rho_\mathrm{n})\quad \text{for $m>0$}~,
 &c&=c_\mathrm{s}(\rho_\mathrm{s})\quad \text{for $m<0$}~.
\end{align}
There is no cap-off point in the patch in which we start, so the branes wrap it entirely.
In the second patch they do not stretch out to the conformal boundary, but rather cap off smoothly at $\rho_\mathrm{n/s}$
for positive/negative $m$.
This is what we call a long embedding, as illustrated in Fig.~\ref{fig:D7-long-embedding}.
The maximal mass $m_\ell$ translates to a maximal depth up to which the branes can extend into the second patch,
as shown in Fig.~\ref{fig:slipping-gauge-field-long}.
In Fig.~\ref{fig:m-c-diagram} the long embeddings correspond to the dashed thick lines,
and for $c_\mathrm{s}(\rho_\mathrm{s})\,{<}\,c\,{<}\,c_\mathrm{n}(\rho_\mathrm{n})$ we get connected embeddings
which we discuss in the next section.

For holographic applications, this offers interesting possibilities to add flavors in both patches.
In addition to the short-short embeddings discussed above, which can be realized for arbitrary combinations of masses, we now
get the option to combine a short embedding in one patch with a long one in the other.
Fig.~\ref{fig:connected-long-embeddings} shows as thick black lines particular long-short combinations with the same value of $m$ 
in the two patches, which corresponds to flavor masses of opposite sign in each of the two copies of AdS$_4$ on the boundary.
Moreover, we could also combine two long embeddings, which would realize partly overlapping stacks of D7-branes from the AdS$_5$ perspective.
Whether the branes actually intersect would depend on the chosen $m$ in each of the patches:
For $m$ of the same sign they can avoid each other, as they cap off at different poles on the S$^5$.
But for $m$ of opposite sign they would intersect.

\subsection{Connected embeddings}
The last class of embeddings we want to discuss are the connected ones, which cover all of AdS$_5$, including both parts of the conformal boundary.
In contrast to Poincar\'{e} AdS$_5$, where finite-mass embeddings always cap off, such embeddings exist for non-zero masses.
The critical value is the same $m_\ell$ given in (\ref{eqn:critical-mass}) for the long embeddings.
As discussed in the section above, for $m<m_\ell$ 
there are choices of $c$ for which there is no $\rho_\star$ to satisfy (\ref{eqn:disconnected-c-from-rho-s}),
and thus no cap-off points. These are given by
\begin{align}\label{eqn:connected-embeddings}
 c_\mathrm{s}(\rho_\mathrm{s})<c<c_\mathrm{n}(\rho_\mathrm{n})~,
\end{align}
where $c_\mathrm{n/s}$ were defined in (\ref{eqn:disconnected-c-from-rho-s}) and $\rho_\mathrm{n/s}$ in (\ref{eqn:disconnected-m-rhos}).
With no cap-off points there are no regularity constraints either, and these are accepted as legitimate embeddings right away.
Due to the very fact that the embeddings are connected,
we immediately get a relation between the masses in the two patches: they have the same modulus but opposite signs.

\begin{figure}[ht]
\center
  \includegraphics[width=0.6\linewidth]{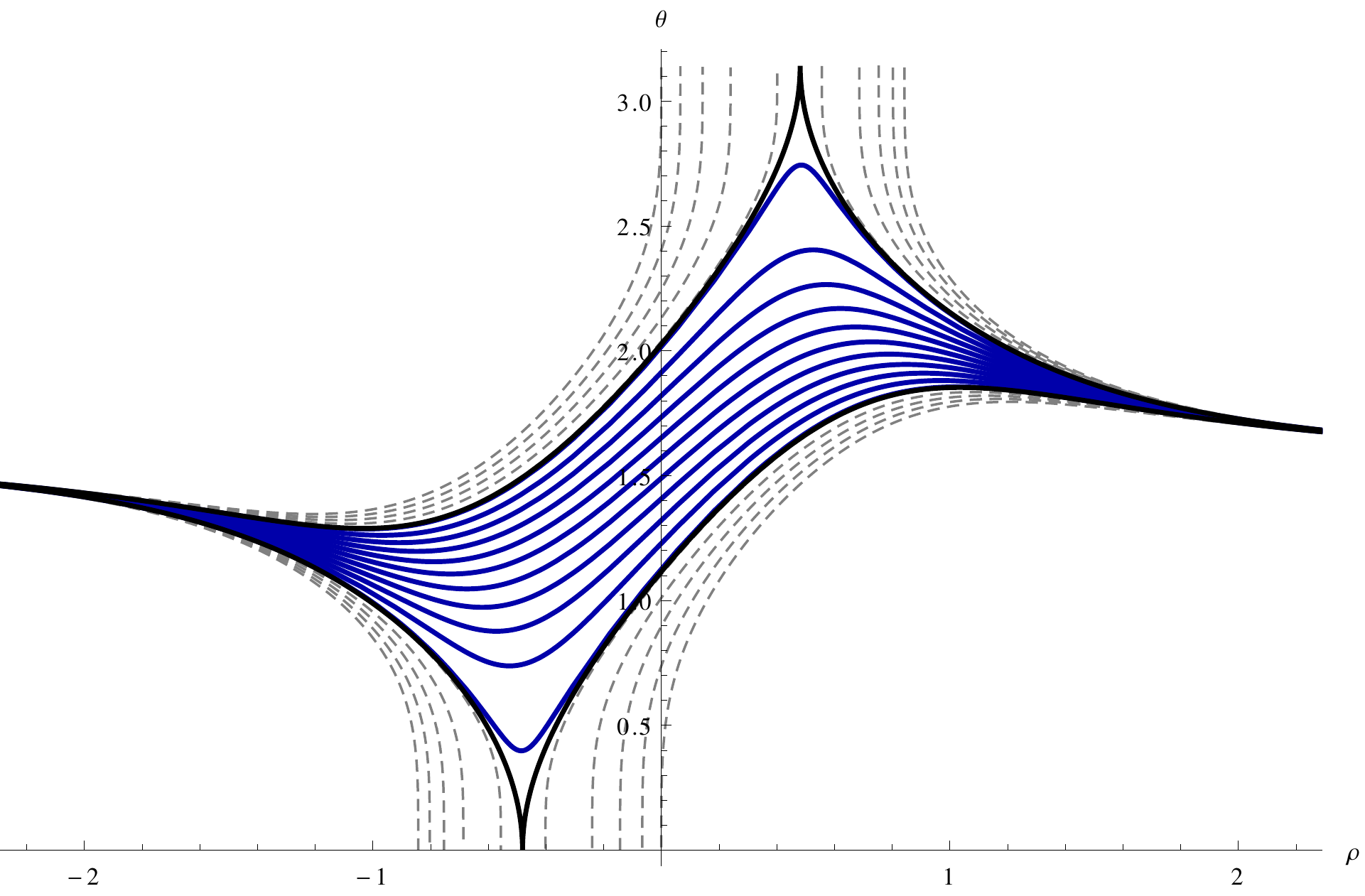}
\caption{
Connected, long and short embeddings with $m=\frac{1}{2}$.
The shown configurations correspond to a vertical section through the phase diagram shown in Fig.~\ref{fig:m-c-diagram}.
The thick blue lines are a family of embeddings with values of $c$ satisfying (\ref{eqn:connected-embeddings}).
The solid black lines show the disconnected limiting embeddings:
the upper/lower ones correspond to positive/negative $c$ saturating the inequalities in (\ref{eqn:connected-embeddings}).
The gray dashed lines correspond to embeddings with irregular cap off.
For the limiting case where the branes cap off, we get to choose among short/short, long/short, short/long and long/long embeddings.
We see how the branes do not intersect for this choice of masses, even for the long/long embedding.
\label{fig:connected-long-embeddings}
}
\end{figure}

An open question at this point is how the massive embeddings for AdS$_4$-sliced AdS$_5$
connect to the massless $\theta\equiv\pi/2$ embedding with $f\equiv 0$, which we know as solution from Poincar\'{e}-AdS$_5$.
So far, we have only seen massless embeddings capping off at either of the poles
(see Fig.~\ref{fig:slipping-gauge-field-short} and \ref{fig:slipping-gauge-field-long}).
The $\theta\equiv\pi/2$ embedding is at the origin in Fig.~\ref{fig:m-c-diagram}, and the blue-shaded region around it are the connected embeddings.
The embeddings corresponding to a vertical section through Fig.~\ref{fig:m-c-diagram} are shown in Fig.~\ref{fig:connected-long-embeddings},
where we chose $m=\frac{1}{2}$, such that the connected embeddings exist.
As $m$ is decreased, the embeddings become more and more symmetric around $\rho=0$, and we eventually find the $\theta\equiv \pi/2$ embedding with $m=c=0$.
This will be seen more explicitly in the next section.

\begin{figure}[ht]
\center
  \includegraphics[width=0.45\linewidth]{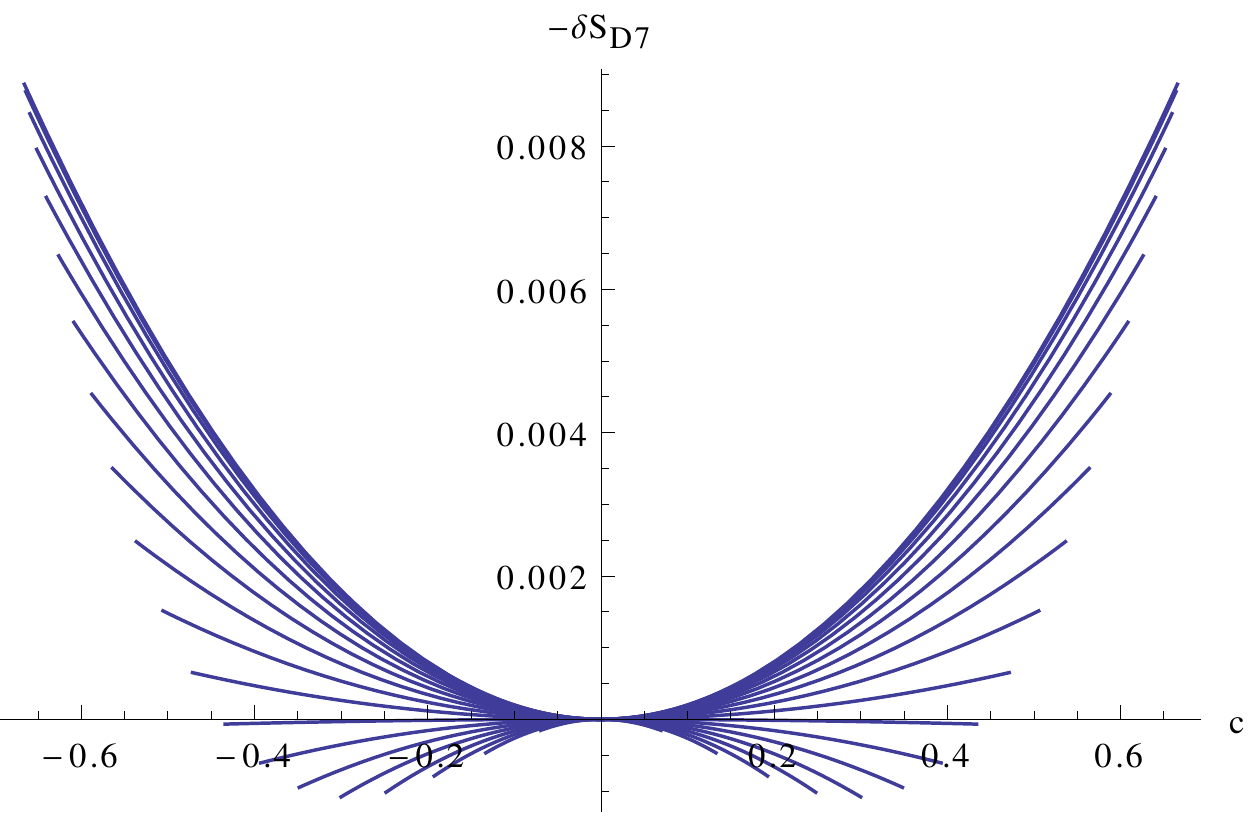}
\caption{
The plot shows $-\delta S_\mathrm{D7}$ defined in (\ref{eqn:delta-SD7}), as function of the parameter $c$ controlling the chiral condensate.
This corresponds to the free energy with the $c=0$ value subtracted off.
The quantity $\delta S_\mathrm{D7}$ then is independent of the chosen renormalization scheme.
From top to bottom the curves correspond to increasing $|m|$.
For $m=0$, corresponding to the top curve, the $\theta\equiv\pi/2$ embedding with $c=0$ is the one with lowest energy.
As $|m|$ is increased, this changes,
seemingly through a $1^\mathrm{st}$-order phase transition,
and the ``marginal'' embeddings with maximal allowed $c$ become the ones with minimal free energy.
\label{fig:free-energy}
}
\end{figure}

For each $m$, one can ask which embedding among those with different $c$ has the minimal action.
The on-shell action as given in (\ref{eqn:D7-action-simplified}) is divergent, as common for this quantity on asymptotically-AdS spaces.
But the divergences do not depend on $c$, and so a simple background subtraction will be sufficient to renormalize when $m$ is held fixed.
We come back to the holographic renormalization with all the bells and whistles in Sec.~\ref{sec:N4SYM}.
For now we simply define the finite quantity
\begin{align}\label{eqn:delta-SD7}
 \delta S_\mathrm{D7}(m,c)&=S_\mathrm{D7}(m,c)-S_\mathrm{D7}(m,0)~.
\end{align}
Strictly speaking, $\delta S_\mathrm{D7}$ is still divergent due to the infinite volume of the AdS$_4$ slices.
But this is a simple overall factor and we can just look at the ``action density'',
with the volume of AdS$_4$ divided out.
Using $\lambda^2=-1$, the explicit expression for $h$ in (\ref{eqn:h-from-kappa})
and integration by parts,
we can further simplify the action (\ref{eqn:D7-action-simplified}) to
\begin{align}\label{eqn:D7-action-simplified-2}
 S_\mathrm{D7}&=-T_7\int d^7\xi \sqrt{g^{}_{\mathrm{AdS}_4}}\sqrt{g^{}_{\mathrm{S}^3}}
 \left[\left[4\zeta f^2\right]
 +\int d\rho\cosh^4\rho \left(\sin^2\theta+4f^2\cot^2\theta\right)
 \right]~.
\end{align}
We introduce $\sigma=\cos\theta$,
such that $\sigma=2\cos\frac{4\pi+\cos^{-1}\tau}{3}$.
This isolates the volume divergence, which is independent of $m$ and $c$, and we find
\begin{align}\label{eqn:D7-action-simplified-3}
 S_\mathrm{D7}&=-T_7\int d^7\xi \sqrt{g^{}_{\mathrm{AdS}_4}}\sqrt{g^{}_{\mathrm{S}^3}}
 \left[\left[\zeta(1+4 f^2)\right]
 -\int d\rho\,\zeta^\prime \sigma^2\left(1-\left(1-\sigma^2\right){\left(\sigma\cosh\rho\right)^\prime}^2\right)
 \right]~.
\end{align}
The prime in the last term denotes a derivative with respect to $\rho$.
$\delta S_\mathrm{D7}$ as defined in (\ref{eqn:delta-SD7}) is then easily evaluated numerically,
and the results are shown in Fig.~\ref{fig:free-energy}.

\subsection{\texorpdfstring{$\mathbb{Z}_2$}{Z2}-symmetric configurations}
In the last section for this part we look at a special class of configurations with an extra $\mathbb{Z}_2$ symmetry relating the two patches.
The slipping mode may be chosen either even or odd under the $\mathbb{Z}_2$ transformation exchanging the two patches,
and we can see from (\ref{eqn:f-fprime-solutions}) that the gauge field $f$ consequently will
have the opposite parity, i.e.\ odd and even, respectively.
For the disconnected embeddings, the extra symmetry simply fixes how the embeddings have to be combined for the two patches.
It narrows the choices down to either short/short or long/long, and depending on the parity the long/long embeddings will be intersecting or not.
For the connected embeddings, we use that for the $k=2$ solutions in (\ref{eqn:soltheta})  we have
\begin{align}\label{eqn:connected-embeddings-patch-relation}
 \theta(\rho)&=\theta(-\rho)\big\vert_{m\rightarrow -m}=\pi-\theta(-\rho)\big\vert_{c\rightarrow -c}~,&
 f(\rho)&=f(-\rho)\big\vert_{c\rightarrow -c}=-f(-\rho)\big\vert_{m\rightarrow -m}~.
\end{align}
So imposing even $\mathbb{Z}_2$ parity fixes $m=0$, and imposing odd $\mathbb{Z}_2$ parity implies $c=0$.
From the connected $\mathbb{Z}_2$-even configurations we therefore get an entire family of massless solutions.
They correspond to the vertical axis in Fig.~\ref{fig:m-c-diagram}.
The $\mathbb{Z}_2$-odd solutions with $c=0$ loosely correspond to vanishing chiral condensate in \N{4} SYM,
but that statement depends on the chosen renormalization scheme, as we will see below.
\begin{figure}[htb]
\center
\subfigure[][]{
  \includegraphics[width=0.40\linewidth]{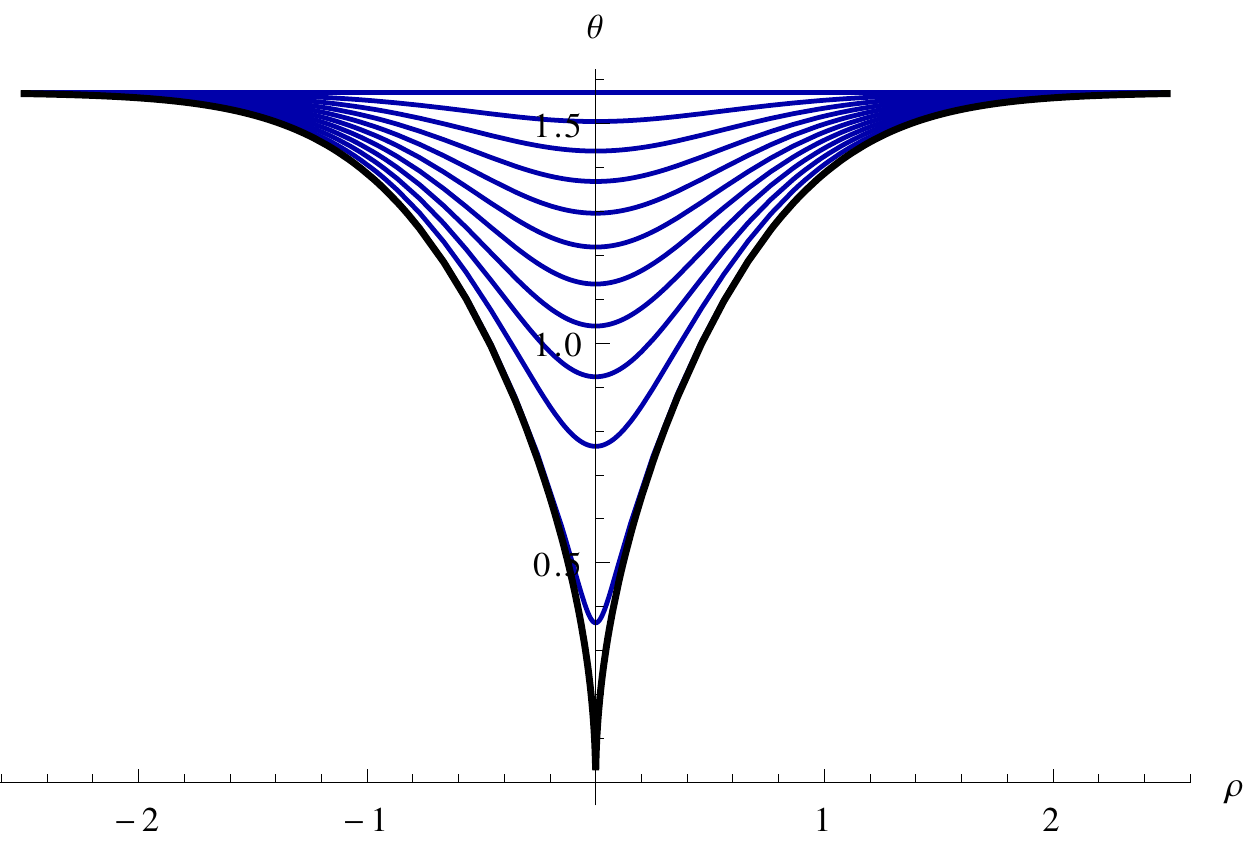} \label{fig:slipping-mode-Z2-even}
}\qquad
\subfigure[][]{
    \includegraphics[width=0.40\linewidth]{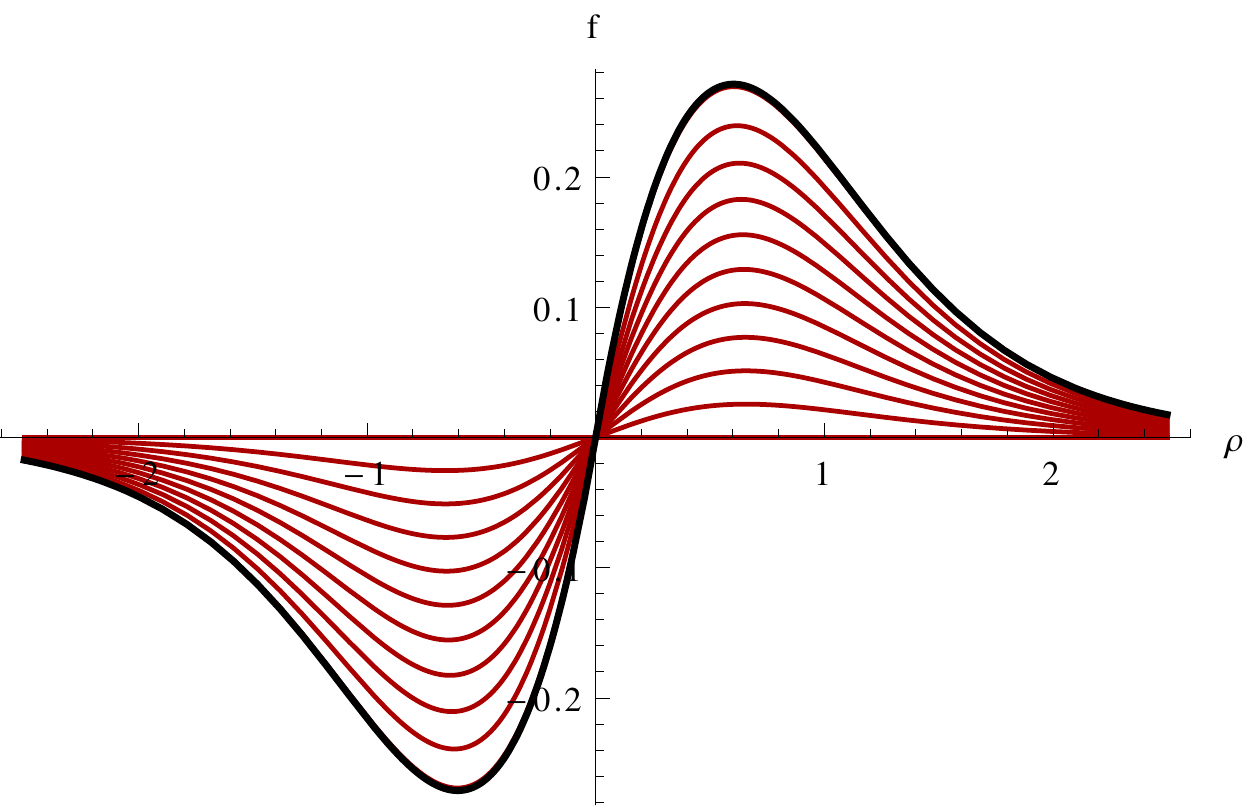}
}
\caption{
Embeddings with $\mathbb{Z}_2$-even slipping mode on the left hand side and the accompanying $\mathbb{Z}_2$-odd gauge field on the right hand side,
as function of the radial coordinate $\rho$.
The $\theta\equiv\frac{\pi}{2}$ solution corresponds to the flat slipping mode and $f\equiv 0$.
The lower the curve on the left hand side, the larger $c$.
On the right hand side larger $c$ corresponds to a larger peak.
We have restricted to $c>0$ for the plot,
the $c<0$ embeddings are obtained by a reflection on the equator of the S$^5$ for the embedding and a sign flip on the gauge field.
Note how the solutions interpolate between $\theta\equiv\frac{\pi}{2}$ and the massless solution capping off at $\rho_\star=0$
discussed in Sec.~\ref{sec:embeddings-short}, which is shown as thick black line in both plots.
\label{fig:slipping-gauge-field-Z2-even}
}
\end{figure}

\begin{figure}[htb]
\center
\subfigure[][]{
  \includegraphics[width=0.40\linewidth]{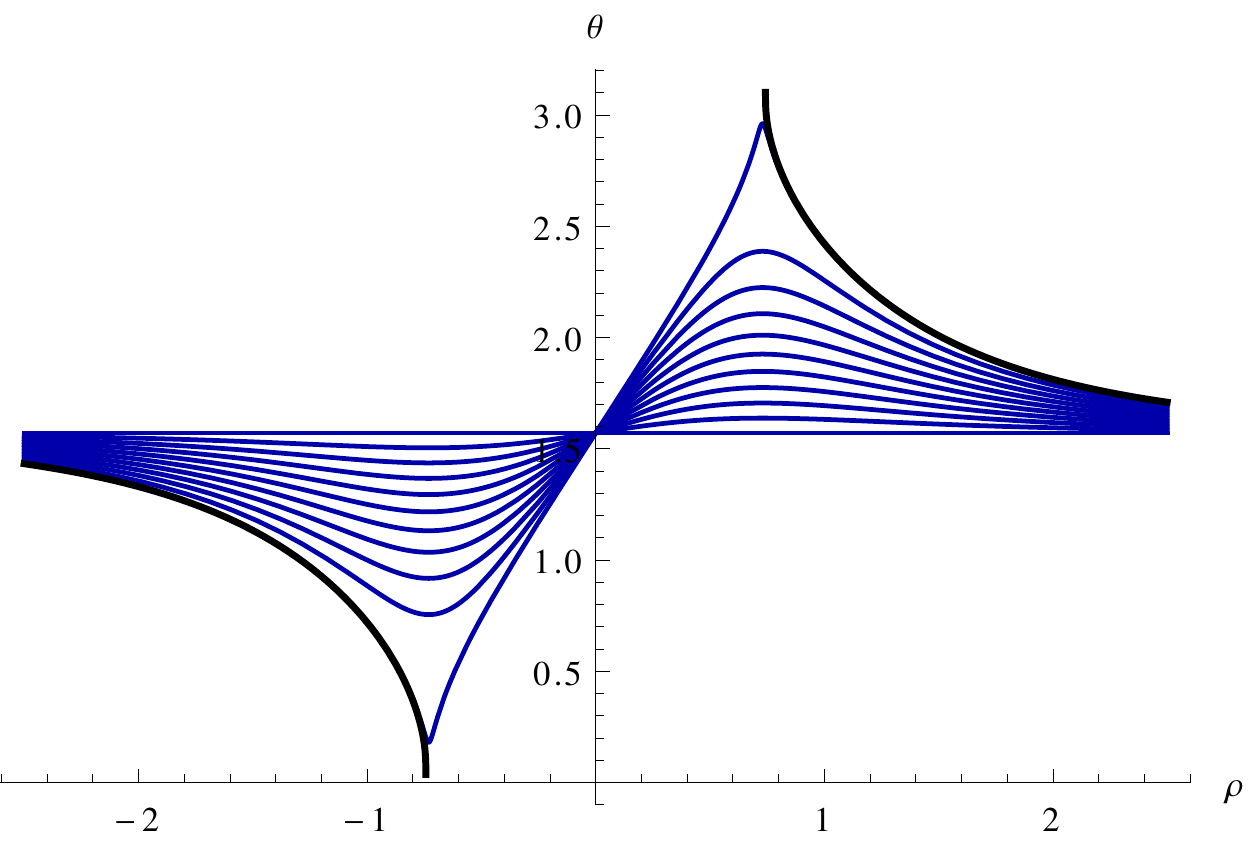} \label{fig:slipping-mode-Z2-odd}
}\qquad
\subfigure[][]{
    \includegraphics[width=0.40\linewidth]{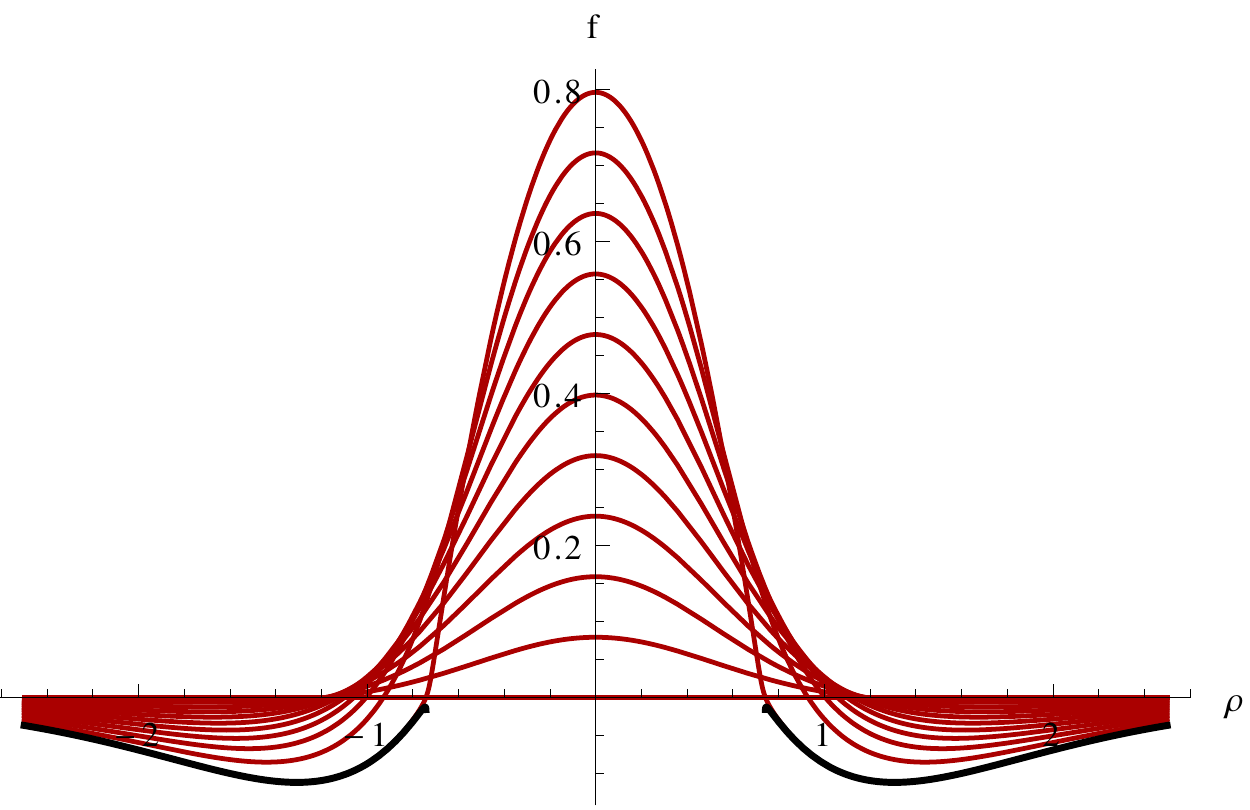}
}
\caption{
Embeddings with $\mathbb{Z}_2$-odd slipping mode on the left hand side and the accompanying $\mathbb{Z}_2$-even gauge field on the right hand side,
as function of the radial coordinate $\rho$.
$m=0$ corresponds to $\theta\equiv\frac{\pi}{2}$ and the further the embedding departs from that, the larger is $m$.
The $\theta\equiv\frac{\pi}{2}$ solution has $f=0$, and the higher the peak of the curves on the right hand side, the larger is $m$.
Note how these curves interpolate between the $\theta\equiv\frac{\pi}{2}$ solution and the massive embedding capping off at
$\rho_\star=\sinh^{-1}\!m_\ell$, shown as thick black line.
The critical mass $m_\ell$ was defined in (\ref{eqn:critical-mass}), and corresponds to $\rho_\star\approx 0.73$.
\label{fig:slipping-gauge-field-Z2-odd}
}
\end{figure}

We can now understand how the connected embeddings connect to the short or long disconnected embeddings discussed above.
Say we assume $\mathbb{Z}_2$ symmetry for a start, which for the connected embeddings confines us to the axes in Fig.~\ref{fig:m-c-diagram}.
That still leaves various possible trajectories through the $(m,c)$ diagram.
For even slipping modes we are restricted to the vertical axis for connected embeddings.
Starting out from large mass and the short embeddings,
one could follow the thick lines in Fig.~\ref{fig:m-c-diagram} all the way to $m=0$, where the cap-off point approaches $\rho_\star=0$.
Another option would be to change to the dashed line at $m=m_\ell$, corrsponding to a long embedding.
In either case, once we hit the vertical axis in Fig.~\ref{fig:m-c-diagram}, this corresponds to the massless embedding with maximal $|c|$.
From there one can then go along the vertical axis to the $\theta\equiv\pi/2$ embedding at the origin.
This last interpolation is shown in Fig.~\ref{fig:slipping-gauge-field-Z2-even}.
For odd slipping modes, we are restricted to the horizontal axis in Fig.~\ref{fig:m-c-diagram} for connected embeddings.
Coming in again from large mass and a short embedding, the thick line eventually hits $c=0$ as the mass is decreased.
From there the branes can immediately go over to a connected embedding,
which corresponds to going over from the thick black line in Fig.~\ref{fig:slipping-mode-Z2-odd} to the blue ones.
These connect to the $\theta\equiv\pi/2$ embedding along the horizontal axis in Fig.~\ref{fig:m-c-diagram}.
Another option would be to make the transition to a long/long embedding.
If we decide to not impose parity at all, the transition to the $\theta\equiv\pi/2$ embedding does not have to proceed along one of the axes.
The transition from connected to disconnected embeddings may then happen at any value of $m$,
small enough to allow for connected embeddings.
Fig.~\ref{fig:connected-long-embeddings} shows an example for the transition,
corresponding to following a vertical line in Fig.~\ref{fig:m-c-diagram} at $m=\frac{1}{2}$.

\section{Flavored \texorpdfstring{\N{4}}{N=4} SYM on (two copies of) \texorpdfstring{A\lowercase{d}S$_4$}{AdS4}}\label{sec:N4SYM}
In the previous section we studied the various classes of brane embeddings, to get a catalog of
allowed embeddings and of how they may be combined for the two coordinate patches needed to cover AdS$_5$.
In this section we take first steps to understanding what these results mean for \N{4} SYM on two copies of AdS$_4$.
In Sec.~\ref{sec:curved-susy-QFT} we discuss relevant aspects of supersymmetry on curved space and
how these are reflected in our embeddings.
We also discuss the boundary conditions that are available to link the two copies of AdS$_4$.
In Sec.~\ref{sec:holo-ren} and \ref{sec:one-point} we carry out the holographic renormalization and compute the one point functions
for the chiral and scalar condensates.
With these results in hand, we come back to the question of boundary conditions and attempt an interpretation of the 
various embeddings in Sec.~\ref{sec:interpretation}.

\subsection{Supersymmetric field theories in curved space}\label{sec:curved-susy-QFT}

While it is well understood how to construct supersymmetric Lagrangians on Minkowski space, for example using superfields,
the study of supersymmetric gauge theories on curved spaces needs a bit more care. 
Generically, supersymmetry is completely broken by the connection terms in the covariant derivatives when 
naively formulating the theory on a curved background.
One simple class of supersymmetric curved space theories is provided by superconformal field theories on spacetimes that are conformal to flat space.
That is, once we have constructed a superconformal field theory on flat space, such as \N{4} SYM without any flavors or with massless quenched flavors,
we can simply obtain the same theory on any curved space with metric
\begin{align}
\label{omegatrafo}
ds^2 = \Omega(t, \vec{x}) \left ( -dt^2 + d \vec{x}^2 \right ).
\end{align}
The crucial ingredient for this extension to work is that all fields need to have their conformal curvature couplings.
These can be constructed systematically by consistently coupling to a conformal supergravity background (see \cite{Fradkin:1985am} for a nice review).
For \N{4} SYM this boils down to adding the usual conformal coupling for the scalars.

For $\Omega=z^{-2}$, where $z$ is one of the spatial directions, the resulting metric is locally AdS$_4$ in Poincare coordinates. 
In fact, the resulting geometry is two copies of AdS$_4$, one for $z>0$ and one for $z<0$.
The two AdS$_4$ spaces are linked with each other along their common boundary at $z=0$ via boundary conditions,
which we can derive from the conformal transformation. 
In Minkowski space, $z=0$ is not a special place. 
All fields as well as their $z$-derivatives, which we denote by primes, have to be continuous at this codimension one locus. 
Denoting the fields at positive (negative) values of $z$ with a subscript R (L), the boundary conditions for a massless scalar field $X$ therefore read
\begin{align}\label{eqn:transparent-bc}
X_L(z=0) = X_R(z=0)~, \quad \quad X'_L(z=0) = X'_R(z=0)~.
\end{align}
The generalization to fermions and vector fields is straightforward.
Under a conformal transformation, the left and right hand sides of these conditions change in the same way,
and so these boundary conditions have to be kept in place when studying the same field theory on the conformally related two copies of AdS.
These ``transparent" boundary conditions were discussed as very natural from the point of view of holography in \cite{Karch:2000ct} and many subsequent works.
They preserve the full supersymmetry of the field theory, as is obvious from their flat space origin.
From the point of view of the field theory on AdS they are unusual.
For physics to be well defined on one copy of AdS$_4$, we need one boundary condition on (say) $X_L$ and $X_L'$ alone.
Typical examples are the standard Dirichlet or Neumann boundary conditions.
For two separate copies of AdS$_4$ we do need two sets of boundary conditions in total, what is unusual is that the transparent boundary conditions relate L and R fields to each other.
A different set of boundary conditions would for example be $X_L(z=0) = X_R(z=0)=0$ with no restrictions on the derivatives.
With these double Dirichlet boundary conditions the two copies of AdS$_4$ are entirely decoupled, and these
are the boundary conditions typically used for field theories on AdS$_4$.
Generic boundary conditions will break all supersymmetries, but it is well known how to impose boundary conditions on \N{4} SYM on a single AdS$_4$ space in a way that preserves half of the supersymmetries.
These boundary conditions follow from the analysis in \cite{Gaiotto:2008sa} and correspond to the field theory living on D3 branes ending on stacks of NS5 or D5 branes.
The detailed choice of boundary conditions dramatically changes the dynamics of field theory on AdS$_4$, as comprehensively discussed in \cite{Aharony:2010ay}.
While \cite{Gaiotto:2008sa,Aharony:2010ay} completely classified the boundary conditions preserving at least half of the supersymmetries 
for a single copy of AdS$_4$, more general supersymmetry preserving boundary conditions are possible on two copies.
We already saw one example, the transparent boundary conditions above, which preserve the full supersymmetry.
It is straightforward to formulate boundary conditions that interpolate between transparent and 
double Dirichlet boundary conditions, even though we have not yet attempted a complete classification of supersymmetric boundary conditions 
for \N{4} SYM on two copies of AdS$_4$.

\begin{figure}[htb]
\center
    \includegraphics[width=0.40\linewidth]{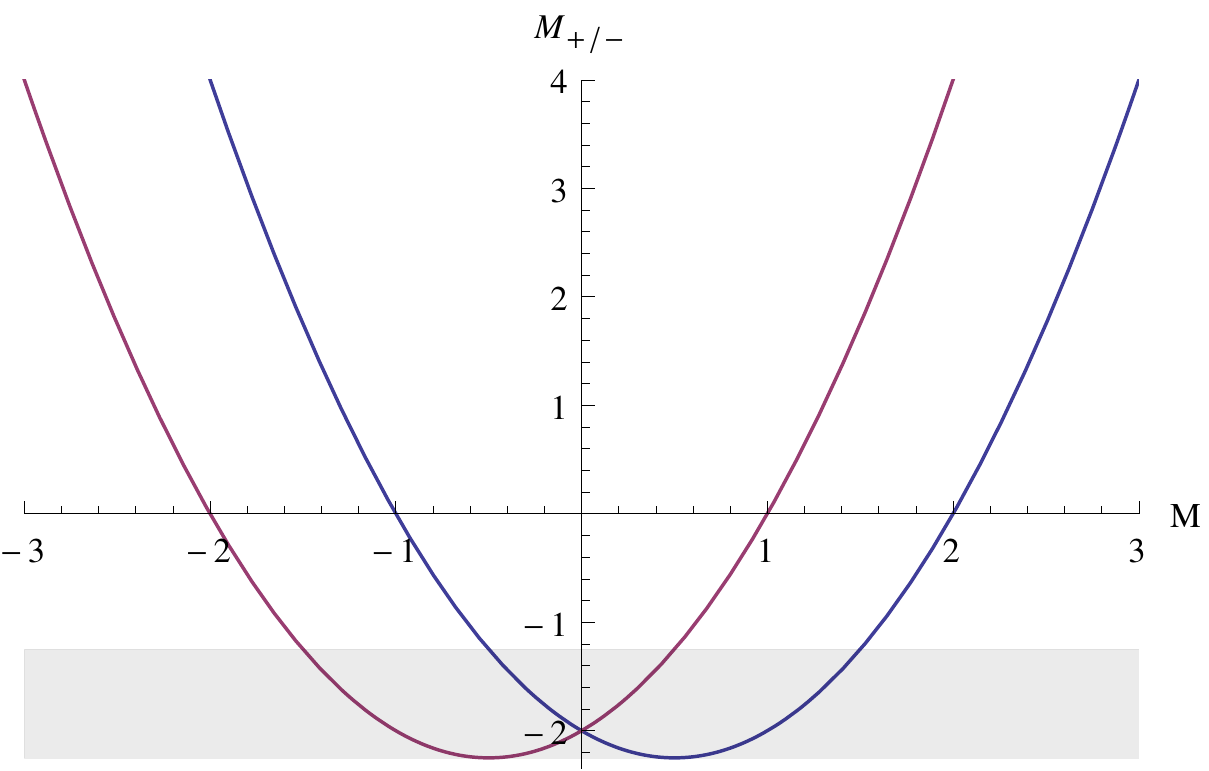}
\caption{Eigenvalues of the scalar mass matrix as a function of superpotential mass $M$ for flavored ${\cal N}=4$ SYM on AdS$_4$.
Alternative quantization is possible for eigenvalues in the gray-shaded region.
\label{evalues}
}
\end{figure}

While conformal theories are the simplest supersymmetric field theories to formulate on curved space, one can also formulate
non-conformal supersymmetric field theories, e.g.\ with masses for at least some of the fields, on curved spaces.
The non-invariance of the connection terms can be compensated by adding additional terms to the Lagrangian.
In the simple case of the 4-sphere, it was shown in \cite{Pestun:2007rz} that for an \N{2} supersymmetric gauge theory with massive hypermultiplets a simple scalar mass can act as a compensating term.
Denoting the two complex scalars in a hypermultiplet by $Q$ and $\tilde{Q}$, which in a common abuse of notation we also use for the entire chiral multiplet they are part of,
one recalls that the superpotential term
\begin{align}
W = M Q \tilde{Q}
\end{align}
gives rise to a fermion mass $m$ as well as an F-term scalar mass term in the potential,
\begin{align}
V_m = |M|^2 \left ( |Q|^2 + |\tilde{Q}|^2 \right )~.
\end{align}
This theory as it stands is not supersymmetric, but can be made supersymmetric by adding a particular dimension-2 operator to the Lagrangian.
The full scalar mass term then reads $V=V_m+V_c^{\mathrm{S}^4}$, with
\begin{align}
\label{compterms}
V_c^{\mathrm{S}^4} = i \frac{M}{R} \left ( Q \tilde{Q} + \mathrm{c.c.} \right )\,,
\end{align}
where $R$ is the curvature radius of the sphere.
Since the compensating mass term is imaginary, the resulting action is not real.
We construct supersymmetric D7-brane embeddings to holographically realize this supersymmetric combination of mass terms
on S$^4$ in App.~\ref{app:global-Euclidean-kappa}, and correspondingly find an imaginary gauge field.
The compensating terms have been understood systematically in \cite{Festuccia:2011ws}.
Again, the natural way to construct supersymmetric field theories in curved space is to couple to a background
supergravity multiplet. One then obtains a supersymmetric field theory for every supersymmetric configuration of the background supergravity.
To have a non-trivial curved-space configuration preserving supersymmetry, the supergravity background has not just the metric turned on but 
also additional dynamical or auxiliary fields. 
The expectation values of these extra fields then appear as the desired compensating terms in the field theory Lagrangian.
Following this logic, the simple compensating term \eqref{compterms} for the $S^4$ can easily be generalized to AdS$_4$, 
which now yields a real coefficient
\begin{align}
\label{compterm}
V_c^{\mathrm{AdS}_4} =  \frac{M}{R} \left ( Q \tilde{Q} + \mathrm{c.c.} \right ).
\end{align}
Including the superpotential mass term, the conformal curvature coupling, as well as the supersymmetry restoring compensating term, the full scalar mass matrix for a field theory in AdS$_{d+1}$ with 8 supercharges reads
\begin{align}
M_{Q\tilde{Q}} = \begin{pmatrix} -\frac{1}{4} (d^2-1) + M^2 & M \cr M & -\frac{1}{4} (d^2-1) + M^2 \end{pmatrix}.
\end{align}
For this work we are of course most interested in the AdS$_4$ case, that is $d=3$. The eigenvalues of this mass matrix are given by
\begin{align}
\label{masseigenvalues}
M_{\pm} = \frac{1}{4} (1-d^2 + 4 M^2 \pm 4M)~.
\end{align}
The full spectrum is symmetric under $M \rightarrow -M$ with the two branches being exchanged,
and the minimal eigenvalue ever reached is $M_\mathrm{min} = -d^2/4$ for $M=\pm 1/2$.
This is exactly the BF bound in $d$ dimensions \cite{Breitenlohner:1982bm,Breitenlohner:1982jf}.
So, reassuringly, the supersymmetric theory never becomes unstable.
The full scalar mass spectrum is depicted in Fig.~\ref{evalues}.
From this discussion we also see that the embeddings we found  and studied in Sec.~\ref{sec:susy-d3d7}, \ref{sec:embedding-classes} exhaust the entire scalar 
mass spectrum corresponding to stable theories. 
This is despite $m^2$ always being positive -- the stable negative-mass theories that are possible 
for AdS$_4$ arise from the combination of all mass-like terms as we have just seen.
We should note, however, that $m$ is related to the field theory mass $M$ via $M=\frac{\sqrt{\lambda}}{2\pi}m$ \cite{Kruczenski:2003be}, so
in the limit where classical supergravity is a good approximation we always deal with large $M$ and positive eigenvalues.

While the compensating term restores supersymmetry, it breaks some of the global symmetries.
Let us discuss this in detail for the case of \N{4} SYM with $N_f$ flavors.
The massless theory has a global SU$(2)_{\Phi} \times $SU$(2)_R \times $U$(1)_R \times $U$(N_f)$ symmetry.
Holographically, the first 3 manifest themselves as the preserved SO$(4) \times $U$(1)$ isometry of the D7-brane embedding 
with $\theta\equiv\pi/2$, 
whereas the U$(N_f)$ global symmetry corresponds to the worldvolume gauge field.
All flavor fields are invariant under SU$(2)_{\Phi}$, which acts only on the adjoint hypermultiplet that 
is part of the \N{4} Lagrangian written in \N{2} language.
So this symmetry will not play a role in the discussion below.
Under SU$(2)_R$ the fundamental scalars transform as a doublet, but the fundamental fermions are neutral.
A superpotential mass breaks the U$(1)_R$.
In the holographic dual, U$(1)_R$ corresponds to shifts in $\psi$, and these are manifestly broken by the massive D7-brane localized at a fixed value of $\psi$.
The superpotential mass term, however, preserves the full SO$(4)$ symmetry since $|Q|^2 + |\tilde{Q}|^2$ 
is the SU$(2)$ invariant combination. 
Not so for the compensating term, which explicitly breaks SU$(2)_R$ down to a $U(1)$ subgroup \cite{Festuccia:2011ws}. 
This pattern mirrors exactly the symmetry breaking pattern we found in our supersymmetric probe branes.

The structure of supersymmetry being restored by a compensating term is reminiscent of supersymmetric Janus solutions, 
that is field theories with varying coupling constants.
Since under a supersymmetry variation the Lagrangian typically transforms into a total derivative, position-dependent coupling constants 
generically break supersymmetry.
It was found in \cite{Clark:2004sb,D'Hoker:2006uv} that supersymmetry can again be restored by compensating terms.
This discussion is in fact related to field theories on AdS 
by a conformal transformation. 
We will find this picture useful for some of our analysis, so we briefly introduce it.
We start with a massive theory on AdS.
In the presence of a mass term, the conformal transformation from AdS to flat space is not a symmetry, but is explicitly broken by the mass term.
If, however, the mass is the only source of conformal symmetry breaking (as is the case in \N{2}$^*$ or flavored \N{4} SYM),
we can restore the conformal symmetry by treating the mass $M$ as a spurion. 
That is, by letting it transform explicitly.
This way a field theory with constant $M$ on AdS can be mapped to a field theory on flat space with a position-dependent 
superpotential mass $M \sim 1/z$. 
Correspondingly, the fermion mass goes as $1/z$, whereas the superpotential induced scalar mass as $1/z^2$. 
Such a position-dependent $M(z) Q \tilde{Q}$ superpotential is precisely the framework discussed in \cite{Clark:2004sb}. 
It was shown that supersymmetry can be restored by a compensating term proportional to $M'(z) Q \tilde{Q}$. 
This $1/z^2$ compensating term is exactly the conformally transformed AdS compensating term from \eqref{compterm}. 
So supersymmetric field theories on AdS are indeed conformally related to Janus like configuration with $1/z$ mass terms.

\subsection{Holographic renormalization}\label{sec:holo-ren}

A puzzling aspect of our supersymmetric probe configurations is that for a given leading term, 
that is for a given mass $M$ in the field theory, we found families of solutions that differed in the subleading term
(see again (\ref{eqn:theta-near-boundary}) below). 
Similar ambiguities were previously found in numerical studies in \cite{Clark:2013mfa}.
Holographic intuition suggests that this corresponds to different allowed vacuum expectation values for a given mass,
and we map that out in detail before attempting an interpretation.

The D7-brane action (\ref{eqn:DBI-action}) is divergent as usual, and we have to carry out its holographic renormalization before extracting
information about the flavor sector of the dual theory \cite{deHaro:2000xn,Bianchi:2001kw}.
The counterterms for the slipping mode $\theta$ have been given in \cite{Karch:2005ms},
so we only need to construct those for the gauge field.
In many regards, $f$ can be seen as a scalar at the BF bound, and it is tempting to just take those counterterms from \cite{Bianchi:2001kw}.
There are some subtleties, however, as we will discuss momentarily.
The counterterms for the slipping mode as given in \cite{Karch:2005ms} are
\begin{align}\label{eqn:counterterms}
\begin{split}
 L_{\mathrm{ct},\theta}=&-\frac{1}{4}\Big[1-\frac{1}{12}R\Big]+\frac{1}{2}\tilde\theta^2-\log\epsilon\, \left[\frac{1}{32}\big(R_{ij}R^{ij}-\frac{1}{3}R^2\big)+\frac{1}{2}\tilde\theta\square_W\tilde\theta\right]
 \\&+\alpha_1 \tilde\theta^4+\alpha_2 \tilde\theta\square_W\tilde\theta+\frac{\alpha_3}{32}\big(R_{ij}R^{ij}-\frac{1}{3}R^2\big)~.
\end{split}
\end{align}
Note that $g_\epsilon$ denotes the metric on the cut-off surface induced from the background metric,
as opposed to the worldvolume metric, and $R$ denotes the curvature of $g_\epsilon$.
$\square_W$ is the Weyl-covariant Laplacian, $\square_W=\square+\frac{1}{6}R$.
We have also defined $\tilde\theta:=\theta-\frac{\pi}{2}$, since our $\theta$ is shifted by $\frac{\pi}{2}$ compared to the coordinates used in \cite{Karch:2005ms}.
In the curvature conventions of \cite{Karch:2005ms}, the AdS$_4$ slices have $R_{ij}=3g_{\epsilon\,ij}$.
The coefficients of the finite counterterms $\alpha_i$ are not determined by the renormalization and reflect a scheme dependence.
For $\alpha_1$, demanding the free energy to vanish for Poincar\'e AdS, as required by supersymmetry, fixes \cite{Karch:2005ms}
\begin{align}\label{eqn:holo-ren-alpha1-susy}
 \alpha_1&=-\frac{5}{12}~.
\end{align}
The holographic counterterms are universal,
in the sense that they should be fixed once and for all, regardless of the background.
So the same argument for a supersymmetry-preserving scheme in Poincar\'e AdS also fixes $\alpha_1$
for us -- it is still the same theory, just evaluated on a different background.
That leaves the scheme dependence coming from $\alpha_2$ and $\alpha_3$, which 
can not be fixed from flat-space considerations since the counterterms vanish then.

We now come back to the extra terms for the S$^3$ gauge field.
There are two possible ways to look at it. 
The first one is the approach taken in \cite{Karch:2005ms} to deal with the slipping mode.
For fixed $\omega$ as given in (\ref{eqn:omega}), the gauge field $A=f\omega$ from the AdS perspective reduces to the radial profile $f$, which
can be treated as a scalar field.
The other one is to take the $6\,{+}\,1$-dimensional boundary introduced by the bulk radial cut-off on the worldvolume of the D7-branes
for what it is.
We would then only allow covariant and gauge-invariant counterterms like $\sqrt{g_\epsilon}F(A_\star)^2$  etc.,
where the star denotes pullback to the
cut-off surface, and determine their coefficients.
Since we already have $\omega$ and the two approaches are equivalent for our purposes, we follow the first one.

The radial profile of the S$^3$ gauge field $f$ is almost like a scalar at the BF bound.
After integrating the WZ term in (\ref{eqn:DBI-action}) by parts, the bulk action takes exactly the same form.
But due to this integration by parts, the action picks up an extra boundary term as compared to a scalar at the BF bound.
Such boundary terms can not be ignored on AdS, and here the extra boundary term cancels a $\log^2$ divergence usually expected for
scalars at the BF bound.
The counterterms are therefore slightly different from those given e.g.\ in \cite{Bianchi:2001kw,Karch:2005ms,Ohl:2010au},
and we find
\begin{align}\label{eqn:f-ct}
 L_{\mathrm{ct},f}&=-\lambda^2f^2\left(\frac{1}{2\log\epsilon}+\frac{\alpha_4}{(\log\epsilon)^2}\right)~.
\end{align}
This introduces an additional scheme-dependent counterterm with coefficient $\alpha_4$.
For the $\kappa$-symmetric embeddings, where $\theta$ and $f$ are related by (\ref{eqn:f-fprime-solutions}),
these finite counterterms are related as well.
So if one stays within this family of supersymmetric embeddings, $\alpha_4$ could be absorbed in a redefinition of $\alpha_2$.
But we will not do that.
For the renormalized D7-brane action corresponding to (\ref{eqn:DBI-action}) we then have
\begin{align}\label{eqn:S-D7-ren}
 S_\mathrm{D7,ren}&=S_\mathrm{D7}-T_7\int_{\rho=-\log(\epsilon/2)}d^7\xi\sqrt{g_\epsilon}\left[L_{\mathrm{ct},\theta}+L_{\mathrm{ct},f}\right]~.
\end{align}
To transform to Fefferman-Graham coordinates where the boundary metric is AdS$_4$ with unit curvature radius,
we have set $\rho=-\log(z/2)$.
The cut-off surface is then given by $\rho=-\log(\epsilon/2)$.
For the covariant counterterms it does not matter how we parametrize the cut-off surface,
but for those involving explicit logarithms a change in the parametrization results in a change of the finite counterterms.
The $\log$-terms in (\ref{eqn:counterterms}), (\ref{eqn:f-ct}) are chosen such that the coefficients of the finite counterterms
agree with the usual Fefferman-Graham gauge conventions.

\subsection{One-point functions}\label{sec:one-point}
With the holographic renormalization carried out we can now compute the renormalized one-point functions for the chiral and scalar condensates
in \N{4} SYM.
To get the near-boundary expansion for the embedding (\ref{eqn:soltheta}) and gauge field (\ref{eqn:f-fprime-solutions}),
we change to FG gauge.
As explained above, this amounts to setting $\rho=-\log(z/2)$.
Expanding then in small $z$ yields
\begin{subequations}\label{eqn:theta-near-boundary}
\begin{align}
 \theta&=
 \frac{\pi }{2}+m z- m z^3\log (z)+
  z^3 \left[\frac{m}{4} \left(2 m^2+4\log 2-3\right)-c\right]
 +\mathcal O\left(z^5\log z\right)~,
 \\
 f&=\frac{m}{i\lambda}z^2\log z+\frac{3c-m^3+3m(1-\log 2)}{3i \lambda}z^2+\mathcal O(z^4\log z)~.
\end{align}
\end{subequations}
The $\mathcal O(z)$ term of the slipping mode as usual sets the mass, while the $\mathcal O(z^3)$ term loosely corresponds to the chiral condensate.
The leading term of the gauge field expansion sets the extra scalar mass, and is related to the $\mathcal O(z)$ term in the slipping mode.
This reflects the relation between the superpotential mass and compensating mass terms discussed in \ref{sec:curved-susy-QFT}.
The $\mathcal O(z^2)$ term encodes the corresponding scalar condensate.

To actually get the one-point functions, we have to compute the variation of the action (\ref{eqn:DBI-action}) and evaluate it on shell.
Going on shell and varying does not necessarily commute, so we will not use any of the $\kappa$-symmetry relations to simplify the action.
The starting point will be (\ref{eqn:D7-action-simplified}), where $h$ was computed in (\ref{eqn:h-def}).
We can, however, use relations like (\ref{eqn:h-from-kappa}) after the variation. That gives
\begin{subequations}
\begin{align}
 \delta_\theta S_\mathrm{D7}&=-T_7V_{\mathrm{S}^3}\int d^5\xi\, \partial_\rho\left[\sqrt{g^{}_{\mathrm{AdS}_4}}
   \zeta^\prime\theta^\prime\sin^4\theta\,\delta\theta\right]+\mathrm{EOM}~,\\
 \delta_f S_\mathrm{D7}&=-T_7V_{\mathrm{S}^3}\int d^5\xi\, \partial_\rho\left[\sqrt{g^{}_{\mathrm{AdS}_4}}
   \left(\zeta^\prime\sin^2\theta f^\prime+8\zeta f\right)\delta f\right]+\mathrm{EOM}~,
\end{align}
\end{subequations}
where EOM denotes contributions which vanish when evaluated on shell.
Combining that with the variation of the counterterms yields
\begin{subequations}
\begin{align}
 \langle\mathcal O_\theta\rangle &=-\frac{1}{\sqrt{-g^{}_{\mathrm{AdS}_4}}}\,\frac{\delta}{\delta\theta^{(0)}} S_\mathrm{D7,ren}
 =T_7 V_{\mathrm{S}^3}\left[2c+(1+4\alpha_1)m^3+\frac{5+8\alpha_2-4\log2}{2}m\right]~,\\
 \langle\mathcal O_f\rangle &=-\frac{1}{\sqrt{-g^{}_{\mathrm{AdS}_4}}}\,\frac{\delta}{\delta f^{(0)}} S_\mathrm{D7,ren}
 =T_7 V_{\mathrm{S}^3}i\lambda\left[c-\frac{m}{3} \left(m^2-3+\log (8)-6 \alpha _4\right)\right]~,
\end{align}
\end{subequations}
where $\theta^{(0)}$ and $f^{(0)}$ denote the leading coefficients in the near-boundary expansion, i.e.\
$\theta=\pi/2+\theta^{(0)}z+\dots$ and $f=z^2\log z\,f^{(0)}+\dots$.
As expected, the scheme-dependent terms only contribute parts proportional to $m$,
and for $m=0$ there is no scheme dependence.
The precise value of $c$ depends on the embedding we choose. 
We recall the explicit values for $m\geq0$ in the $\rho>0$ patch.
For the short and long embeddings, we find from (\ref{eqn:short-embedding}) and (\ref{eqn:long-embedding})
\begin{align}
 c_\mathrm{short}&=\frac{m^2-2}{3}\sqrt{m^2+1}+ m \sinh ^{-1}(m)~,&
 c_\mathrm{long}&=-c_\mathrm{short}~.
\end{align}
Recall that the long embeddings are possible only for $m<m_\ell$, with $m_\ell$ given in (\ref{eqn:critical-mass}).
For the connected embeddings, which also only exist in this mass range, $c$ is free to vary within the ranges given by (\ref{eqn:connected-embeddings}), that is between $c_\mathrm{short}$ and $c_\mathrm{long}$.

\subsection{Interpretation}\label{sec:interpretation}

We found that, for a range of masses, a one-parameter family of solutions exists with different values of the
chiral and scalar condensates for one and the same mass.
We now attempt to interpret what these solutions mean in the field theory.
Our interpretation will be somewhat similar to the one offered in \cite{Clark:2013mfa} for the case of non-supersymmetric flavors on AdS$_4$.
The family of massless solutions should be easiest to understand.
Recall that there is one solution among this family that is singled out: the $m=c=0$ connected solution,
which is the only massless solution where chiral symmetry is not spontaneously broken.
This solution can be conformally mapped to massless flavors on flat space, as described at the beginning of this section.
Correspondingly, this particular solution should correspond to flavored \N{4} SYM with transparent boundary conditions.

Under the same conformal transformations our other solutions also map into massless flavors on all of flat space, but now with a position dependent 
chiral condensate that falls of as $1/z^3$ as a function of distance to the plane at $z=0$.
Our basic suggestion is that these other embeddings should correspond to supersymmetric flavors in the field theory with different boundary
conditions imposed on the flavor fields at $z=0$.
Only the standard transparent boundary conditions will yield a vanishing chiral condensate.
This case is easiest to make for the other extreme case: the disconnected embedding with $m=0$ and $c=-2/3$,
shown as thick black line in Fig.~\ref{fig:slipping-mode-Z2-even}.
In this case, we can decide to only study flavors in one of the two asymptotic AdS spaces (or, alternatively, on one half of Minkowski space).
The disconnected embedding at positive $\rho$ is perfectly smooth and well behaved without the second disconnected embedding at negative $\rho$.
Since we now only added flavors in the $z>0$ half of Minkowski space, this embedding can not correspond to a field theory with transparent boundary conditions. In the notation introduced above (\ref{eqn:transparent-bc}), only the R fields exist, there are no L fields we could relate them to at $z=0$.
So one has to impose boundary conditions at $z=0$ on the R flavor fields alone, presumably either Dirichlet
or Neumann conditions.
Either of these choices is expected to give rise to a position dependent condensate, just as we observed in our brane embedding.
This happens already in the free field theory, as can be seen by employing the standard method of images, as was e.g.\ done in this context in \cite{Clark:2004sb}. For simplicity, let us consider the case of a scalar field. The propagator of a scalar field on
half space is given by
\begin{align}
G(x,y) = \frac{1}{4 \pi} \left ( \frac{1}{(x-y)^2} \pm \frac{1}{(x-Ry)^2} \right ),
\end{align}
where $(Ry)_{\mu} = (t_y,x_y,y_y,-z_y)$ for $y_{\mu} = (t_y,x_y,y_y,z_y)$ and upper/lower sign corresponds to Neumann/Dirichlet boundary conditions.
To calculate the expectation value $\langle X^2(x) \rangle$, we simply need to evaluate $G(x,x)$.
The first term gives rise to a divergent contribution that needs to be subtracted in order to properly define the composite operator $X^2$.
In \N{4} SYM, the corresponding divergence was even shown to cancel between contributions from different fields in \cite{Clark:2004sb}.
We get a non-vanishing expectation value from the mirror charge term that goes as $1/z^2$, as appropriate for a dimension-2 operator.
For a fermion a similar calculation gives $1/z^3$, and these are indeed the expectation values as we found here.
So in principle, at least for this special configuration, the behavior we found holographically makes qualitative sense.
It would be nice to see whether non-renormalization theorems could allow a quantitative comparison between free and
strongly coupled field theories, as was done in the context of Janus solutions in \cite{Clark:2004sb}.

For the other embeddings, that is the connected embeddings with $m=0$, $c \neq 0$
and the disconnected embeddings with $m=0$ in both halves of spacetime, we have no strong argument for
what the field theory boundary conditions should be.
But we suspect that they interpolate between the transparent boundary conditions at $m=c=0$ and Dirichlet (or Neumann)
boundary conditions for the maximal $c$.
It is straightforward to formulate such interpolating boundary conditions that give a $1/z^3$ condensate with growing coefficient,
but the choice is not unique.
Potentially, a careful study of supersymmetry together with our results for the expectation values could help pin this down.
But at least our results for the massless embeddings appear to be consistent with this interpretation.

For massive embeddings there is no such simple argument.
Mapping to flat space gives position-dependent masses that diverge at $z=0$, and so any discussion of boundary conditions is more involved.
But it is tempting to relate the presence of connected and long embeddings for small mass to boundary conditions in a similar way.
In the window $m_\mathrm{BF}^2\leq m^2< m^2_\mathrm{BF}+1$ we can do standard and alternative quantization for scalar fields on AdS$_4$ \cite{Breitenlohner:1982bm,Breitenlohner:1982jf}.
That similarly allows for a family of boundary conditions, and hence presumably a family of expectation values.
But which mass is it that approaches $m^2_\mathrm{BF}+1$ when we dial $m$ from zero to its critical value given in (\ref{eqn:critical-mass})?
Note that, for a given leading coefficient $m$ in our expansion of the fluctuating field $\theta$ in the bulk,
the field theory mass $M$ is actually given by $M = \sqrt{\lambda}/(2\pi) m$ \cite{Kruczenski:2003be}.
Since this $M$ is thus much larger than 1 except for infinitesimally small values of $m$, we find that both mass
eigenvalues in \eqref{masseigenvalues} are large and positive for any finite $m$.
So none of the fundamental fields is even close to the window in which two different boundary conditions are allowed.
However, while the fundamental fields have masses of order $\sqrt{\lambda}$ in all our embeddings, the gauge invariant meson fields
actually have order one masses \cite{Kruczenski:2003be}.
This makes the mesons a natural candidate for a field that obeys different boundary conditions in our different embeddings for one and the same mass.
We can indeed see directly from the geometry that the meson spectrum is strongly affected by the difference of embeddings.
Corresponding to the different classes of brane embeddings, we get different classes of mesons:
those built from pairs of L quarks and R quarks separately (again in the notation introduced above (\ref{eqn:transparent-bc})),
and those with mixed content.
Denoting the mesons by their quark content, we see that for connected embeddings
both LL, RR as well as LR and RL mesons are light, as they all correspond to fluctuations of the brane.
For disconnected embeddings, only LL and RR mesons can be light, whereas LR and RL mesons correspond to semi-classical strings stretched
between the two disconnected branes, and hence to order $\sqrt{\lambda}$ masses.
A more quantitative discussion of this suggestion requires an analysis of the meson spectrum, 
encoded in the spectrum of linearized fluctuations around our embeddings, and is beyond the scope of the present work.

\section{Discussion}\label{sec:discussion}

The main result of this work is a class of supersymmetric D7-brane embeddings into AdS$_5\times$S$^5$, which allow
to holographically describe \N{4} SYM coupled to massive \N{2} supersymmetric flavor hypermultiplets on spaces of constant curvature.
For AdS$_4$-sliced AdS$_5\times$S$^5$, which corresponds to a field theory on two copies of AdS$_4$, 
the embeddings are given in Sec.~\ref{sec:fully-massive-embedding-solution}, 
and for S$^4$ and dS$_4$ slices in App.~\ref{app:global-Euclidean-kappa}.
Preserving supersymmetry in the transition from flat to curved space needs additional care for non-conformal theories,
and in particular requires extra compensating terms to make mass terms supersymmetric \cite{Festuccia:2011ws}.
This has to be taken into account in the construction of holographic duals as well, and translates to
non-trivial profiles for some of the matter fields in the corresponding solutions.
For the D7-branes the compensating mass term on the field theory side translates to non-trivial worldvolume flux.
Finding supersymmetric probe brane embeddings translates to solving for the constraint imposed by 
$\kappa$-symmetry on the background Killing spinors to have non-trivial solutions, and we went through that
discussion systematically in Sec.~\ref{sec:susy-d3d7}.
Isolating necessary conditions for supersymmetric embeddings from the $\kappa$-symmetry analysis, although technically cumbersome,
allowed us to decouple the slipping mode and the gauge field, and find  analytic solutions.

We then focused on a more detailed analysis of AdS$_4$-sliced AdS$_5$.
AdS is a preferred choice among the constant-curvature spaces in Lorentzian signature,
as the corresponding supergroups have unitary representations and realizing supersymmetric QFTs
consequently is more straightforward than on dS.
Holographically we naturally get two copies of AdS$_4$ as boundary geometries,
each one being the boundary in of the two coordinate patches needed to cover AdS$_5$ with AdS$_4$ slices.
In Sec.~\ref{sec:embedding-classes} we discussed in detail the families of regular massive D7-brane embeddings,
and how they can be combined for the two patches.
That revealed a surprisingly rich set of options for small masses.
For generic large masses, the D7-branes cap off in the coordinate patch where they extend to the conformal boundary,
much like they do on Poincar\'{e} AdS.
This feature of the ``short'' embeddings reflects that the massive flavors do not affect the deep IR of the QFT,
as they are simply gapped out.
On Poincar\'{e} AdS this is the generic behavior, regardless of the value of the mass \cite{Karch:2002sh}.
For the AdS$_4$ slices, on the other hand, we found that for small masses there are also ``long'' brane embeddings,
which cover all of the patch in which they extend to the conformal boundary.
They extend into the second patch and cap off at a finite value of the radial coordinate there,
as illustrated in Fig.~\ref{fig:D7-embeddings}.
In addition to that, we found families of connected embeddings, 
which cover both copies of AdS$_4$ on the conformal boundary.
Similar, although non-supersymmetric embeddings had been found numerically in \cite{Clark:2013mfa} before.
The connected embeddings become available below the same critical mass as the long embeddings.
They then come in one-parameter families for each fixed mass, corresponding to different values for the chiral and scalar condensates.
Generically, that family of connected embeddings interpolates between long and short embeddings, as shown in Fig.~\ref{fig:connected-long-embeddings}.
For the particular case of massless flavors, the one-parameter family of connected embeddings includes the one
conformally related to massless flavors on flat space as special case.
A phase diagram summarizing the embeddings can be found in Fig.~\ref{fig:m-c-diagram}.

The embeddings stretching all through at least one of the coordinate patches, which are available for low enough values of the mass,
suggest that those flavors can indeed affect also the deep IR regime of the dual QFT, despite their non-zero mass.
A natural candidate feature of QFT on AdS$_4$, that is suggestive of this behavior,
is the possibility of having stable negative-mass scalars.
We turned to a more detailed discussion focussing on the QFT side in Sec.~\ref{sec:N4SYM}.
After a general discussion of how precisely the embeddings encode the flavor mass
and of the relation to supersymmetric Janus solutions,
we carried out the holographic renormalization and computed the one-point functions.
For the massless embeddings we found that the one familiar from Poincar\'e AdS is the only one
where chiral symmetry is not spontaneously broken. 
The others have non-vanishing (constant) chiral and scalar condensates.
We also gave a simple field theory toy model to explain the one-parameter family of massless embeddings, based on
the possible choices of boundary conditions at the conformal boundary for each of the AdS$_4$ spaces.
For the massless embeddings we could employ a conformal map to two halves of Minkowski space,
mapping the constant condensates on AdS$_4$ to position-dependent condensates on Minkowski space.
For a family of free theories with boundary conditions interpolating between transparent and either
Dirichlet or Neumann boundary conditions,
we found exactly the position-dependent scalar condensates that came out of our holographic calculation.
This suggests that going through the different embeddings available at zero mass might be a holographic analog of the transitions studied in \cite{Shapere:2012wn}.
As discussed in Sec.~\ref{sec:interpretation}, the meson spectrum could give crucial insights into whether this interpretation
can also explain the one-parameter families of massive embeddings.

Our results open up a number of additional directions for future research. 
In a companion paper we use the S$^4$ embeddings to conduct a precision test
for probe brane holography using supersymmetric localization.
In the same spirit, following through the $\kappa$-symmetry analysis laid out in Sec.~\ref{sec:susy-d3d7} 
should  allow to find similar supersymmetric embeddings into S$^1\times$S$^3$-sliced AdS$_5\times$S$^5$, 
and thus to compute the superconformal index.
Another direction can be seen by noting that topological twisting \cite{Witten:1988ze} can also be seen 
as an example where compensating terms restore supersymmetry.
One can think of topological twisting as turning on the background R-charge gauge field.
If it is chosen to be equal to the spin connection\footnote{More precisely,
for \texorpdfstring{\N{4}}{N=4} SYM one picks \cite{Vafa:1994tf} a SU($2$) subgroup of the SU(4)$_R$ symmetry and sets the corresponding 
gauge field equal to the SU$(2)_r$ part of the spin connection,
which transforms in the $(1,3) \oplus (3,1)$ representation of the Spin($4$)=SU($2)_l \times $SU$(2)_r$ Euclidean Lorentz group.}, 
some supercharges can be made to transform as scalars under parallel transport in the gravitational and R-charge background.
Turning on the background gauge fields adds extra terms to the Lagrangian, proportional to the R-current, 
and these serve as the compensating terms that restore supersymmetry.
This twisting procedure puts no constraints on the geometry of the background space,
and in principle it will be straightforward to implement it in the dual bulk description.
The R-charge gauge field is now dynamical,
and the topological twisting implies that its leading (radially independent) piece at the boundary no 
longer is taken to vanish, but is set equal to the spin connection.
It would be very interesting to construct the corresponding supergravity solutions.
As compared to the compensating terms of the topologically twisted \N{4} theory,
the implementation of the compensating terms of \cite{Pestun:2007rz,Festuccia:2011ws,Clark:2004sb} is easier to accomplish holographically,
since we only need to turn on a single extra scalar field.
For \N{2}$^*$ this was done in \cite{Bobev:2013cja}, for super Janus in \cite{Clark:2005te,D'Hoker:2006uu} and for flavored \N{4} SYM in this work.

\section*{Acknowledgments}
The work of AK and BR was supported, in part, by the US Department of Energy under grant number DE-SC0011637.
CFU is supported by {\it Deutsche Forschungsgemeinschaft} through a research fellowship.

\appendix

\section{Clifford algebra identities}\label{sec:s5-gamma-identities}
In this appendix we derive the technical identities needed to explicitly evaluate the $\kappa$-symmetry condition in Sec.~\ref{sec:fully-massive},
and specifically $\RS^{-1}\RAdS^{-1}$(\ref{eqn:kappa-nonlinear-5}).
We use $\tilde\Gamma_{\rho\mathrm{A}}$ defined in (\ref{eqn:Rtilde-R-relation}),
and define the operator $\mathcal R[\Gamma]:=\RS^{-1}\Gamma\RS$.
Noting that
$\hat\Gamma\epsilon=
\left[\Gamma^{\underline{\psi}}\Gamma^{\underline{\theta}}+\theta^\prime\Gamma_{\underline{\theta}}\GammaChi\Gamma_{\underline{\rho}}\GammaAdS\right]\epsilon$,
we then find
\begin{align}\nonumber
 \RS^{-1}\RAdS^{-1}\hat\Gamma\epsilon&=
 \mathcal R[\Gamma^{\underline{\psi}}\Gamma^{\underline{\theta}}]\epsilon_0
 +\theta^\prime\mathcal R[\Gamma^{\underline{\theta}}\GammaChi]\,\tilde\Gamma_{\rho\mathrm{A}}\epsilon_0
 \\&=
 \mathcal R[\Gamma^{\underline{\theta}}\GammaChi]\left[i\cot\theta\cdot\mathds{1}+\theta^\prime\tilde\Gamma_{\rho\mathrm{A}}\right]\epsilon_0
 -i\csc\theta\epsilon_0~.
 \label{eqn:LHS-aux-1}
\end{align}
For the second line we used
$\mathcal R[\Gamma^{\underline{\psi}}\Gamma^{\underline{\theta}}]=
i\cot\theta\mathcal R[\Gamma^{\underline{\theta}}\GammaChi]-i\csc\theta\Gamma^{\underline{\theta}}\GammaChi$.
This will allow us to evaluate the first term on the left hand side in (\ref{eqn:kappa-nonlinear-5}).
For the second term we use the relation for $\tilde R_\mathrm{AdS}$ of (\ref{eqn:Rtilde-R-relation}),
to find
\begin{align}\label{eqn:LHS-aux-2}
 \RS^{-1}\RAdS^{-1}\Gamma_{\underline{\rho}}\GammaAdS\Gamma^{\underline{\psi}}\RS\tilde R_\mathrm{AdS}\Gamma_{\underline{\rho}}\epsilon_0&=
 \mathcal R[\Gamma^{\underline{\theta}}\GammaChi]\left[\cosh\rho\cdot\mathds{1}-i\sinh\rho\,\tilde\Gamma_{\rho\mathrm{A}}\right]\epsilon_0~.
\end{align}
This will allow us to evaluate the first term in the brackets of the second term on the left hand side in (\ref{eqn:kappa-nonlinear-5}).
For the last term we note that
\begin{align}
 \Gamma^{\underline{\theta}}\RS\tilde R_\mathrm{AdS}\Gamma_{\underline{\rho}}\epsilon_0&=
 \Gamma^{\underline{\psi}}\GammaChi\RS\left[\cosh\rho\Gamma_{\underline{\rho}}\GammaAdS+i\sinh\rho\cdot\mathds{1}\right]\RAdS\epsilon_0~.
\end{align}
So we find
\begin{align}
 \RS^{-1}\RAdS^{-1}\Gamma^{\underline{\theta}}\RS\tilde R_\mathrm{AdS}\Gamma_{\underline{\rho}}\epsilon_0
 &=\mathcal R[\Gamma^{\underline{\psi}}\GammaChi]\left[\cosh\rho\,\tilde\Gamma_{\rho\mathrm{A}}+i\sinh\rho\cdot\mathds{1}\right]\epsilon_0~.
\end{align}
We will now use
$\mathcal R[\Gamma^{\underline{\psi}}\GammaChi]=i\csc\theta\,\mathcal R[\Gamma_{\underline{\theta}}\Gamma\chi]\,\Gamma_{\underline{\theta}}\GammaChi-i\cot\theta\cdot\mathds{1}$,
which gives us
\begin{align}
\begin{split}\label{eqn:LHS-aux-3}
 \RS^{-1}\RAdS^{-1}\Gamma^{\underline{\theta}}\RS\tilde R_\mathrm{AdS}\Gamma_{\underline{\rho}}\epsilon_0
 =&\left[\csc\theta\,\mathcal R[\Gamma_{\underline{\theta}}\GammaChi]-\cot\theta\,\mathds{1}\right]\times
  \left[i\cosh\rho\tilde\Gamma_{\rho\mathrm{A}}-\sinh\rho\cdot\mathds{1}\right]\epsilon_0~.
\end{split}
\end{align}
These are the tools needed in Sec.~\ref{sec:fully-massive} to identify the parts with non-trivial dependence on the internal space
on the left hand side of (\ref{eqn:kappa-nonlinear-5}).

\section{Supersymmetric D3/D7 for massive flavors on \texorpdfstring{S$^4$}{S4}} \label{app:global-Euclidean-kappa}
We now go through the derivation of $\kappa$-symmetric D7 embeddings into AdS$_5\times$S$^5$,
where AdS$_5$ is Euclidean and in global coordinates.
The result will allow us to holographically describe \N{4} SYM coupled to (massive) flavor hypermultiplets on S$^4$.
For the metric we take $g=g_{\mathrm{AdS}_5}+g_{\mathrm{S}^5}$,
with the S$^5$ metric given in (\ref{eqn:AdS5S5-metric}).
For the S$^4$ part of AdS$_5$ we use conformally flat coordinates to simplify the explicit computations,
but note that the resulting embeddings are independent of the chosen S$^4$ coordinates.
The AdS$_5$ metric then takes the form
\begin{align}
 g_{\mathrm{AdS}_5}^{}&=dR^2+\sinh^2\!R\, d\Omega_4^2~,&
 d\Omega_4^2&=W^{-2}d\vec{x}^2~,&
 W&=1+\vec{x}^2~.
\end{align}
The Killing-spinor equation for Euclidean AdS differs by a factor of $i$ from the Lorentzian one, and there is a sign convention to be fixed.
For the analytic continuation to be discussed momentarily, we have
\begin{align}\label{eqn:Killing-spinor-eq-E}
 D_\mu\epsilon&=\frac{1}{2}\,\varGammaAdS\varGamma_\mu\epsilon~,\quad \mu=0\dots 4~.
\end{align}
We have denoted by $\varGamma$ (as opposed to $\Gamma$) the Euclidean Clifford-algebra generators.
The conventions for the Euclidean Clifford algebra will be laid out in more detail along with the analytic continuation below.
The Killing spinor equation for the S$^5$ part stays the same, and is given in (\ref{eqn:Killing-spinor-eq}).
The AdS$_5\times$S$^5$ Killing spinors are again of the form (\ref{eqn:Killing-spinors-general}),
i.e.\ $\epsilon=\RAdS \RS \epsilon_0$,
with $R_{\mathrm{S}^5}$ given in (\ref{eqn:Killing-spinors}) and
\begin{align}\label{eqn:RAdSglobal}
 R_\mathrm{AdS}&=W^{-1/2}e^{\frac{1}{2}\rho\varGamma_{\underline{\rho}}\varGammaAdS}
 \left[\mathds{1}+x_i\varGamma_{\underline{\rho}}\varGamma^{\underline{x_i}}\right]~.
\end{align}
For the embedding and gauge field we choose the same ansatz as before and motivated in Sec.~\ref{sec:embedding-ansatz}.
That is, we take a non-trivial slipping mode as function of the
radial coordinate $\rho$ and a worldvolume gauge field $A=f\omega$.
Note that although we have used the same name for the radial coordinate, $\rho$, it does not have the same geometric meaning as in the AdS$_4$-sliced case,
and consequently the embeddings are geometrically different.
The ten-bein pulled back to the worldvolume is again given by (\ref{eqn:D7-vielbein}),
and the induced metric reads
\begin{align}\label{eqn:induced-metric-global-AdS}
 g&=\big(1+{\theta^\prime}^2\big)d\rho^2+\sinh^2\!\rho\, d\Omega_4^2+\sin^2\!\theta\,d\Omega_3^2~.
\end{align}

\subsection{Analytic continuation}
For the S$^3$ mode $\omega$ appearing in the gauge field we again take (\ref{eqn:omega}).
Evaluating the $\kappa$-symmetry projection conditions (\ref{eqn:kappa})
then proceeds almost in the same way as for the
AdS$_4$ slicing in Lorentzian signature all the way up to (\ref{eqn:kappa-nonlinear-2}).
To make this more precise, we will have to set up the analytic continuation to Euclidean signature.
The metric quantities as we set them up are already in Euclidean signature,
and the more subtle step is to implement the analytic continuation on the Clifford algebra and spinors.
The Killing spinors we have given in (\ref{eqn:RAdSglobal}) also assume a Euclidean-signature Clifford algebra already.
But we have to implement the continuation in the $\kappa$-symmetry condition (\ref{eqn:kappa}).
To reflect the change in signature we set
\begin{align}
 \Gamma_{\underline{x_0}}&\rightarrow \varGamma_{\underline{x_0}}=i\Gamma_{\underline{x_0}}~,&
 \Gamma^{\underline{x_0}}&\rightarrow \varGamma^{\underline{x_0}}=-i\Gamma^{\underline{x_0}}~.
\end{align}
With this continuation, we note that $C\varGamma_{\underline{x_0}}^\star=-\varGamma_{\underline{x_0}}C$.
The S$^5$ part is not affected by the analytic continuation, and for the AdS part we define $\varGammaAdS$
using the same expressions as before below (\ref{eqn:Killing-spinor-eq}).
That is, with indices up but $\Gamma$s replaced by $\varGamma$s.
This gives $\varGammaAdS=-i\GammaAdS$ (lowering the indices, however, does not produce a sign for $\varGammaAdS$).
This is the matrix we have used in (\ref{eqn:RAdSglobal}).
For the chirality projector we define $\varGamma_{11}=-i\varGammaAdS\varGammaS=\Gamma_{11}$.

We can now take a closer look at the matrix $\tilde R_\mathrm{AdS}$, defined by $C\RAdS=\tilde R_\mathrm{AdS}C$ as before.
We follow \cite{Wetterich:2010ni} (extending earlier work in \cite{Nicolai:1978vc}), in including a time reflection in the
complex structure on Euclidean space to be compatible with analytic continuation, i.e.\ $x_0^\star=-x_0$.
The charge conjugation matrix is kept as the Lorentzian one.
Noting that now $C\varGammaAdS=-\varGammaAdS C$, we then find
\begin{align}\label{eqn:Rtilde-R-E}
 \tilde R_\mathrm{AdS}&=e^{\rho\varGammaAdS\varGamma_{\underline{\rho}}}\RAdS~.
\end{align}
Comparing to (\ref{eqn:Rtilde-R-relation}), we note that this relation is different from the
one we found for the AdS$_4$ slicing in Lorentzian signature.
That means we will also have to change the projection condition (\ref{eqn:massive-projector}) to solve the $\kappa$-symmetry constraint.

\subsection{\texorpdfstring{$\kappa$}{kappa}-symmetry}
Although we did not have to explicitly unpack the AdS$_4$-slice $\Gamma$-matrices in Sec.~\ref{sec:susy-d3d7},
the analytic continuation still has implications.
We start with the $\kappa$-symmetry condition as spelled out in (\ref{eqn:kappa}).
The lowercase $\gamma$'s are now defined as $\gamma_i=e_i^a\varGamma_a$, i.e.\ with the Euclidean $\varGamma$-matrices, and we have
\begin{align}\label{eqn:Gamma0-Euclidean}
 \varGamma_{(0)}&=\frac{i}{(p+1)!\sqrt{\det g}}\epsilon^{i_1\dots i_{p+1}}\gamma^{}_{i_1\dots i_{p+1}}~.
\end{align}
The extra $i$ is due to the change in sign for the metric determinant.
Since our embedding ansatz is formally the same as for the AdS$_4$ slicing,
we get to an analog of (\ref{eqn:kappa-with-our-ansatz}) in just the same way, and find
\begin{align}\label{eqn:kappa-with-our-ansatz-E}
 \varGamma_\kappa\epsilon&=\frac{-i}{\sqrt{\det(1+X)}}\left[
 \Big(\mathds{1}+\frac{1}{8}\gamma^{ijkl}F_{ij}F_{kl}\Big)\varGamma_{(0)}\epsilon
 +\frac{1}{2}\gamma^{ij}F_{ij}C\left(\varGamma_{(0)}\epsilon\right)^\star\right]~.
\end{align}
Since $F$ does not have timelike components, its continuation is trivial.
The matrices appearing in (\ref{eqn:kappa-with-our-ansatz-E}) are now
\begin{align}\label{eqn:Gamma0-Gammahat-E}
 \varGamma_{(0)}&= \frac{i}{\sqrt{1+{\theta^\prime}^2}}\hat\varGamma~,
 &
 \hat\varGamma&=\left[\mathds{1}+\theta^\prime\varGamma_{\underline{\theta}}\varGamma_{\underline{\rho}}\right]\varGammaAdS\varGammaChi~.
\end{align}
As compared to the Lorentzian case, there is now no relative sign between
(\ref{eqn:Gamma0-Gammahat-E}) and (\ref{eqn:Gamma0-Euclidean}),
since $\varGammaAdS$ is equal to the product of all AdS$_5$ $\varGamma$-matrices with indices down.
To proceed to the analog of (\ref{eqn:kappa-nonlinear-2}), we note that the function $h$ again evaluates to (\ref{eqn:h-def}).
Since now $C(i\varGammaAdS)^\star=i\varGammaAdS C$, we get the same sign in the
term with the charge conjugation matrix as before in (\ref{eqn:kappa-nonlinear-2}), and find
\begin{align}\label{eqn:kappa-globalAdS}
 \Big(\mathds{1}+\frac{1}{8}\gamma^{ijkl}F_{ij}F_{kl}\Big)\hat\varGamma\epsilon
 +\frac{1}{2}\gamma^{ij}F_{ij}\hat\varGamma C\epsilon^\star
 &=h\epsilon~.
\end{align}

\subsubsection{Explicit evaluation}
Since we kept the S$^3$ mode (\ref{eqn:omega}), we again find (\ref{eqn:F2-term}) for the $F^2$-term.
As pointed out above (\ref{eqn:Rtilde-R-E}), we define $R$-matrices with a tilde analogously to Sec.~\ref{sec:susy-d3d7}.
Since $\RS$ does not contain any AdS$_5$ $\Gamma$-matrices, it is not affected by the Wick rotation and
this procedure results in the same matrix as in Sec.~\ref{sec:susy-d3d7}.
Instead of the projection condition (\ref{eqn:massive-projector}), we now use
\begin{align}\label{eqn:massive-projector-E}
  \tilde\varGamma C\epsilon^\star_0&=\lambda\epsilon_0~,&
  \tilde\varGamma&=\varGamma^{\underline{\chi_1}}\varGamma^{\underline{\theta}}~.
\end{align}
We then find that (\ref{eqn:kappa-globalAdS}) evaluates to
\begin{align}\label{eqn:kappa-globalAdS-2}
 \hat\varGamma\epsilon
 +\frac{\lambda}{2}\gamma^{ij}F_{ij}\hat\varGamma \tilde R_{\mathrm{S}^5}\tilde R_\mathrm{AdS}\tilde\varGamma\epsilon_0
 &=2f^\prime f \csc^3\!\theta\,\varGamma_{\underline{\rho}}\varGammaAdS\epsilon+h\epsilon~.
\end{align}
We first evaluate the $\gamma\cdot F$ term. With the definitions above we find
that on the spinor subspace singled out by $P_0$ (which stays the same as in Lorentzian signature) and $(\mathds{1}+\varGamma_{11})$,
\begin{align}
 \frac{1}{2}\gamma^{ij}F_{ij}\hat\varGamma\tilde R_{\mathrm{S}^5}\varGamma^{\underline{\theta}}\varGamma^{\underline{\chi_1}}
 &=
 \csc\theta\left[
 if^\prime\varGamma_{\underline{\rho}}\varGamma^{\underline{\psi}}-2f\csc\theta\left(\varGamma^{\underline{\theta}}-\theta^\prime\varGamma_{\underline{\rho}}\right)\varGammaAdS
 \right]\RS~.
\end{align}
With that in hand we can evaluate (\ref{eqn:kappa-globalAdS-2}), for which we find
\begin{align}\label{eqn:kappa-globalAdS-3}
\begin{split}
 \mathrm{LHS}:=\hat\varGamma\epsilon
 -&\lambda\csc\theta\left[
   if^\prime\varGamma_{\underline{\rho}}\varGamma^{\underline{\psi}}-2f\csc\theta\varGamma^{\underline{\theta}}\varGammaAdS
 \right] \RS\tilde R_\mathrm{AdS}\epsilon_0
 \\
 &=
 2\lambda f\theta^\prime\csc^2\!\theta\,\RS\varGamma_{\underline{\rho}}\varGammaAdS\tilde R_\mathrm{AdS}\epsilon_0
 +2f^\prime f\csc^3\theta\varGamma_{\underline{\rho}}\varGammaAdS\epsilon+h\epsilon=:\mathrm{RHS}~.
\end{split}
\end{align}
The left hand side once again is linear in $f$ and its derivative,
and the right hand side does not involve explicit S$^5$ $\Gamma$-matrices.
To proceed, we need the analogues of (\ref{eqn:LHS-aux-1}), (\ref{eqn:LHS-aux-2}) and (\ref{eqn:LHS-aux-3}).
With $\tilde\varGamma_{\rho\mathrm{A}}=\RAdS^{-1}\varGamma_{\underline{\rho}}\varGammaAdS\RAdS$, we find
\begin{align}\label{eqn:LHS-aux-1-E}
 \RS^{-1}\RAdS^{-1}\hat\varGamma\epsilon&=
 \mathcal R[\varGamma^{\underline{\theta}}\varGammaChi]\left[\theta^\prime\tilde\varGamma_{\rho\mathrm{A}}-\cot\theta\cdot\mathds{1}\right]\epsilon_0
 +\csc\theta\,\epsilon_0~,
\\
\label{eqn:LHS-aux-2-E}
 \RS^{-1}\RAdS^{-1}\varGamma_{\underline{\rho}}\varGamma^{\underline{\psi}}\RS\tilde R_\mathrm{AdS}\epsilon_0&=
 i\mathcal R[\varGamma^{\underline{\theta}}\varGammaChi]\left[\sinh\rho\cdot\mathds{1}-\cosh\rho\,\tilde\varGamma_{\rho\mathrm{A}}\right]\epsilon_0~,
\\
\label{eqn:LHS-aux-3-E}
 \RS^{-1}\RAdS^{-1}\varGamma^{\underline{\theta}}\varGammaAdS\RS\tilde R_\mathrm{AdS}\epsilon_0
 &=\left[\csc\theta\,\mathcal R[\varGamma_{\underline{\theta}}\varGammaChi]-\cot\theta\,\mathds{1}\right]\times
  \left[\cosh\rho\cdot\mathds{1}-\sinh\!\rho\,\tilde\varGamma_{\rho\mathrm{A}}\right]\epsilon_0~.
\end{align}
We now come back to the left hand side of (\ref{eqn:kappa-globalAdS-3}).
Since the right hand side does not involve S$^5$ $\Gamma$-matrices, any terms involving those will have to vanish on the left hand side.
We find
\begin{align}
\begin{split}
 \RS^{-1}\RAdS^{-1}\mathrm{LHS}=&
 \left(\theta^\prime-\lambda f^\prime \csc\theta\cosh\rho-2\lambda f \csc^3\!\theta\,\sinh\rho\right)\mathcal R[\varGamma^{\underline{\theta}}\varGammaChi]\tilde\varGamma_{\rho\mathrm{A}}\\
 &-\left(\cot\theta-\lambda f^\prime \csc\theta\sinh\rho-2\lambda f\csc^3\!\theta\cosh\rho\right)R[\varGamma^{\underline{\theta}}\varGammaChi]\\
 &+\csc\theta\epsilon_0-2\lambda f\csc^2\!\theta\, \cot\theta\left(\cosh\rho\cdot\mathds{1}-\sinh\rho\tilde\varGamma_{\rho\mathrm{A}}\right)\epsilon_0~.
\end{split}
\end{align}
Just like in the AdS$_4$-sliced case, the round brackets of the upper two lines on the right hand have to vanish separately,
since they multiply linearly independent $\Gamma$-matrix structures and nothing on the right hand
side of (\ref{eqn:kappa-globalAdS-3}) can cancel them.
Since $f$ and $f^\prime$ appear linearly, we can solve for them and find
\begin{align}\label{eqn:f-fp-Euclidean}
 f&=\frac{1}{2 \lambda }\sin^3\!\theta \left(\cot\theta \cosh\rho -\theta^\prime \sinh\rho\right)~,&
 f^\prime&=\frac{1}{\lambda}\left(\theta^\prime\sin\theta\cosh\rho - \cos\theta \sinh\rho\right)~.
\end{align}
Since both of these are functions of $\theta$ and $\theta^\prime$ only, we can derive a second-order ODE for $\theta$, which reads
\begin{align}\label{eqn:theta-E}
 \theta^{\prime\prime} + 3 {\theta^\prime}^2 \cot\theta + 4 \theta^\prime\coth\rho - \cot\theta \left(1+2 \csc ^2\theta\right)&=0~.
\end{align}
These will once again be our main results.
With (\ref{eqn:f-fp-Euclidean}), we can simplify (\ref{eqn:kappa-globalAdS-3}) -- after applying $\RS^{-1}\RAdS^{-1}$ --
to find the remaining condition
\begin{align}
\begin{split}
 &\left[\csc\theta-h+2\lambda f\csc^2\!\theta\left(\theta^\prime\sinh\rho-\cot\theta\cosh\rho\right)\right]\epsilon_0\\
 &+2f\csc^2\theta\left[f^\prime\csc\theta+\lambda\left(\theta^\prime\cosh\rho-\cot\theta\sinh\rho\right)\right]\tilde \varGamma_{\rho\mathrm{A}}\epsilon_0=0~.
\end{split}
\end{align}
The terms in brackets mutltiply independent $\varGamma$-matrix structures and have to vanish separately.
Thanks to (\ref{eqn:f-fp-Euclidean}) they do indeed vanish separately if $\lambda^2=-1$.
To complete the analysis, we once again checked that (\ref{eqn:f-fp-Euclidean}) and (\ref{eqn:theta-E}) together imply that
the equations of motion for $f$ and $\theta$ resulting from the DBI action are satisfied.

\subsubsection{Solutions}
The solutions to (\ref{eqn:theta-E}) again come in 3 branches, and the one which gives real slipping mode is
\begin{align}\label{eqn:soltheta-E}
 \theta&=\cos^{-1}\left(2\cos\frac{2\pi k+\cos^{-1}\tau}{3}\right)~,&
 \tau&=\frac{3m \sinh(2\rho)-6c-6m \rho}{4\sinh^3\rho}~,
\end{align}
with $k=2$.
The function $\tau$ is related to that in (\ref{eqn:soltheta}) by a simple analytic
continuation $\rho\rightarrow \rho+i\pi/2$ along with a redefinition of the parameters $m$ and $c$,
and the same applies for the accompanying gauge field.
The gauge field is now imaginary, which had to be expected from the discussion of the field-theory side in Sec.~\ref{sec:curved-susy-QFT},
and specifically the results of \cite{Pestun:2007rz,Festuccia:2011ws}.
We note that even though the final embeddings could have been obtained by a simple analytic continuation
from the AdS solutions,
the evaluation of the $\kappa$-symmetry constraint differs from the AdS case by more than that, due to the change in the Killing spinors and projectors.

Finally, we note that the same embeddings are solutions for D7-brane embeddings into dS$_4$-sliced AdS$_5$
with metric
\begin{align}\label{eqn:dS4-AdS5-metric}
 g_{\mathrm{AdS}_5}^{}&=d\rho^2+\sinh^2\!\rho\, g_{\mathrm{dS}_4}~.
\end{align}
The reason is simple:
Once the S$^3$-mode $\omega$ is fixed, the field equations are only sensitive to
radial dependences and warp factors, and not to the metric on the slices.
Since these parts are the same for global Euclidean and dS$_4$-sliced Lorentzian AdS,
the solutions found here work on dS$_4$-sliced AdS$_5$ just as well.
We have not gone through the $\kappa$-symmetry analysis for that case in detail, but clearly expect them to also be supersymmetric.
As far as geometries where AdS$_5$ is sliced by spaces of constant curvature are concerned, that only leaves
hyperbolic space $\mathbb{H}_4$ (or Euclidean AdS$_4$) as slice geometry.
But that can be obtained by simply Wick-rotating $t\rightarrow it$ in (\ref{eqn:AdS5S5-metric}).
By the same arguments as above, the embeddings found in Sec.~\ref{sec:fully-massive-embedding-solution}
are still solutions with that analytic continuation.
We thus have a comprehensive catalog of analytic, supersymmetric D7-brane embeddings to describe \N{4} SYM with massive flavors on
spacetimes of constant curvature.

\bibliography{AdSFlavors}

\begin{thebibliography}{43}%
\makeatletter
\providecommand \@ifxundefined [1]{%
 \@ifx{#1\undefined}
}%
\providecommand \@ifnum [1]{%
 \ifnum #1\expandafter \@firstoftwo
 \else \expandafter \@secondoftwo
 \fi
}%
\providecommand \@ifx [1]{%
 \ifx #1\expandafter \@firstoftwo
 \else \expandafter \@secondoftwo
 \fi
}%
\providecommand \natexlab [1]{#1}%
\providecommand \enquote  [1]{``#1''}%
\providecommand \bibnamefont  [1]{#1}%
\providecommand \bibfnamefont [1]{#1}%
\providecommand \citenamefont [1]{#1}%
\providecommand \href@noop [0]{\@secondoftwo}%
\providecommand \href [0]{\begingroup \@sanitize@url \@href}%
\providecommand \@href[1]{\@@startlink{#1}\@@href}%
\providecommand \@@href[1]{\endgroup#1\@@endlink}%
\providecommand \@sanitize@url [0]{\catcode `\\12\catcode `\$12\catcode
  `\&12\catcode `\#12\catcode `\^12\catcode `\_12\catcode `\%12\relax}%
\providecommand \@@startlink[1]{}%
\providecommand \@@endlink[0]{}%
\providecommand \url  [0]{\begingroup\@sanitize@url \@url }%
\providecommand \@url [1]{\endgroup\@href {#1}{\urlprefix }}%
\providecommand \urlprefix  [0]{URL }%
\providecommand \Eprint [0]{\href }%
\providecommand \doibase [0]{http://dx.doi.org/}%
\providecommand \selectlanguage [0]{\@gobble}%
\providecommand \bibinfo  [0]{\@secondoftwo}%
\providecommand \bibfield  [0]{\@secondoftwo}%
\providecommand \translation [1]{[#1]}%
\providecommand \BibitemOpen [0]{}%
\providecommand \bibitemStop [0]{}%
\providecommand \bibitemNoStop [0]{.\EOS\space}%
\providecommand \EOS [0]{\spacefactor3000\relax}%
\providecommand \BibitemShut  [1]{\csname bibitem#1\endcsname}%
\let\auto@bib@innerbib\@empty
%</preamble>
\bibitem [{\citenamefont {Karch}\ and\ \citenamefont
  {Randall}(2001{\natexlab{a}})}]{Karch:2000ct}%
  \BibitemOpen
  \bibfield  {author} {\bibinfo {author} {\bibfnamefont {Andreas}\ \bibnamefont
  {Karch}}\ and\ \bibinfo {author} {\bibfnamefont {Lisa}\ \bibnamefont
  {Randall}},\ }\bibfield  {title} {\enquote {\bibinfo {title} {{Locally
  localized gravity}},}\ }\bibfield  {booktitle} {\emph {\bibinfo {booktitle}
  {{Superstrings. Proceedings, International Conference, Strings 2000, Ann
  Arbor, USA, July 10-15, 2000}}},\ }\href {\doibase
  10.1088/1126-6708/2001/05/008} {\bibfield  {journal} {\bibinfo  {journal}
  {JHEP}\ }\textbf {\bibinfo {volume} {05}},\ \bibinfo {pages} {008} (\bibinfo
  {year} {2001}{\natexlab{a}})},\ \bibinfo {note} {[,140(2000)]},\ \Eprint
  {http://arxiv.org/abs/hep-th/0011156} {arXiv:hep-th/0011156 [hep-th]}
  \BibitemShut {NoStop}%
%%CITATION = HEP-TH/0011156;%%
\bibitem [{\citenamefont {Karch}\ and\ \citenamefont
  {Randall}(2001{\natexlab{b}})}]{Karch:2000gx}%
  \BibitemOpen
  \bibfield  {author} {\bibinfo {author} {\bibfnamefont {Andreas}\ \bibnamefont
  {Karch}}\ and\ \bibinfo {author} {\bibfnamefont {Lisa}\ \bibnamefont
  {Randall}},\ }\bibfield  {title} {\enquote {\bibinfo {title} {{Open and
  closed string interpretation of SUSY CFT's on branes with boundaries}},}\
  }\href {\doibase 10.1088/1126-6708/2001/06/063} {\bibfield  {journal}
  {\bibinfo  {journal} {JHEP}\ }\textbf {\bibinfo {volume} {06}},\ \bibinfo
  {pages} {063} (\bibinfo {year} {2001}{\natexlab{b}})},\ \Eprint
  {http://arxiv.org/abs/hep-th/0105132} {arXiv:hep-th/0105132 [hep-th]}
  \BibitemShut {NoStop}%
%%CITATION = HEP-TH/0105132;%%
\bibitem [{\citenamefont {Buchel}(2002)}]{Buchel:2002wf}%
  \BibitemOpen
  \bibfield  {author} {\bibinfo {author} {\bibfnamefont {Alex}\ \bibnamefont
  {Buchel}},\ }\bibfield  {title} {\enquote {\bibinfo {title} {{Gauge / gravity
  correspondence in accelerating universe}},}\ }\href {\doibase
  10.1103/PhysRevD.65.125015} {\bibfield  {journal} {\bibinfo  {journal} {Phys.
  Rev.}\ }\textbf {\bibinfo {volume} {D65}},\ \bibinfo {pages} {125015}
  (\bibinfo {year} {2002})},\ \Eprint {http://arxiv.org/abs/hep-th/0203041}
  {arXiv:hep-th/0203041 [hep-th]} \BibitemShut {NoStop}%
%%CITATION = HEP-TH/0203041;%%
\bibitem [{\citenamefont {Marolf}\ \emph {et~al.}(2014)\citenamefont {Marolf},
  \citenamefont {Rangamani},\ and\ \citenamefont {Wiseman}}]{Marolf:2013ioa}%
  \BibitemOpen
  \bibfield  {author} {\bibinfo {author} {\bibfnamefont {Donald}\ \bibnamefont
  {Marolf}}, \bibinfo {author} {\bibfnamefont {Mukund}\ \bibnamefont
  {Rangamani}}, \ and\ \bibinfo {author} {\bibfnamefont {Toby}\ \bibnamefont
  {Wiseman}},\ }\bibfield  {title} {\enquote {\bibinfo {title} {{Holographic
  thermal field theory on curved spacetimes}},}\ }\href {\doibase
  10.1088/0264-9381/31/6/063001} {\bibfield  {journal} {\bibinfo  {journal}
  {Class. Quant. Grav.}\ }\textbf {\bibinfo {volume} {31}},\ \bibinfo {pages}
  {063001} (\bibinfo {year} {2014})},\ \Eprint {http://arxiv.org/abs/1312.0612}
  {arXiv:1312.0612 [hep-th]} \BibitemShut {NoStop}%
%%CITATION = ARXIV:1312.0612;%%
\bibitem [{\citenamefont {Rangamani}\ \emph {et~al.}(2015)\citenamefont
  {Rangamani}, \citenamefont {Rozali},\ and\ \citenamefont
  {Van~Raamsdonk}}]{Rangamani:2015qga}%
  \BibitemOpen
  \bibfield  {author} {\bibinfo {author} {\bibfnamefont {Mukund}\ \bibnamefont
  {Rangamani}}, \bibinfo {author} {\bibfnamefont {Moshe}\ \bibnamefont
  {Rozali}}, \ and\ \bibinfo {author} {\bibfnamefont {Mark}\ \bibnamefont
  {Van~Raamsdonk}},\ }\bibfield  {title} {\enquote {\bibinfo {title}
  {{Cosmological Particle Production at Strong Coupling}},}\ }\href@noop {} {\
  (\bibinfo {year} {2015})},\ \Eprint {http://arxiv.org/abs/1505.03901}
  {arXiv:1505.03901 [hep-th]} \BibitemShut {NoStop}%
%%CITATION = ARXIV:1505.03901;%%
\bibitem [{\citenamefont {Marolf}\ \emph {et~al.}(2011)\citenamefont {Marolf},
  \citenamefont {Rangamani},\ and\ \citenamefont
  {Van~Raamsdonk}}]{Marolf:2010tg}%
  \BibitemOpen
  \bibfield  {author} {\bibinfo {author} {\bibfnamefont {Donald}\ \bibnamefont
  {Marolf}}, \bibinfo {author} {\bibfnamefont {Mukund}\ \bibnamefont
  {Rangamani}}, \ and\ \bibinfo {author} {\bibfnamefont {Mark}\ \bibnamefont
  {Van~Raamsdonk}},\ }\bibfield  {title} {\enquote {\bibinfo {title}
  {{Holographic models of de Sitter QFTs}},}\ }\href {\doibase
  10.1088/0264-9381/28/10/105015} {\bibfield  {journal} {\bibinfo  {journal}
  {Class. Quant. Grav.}\ }\textbf {\bibinfo {volume} {28}},\ \bibinfo {pages}
  {105015} (\bibinfo {year} {2011})},\ \Eprint {http://arxiv.org/abs/1007.3996}
  {arXiv:1007.3996 [hep-th]} \BibitemShut {NoStop}%
%%CITATION = ARXIV:1007.3996;%%
\bibitem [{\citenamefont {Aharony}\ \emph {et~al.}(2011)\citenamefont
  {Aharony}, \citenamefont {Marolf},\ and\ \citenamefont
  {Rangamani}}]{Aharony:2010ay}%
  \BibitemOpen
  \bibfield  {author} {\bibinfo {author} {\bibfnamefont {Ofer}\ \bibnamefont
  {Aharony}}, \bibinfo {author} {\bibfnamefont {Donald}\ \bibnamefont
  {Marolf}}, \ and\ \bibinfo {author} {\bibfnamefont {Mukund}\ \bibnamefont
  {Rangamani}},\ }\bibfield  {title} {\enquote {\bibinfo {title} {{Conformal
  field theories in anti-de Sitter space}},}\ }\href {\doibase
  10.1007/JHEP02(2011)041} {\bibfield  {journal} {\bibinfo  {journal} {JHEP}\
  }\textbf {\bibinfo {volume} {1102}},\ \bibinfo {pages} {041} (\bibinfo {year}
  {2011})},\ \Eprint {http://arxiv.org/abs/1011.6144} {arXiv:1011.6144
  [hep-th]} \BibitemShut {NoStop}%
%%CITATION = ARXIV:1011.6144;%%
\bibitem [{\citenamefont {Karch}\ and\ \citenamefont
  {Katz}(2002)}]{Karch:2002sh}%
  \BibitemOpen
  \bibfield  {author} {\bibinfo {author} {\bibfnamefont {Andreas}\ \bibnamefont
  {Karch}}\ and\ \bibinfo {author} {\bibfnamefont {Emanuel}\ \bibnamefont
  {Katz}},\ }\bibfield  {title} {\enquote {\bibinfo {title} {{Adding flavor to
  AdS / CFT}},}\ }\href {\doibase 10.1088/1126-6708/2002/06/043} {\bibfield
  {journal} {\bibinfo  {journal} {JHEP}\ }\textbf {\bibinfo {volume} {0206}},\
  \bibinfo {pages} {043} (\bibinfo {year} {2002})},\ \Eprint
  {http://arxiv.org/abs/hep-th/0205236} {arXiv:hep-th/0205236 [hep-th]}
  \BibitemShut {NoStop}%
%%CITATION = HEP-TH/0205236;%%
\bibitem [{\citenamefont {Karch}\ \emph {et~al.}(2009)\citenamefont {Karch},
  \citenamefont {O'Bannon},\ and\ \citenamefont {Yaffe}}]{Karch:2009ph}%
  \BibitemOpen
  \bibfield  {author} {\bibinfo {author} {\bibfnamefont {Andreas}\ \bibnamefont
  {Karch}}, \bibinfo {author} {\bibfnamefont {Andy}\ \bibnamefont {O'Bannon}},
  \ and\ \bibinfo {author} {\bibfnamefont {Laurence~G.}\ \bibnamefont
  {Yaffe}},\ }\bibfield  {title} {\enquote {\bibinfo {title} {{Critical
  Exponents from AdS/CFT with Flavor}},}\ }\href {\doibase
  10.1088/1126-6708/2009/09/042} {\bibfield  {journal} {\bibinfo  {journal}
  {JHEP}\ }\textbf {\bibinfo {volume} {0909}},\ \bibinfo {pages} {042}
  (\bibinfo {year} {2009})},\ \Eprint {http://arxiv.org/abs/0906.4959}
  {arXiv:0906.4959 [hep-th]} \BibitemShut {NoStop}%
%%CITATION = ARXIV:0906.4959;%%
\bibitem [{\citenamefont {Clark}\ \emph {et~al.}(2014)\citenamefont {Clark},
  \citenamefont {Crossette}, \citenamefont {Newman},\ and\ \citenamefont
  {Rommal}}]{Clark:2013mfa}%
  \BibitemOpen
  \bibfield  {author} {\bibinfo {author} {\bibfnamefont {Adam~B.}\ \bibnamefont
  {Clark}}, \bibinfo {author} {\bibfnamefont {Nathan}\ \bibnamefont
  {Crossette}}, \bibinfo {author} {\bibfnamefont {George~M.}\ \bibnamefont
  {Newman}}, \ and\ \bibinfo {author} {\bibfnamefont {Andrea}\ \bibnamefont
  {Rommal}},\ }\bibfield  {title} {\enquote {\bibinfo {title} {{AdS-Sliced
  Flavor Branes and Adding Flavor to the Janus Solution}},}\ }\href {\doibase
  10.1103/PhysRevD.89.026014} {\bibfield  {journal} {\bibinfo  {journal}
  {Phys.Rev.}\ }\textbf {\bibinfo {volume} {D89}},\ \bibinfo {pages} {026014}
  (\bibinfo {year} {2014})},\ \Eprint {http://arxiv.org/abs/1309.7872}
  {arXiv:1309.7872 [hep-th]} \BibitemShut {NoStop}%
%%CITATION = ARXIV:1309.7872;%%
\bibitem [{\citenamefont {Pestun}(2012)}]{Pestun:2007rz}%
  \BibitemOpen
  \bibfield  {author} {\bibinfo {author} {\bibfnamefont {Vasily}\ \bibnamefont
  {Pestun}},\ }\bibfield  {title} {\enquote {\bibinfo {title} {{Localization of
  gauge theory on a four-sphere and supersymmetric Wilson loops}},}\ }\href
  {\doibase 10.1007/s00220-012-1485-0} {\bibfield  {journal} {\bibinfo
  {journal} {Commun.Math.Phys.}\ }\textbf {\bibinfo {volume} {313}},\ \bibinfo
  {pages} {71--129} (\bibinfo {year} {2012})},\ \Eprint
  {http://arxiv.org/abs/0712.2824} {arXiv:0712.2824 [hep-th]} \BibitemShut
  {NoStop}%
%%CITATION = ARXIV:0712.2824;%%
\bibitem [{\citenamefont {Festuccia}\ and\ \citenamefont
  {Seiberg}(2011)}]{Festuccia:2011ws}%
  \BibitemOpen
  \bibfield  {author} {\bibinfo {author} {\bibfnamefont {Guido}\ \bibnamefont
  {Festuccia}}\ and\ \bibinfo {author} {\bibfnamefont {Nathan}\ \bibnamefont
  {Seiberg}},\ }\bibfield  {title} {\enquote {\bibinfo {title} {{Rigid
  Supersymmetric Theories in Curved Superspace}},}\ }\href {\doibase
  10.1007/JHEP06(2011)114} {\bibfield  {journal} {\bibinfo  {journal} {JHEP}\
  }\textbf {\bibinfo {volume} {1106}},\ \bibinfo {pages} {114} (\bibinfo {year}
  {2011})},\ \Eprint {http://arxiv.org/abs/1105.0689} {arXiv:1105.0689
  [hep-th]} \BibitemShut {NoStop}%
%%CITATION = ARXIV:1105.0689;%%
\bibitem [{\citenamefont {Anous}\ \emph {et~al.}(2014)\citenamefont {Anous},
  \citenamefont {Freedman},\ and\ \citenamefont {Maloney}}]{Anous:2014lia}%
  \BibitemOpen
  \bibfield  {author} {\bibinfo {author} {\bibfnamefont {Tarek}\ \bibnamefont
  {Anous}}, \bibinfo {author} {\bibfnamefont {Daniel~Z.}\ \bibnamefont
  {Freedman}}, \ and\ \bibinfo {author} {\bibfnamefont {Alexander}\
  \bibnamefont {Maloney}},\ }\bibfield  {title} {\enquote {\bibinfo {title}
  {{de Sitter Supersymmetry Revisited}},}\ }\href {\doibase
  10.1007/JHEP07(2014)119} {\bibfield  {journal} {\bibinfo  {journal} {JHEP}\
  }\textbf {\bibinfo {volume} {1407}},\ \bibinfo {pages} {119} (\bibinfo {year}
  {2014})},\ \Eprint {http://arxiv.org/abs/1403.5038} {arXiv:1403.5038
  [hep-th]} \BibitemShut {NoStop}%
%%CITATION = ARXIV:1403.5038;%%
\bibitem [{\citenamefont {Bergshoeff}\ and\ \citenamefont
  {Townsend}(1997)}]{Bergshoeff:1996tu}%
  \BibitemOpen
  \bibfield  {author} {\bibinfo {author} {\bibfnamefont {E.}~\bibnamefont
  {Bergshoeff}}\ and\ \bibinfo {author} {\bibfnamefont {P.K.}\ \bibnamefont
  {Townsend}},\ }\bibfield  {title} {\enquote {\bibinfo {title} {{Super
  D-branes}},}\ }\href {\doibase 10.1016/S0550-3213(97)00072-2} {\bibfield
  {journal} {\bibinfo  {journal} {Nucl.Phys.}\ }\textbf {\bibinfo {volume}
  {B490}},\ \bibinfo {pages} {145--162} (\bibinfo {year} {1997})},\ \Eprint
  {http://arxiv.org/abs/hep-th/9611173} {arXiv:hep-th/9611173 [hep-th]}
  \BibitemShut {NoStop}%
%%CITATION = HEP-TH/9611173;%%
\bibitem [{\citenamefont {Cederwall}\ \emph
  {et~al.}(1997{\natexlab{a}})\citenamefont {Cederwall}, \citenamefont {von
  Gussich}, \citenamefont {Nilsson},\ and\ \citenamefont
  {Westerberg}}]{Cederwall:1996pv}%
  \BibitemOpen
  \bibfield  {author} {\bibinfo {author} {\bibfnamefont {Martin}\ \bibnamefont
  {Cederwall}}, \bibinfo {author} {\bibfnamefont {Alexander}\ \bibnamefont {von
  Gussich}}, \bibinfo {author} {\bibfnamefont {Bengt~E.W.}\ \bibnamefont
  {Nilsson}}, \ and\ \bibinfo {author} {\bibfnamefont {Anders}\ \bibnamefont
  {Westerberg}},\ }\bibfield  {title} {\enquote {\bibinfo {title} {{The
  Dirichlet super three-brane in ten-dimensional type IIB supergravity}},}\
  }\href {\doibase 10.1016/S0550-3213(97)00071-0} {\bibfield  {journal}
  {\bibinfo  {journal} {Nucl.Phys.}\ }\textbf {\bibinfo {volume} {B490}},\
  \bibinfo {pages} {163--178} (\bibinfo {year} {1997}{\natexlab{a}})},\ \Eprint
  {http://arxiv.org/abs/hep-th/9610148} {arXiv:hep-th/9610148 [hep-th]}
  \BibitemShut {NoStop}%
%%CITATION = HEP-TH/9610148;%%
\bibitem [{\citenamefont {Cederwall}\ \emph
  {et~al.}(1997{\natexlab{b}})\citenamefont {Cederwall}, \citenamefont {von
  Gussich}, \citenamefont {Nilsson}, \citenamefont {Sundell},\ and\
  \citenamefont {Westerberg}}]{Cederwall:1996ri}%
  \BibitemOpen
  \bibfield  {author} {\bibinfo {author} {\bibfnamefont {Martin}\ \bibnamefont
  {Cederwall}}, \bibinfo {author} {\bibfnamefont {Alexander}\ \bibnamefont {von
  Gussich}}, \bibinfo {author} {\bibfnamefont {Bengt~E.W.}\ \bibnamefont
  {Nilsson}}, \bibinfo {author} {\bibfnamefont {Per}\ \bibnamefont {Sundell}},
  \ and\ \bibinfo {author} {\bibfnamefont {Anders}\ \bibnamefont
  {Westerberg}},\ }\bibfield  {title} {\enquote {\bibinfo {title} {{The
  Dirichlet super p-branes in ten-dimensional type IIA and IIB
  supergravity}},}\ }\href {\doibase 10.1016/S0550-3213(97)00075-8} {\bibfield
  {journal} {\bibinfo  {journal} {Nucl.Phys.}\ }\textbf {\bibinfo {volume}
  {B490}},\ \bibinfo {pages} {179--201} (\bibinfo {year}
  {1997}{\natexlab{b}})},\ \Eprint {http://arxiv.org/abs/hep-th/9611159}
  {arXiv:hep-th/9611159 [hep-th]} \BibitemShut {NoStop}%
%%CITATION = HEP-TH/9611159;%%
\bibitem [{\citenamefont {Grisaru}\ \emph {et~al.}(2000)\citenamefont
  {Grisaru}, \citenamefont {Myers},\ and\ \citenamefont
  {Tafjord}}]{Grisaru:2000zn}%
  \BibitemOpen
  \bibfield  {author} {\bibinfo {author} {\bibfnamefont {Marcus~T.}\
  \bibnamefont {Grisaru}}, \bibinfo {author} {\bibfnamefont {Robert~C.}\
  \bibnamefont {Myers}}, \ and\ \bibinfo {author} {\bibfnamefont {Oyvind}\
  \bibnamefont {Tafjord}},\ }\bibfield  {title} {\enquote {\bibinfo {title}
  {{SUSY and goliath}},}\ }\href {\doibase 10.1088/1126-6708/2000/08/040}
  {\bibfield  {journal} {\bibinfo  {journal} {JHEP}\ }\textbf {\bibinfo
  {volume} {0008}},\ \bibinfo {pages} {040} (\bibinfo {year} {2000})},\ \Eprint
  {http://arxiv.org/abs/hep-th/0008015} {arXiv:hep-th/0008015 [hep-th]}
  \BibitemShut {NoStop}%
%%CITATION = HEP-TH/0008015;%%
\bibitem [{\citenamefont {Lu}\ \emph {et~al.}(1999)\citenamefont {Lu},
  \citenamefont {Pope},\ and\ \citenamefont {Rahmfeld}}]{Lu:1998nu}%
  \BibitemOpen
  \bibfield  {author} {\bibinfo {author} {\bibfnamefont {Hong}\ \bibnamefont
  {Lu}}, \bibinfo {author} {\bibfnamefont {C.N.}\ \bibnamefont {Pope}}, \ and\
  \bibinfo {author} {\bibfnamefont {J.}~\bibnamefont {Rahmfeld}},\ }\bibfield
  {title} {\enquote {\bibinfo {title} {{A Construction of Killing spinors on
  S**n}},}\ }\href {\doibase 10.1063/1.532983} {\bibfield  {journal} {\bibinfo
  {journal} {J.Math.Phys.}\ }\textbf {\bibinfo {volume} {40}},\ \bibinfo
  {pages} {4518--4526} (\bibinfo {year} {1999})},\ \Eprint
  {http://arxiv.org/abs/hep-th/9805151} {arXiv:hep-th/9805151 [hep-th]}
  \BibitemShut {NoStop}%
%%CITATION = HEP-TH/9805151;%%
\bibitem [{\citenamefont {Lu}\ \emph {et~al.}(1997)\citenamefont {Lu},
  \citenamefont {Pope},\ and\ \citenamefont {Townsend}}]{Lu:1996rhb}%
  \BibitemOpen
  \bibfield  {author} {\bibinfo {author} {\bibfnamefont {Hong}\ \bibnamefont
  {Lu}}, \bibinfo {author} {\bibfnamefont {C.N.}\ \bibnamefont {Pope}}, \ and\
  \bibinfo {author} {\bibfnamefont {P.K.}\ \bibnamefont {Townsend}},\
  }\bibfield  {title} {\enquote {\bibinfo {title} {{Domain walls from anti-de
  Sitter space-time}},}\ }\href {\doibase 10.1016/S0370-2693(96)01443-8}
  {\bibfield  {journal} {\bibinfo  {journal} {Phys.Lett.}\ }\textbf {\bibinfo
  {volume} {B391}},\ \bibinfo {pages} {39--46} (\bibinfo {year} {1997})},\
  \Eprint {http://arxiv.org/abs/hep-th/9607164} {arXiv:hep-th/9607164 [hep-th]}
  \BibitemShut {NoStop}%
%%CITATION = HEP-TH/9607164;%%
\bibitem [{\citenamefont {Kim}\ \emph {et~al.}(1985)\citenamefont {Kim},
  \citenamefont {Romans},\ and\ \citenamefont {van
  Nieuwenhuizen}}]{Kim:1985ez}%
  \BibitemOpen
  \bibfield  {author} {\bibinfo {author} {\bibfnamefont {H.J.}\ \bibnamefont
  {Kim}}, \bibinfo {author} {\bibfnamefont {L.J.}\ \bibnamefont {Romans}}, \
  and\ \bibinfo {author} {\bibfnamefont {P.}~\bibnamefont {van
  Nieuwenhuizen}},\ }\bibfield  {title} {\enquote {\bibinfo {title} {{The Mass
  Spectrum of Chiral N=2 D=10 Supergravity on S**5}},}\ }\href {\doibase
  10.1103/PhysRevD.32.389} {\bibfield  {journal} {\bibinfo  {journal}
  {Phys.Rev.}\ }\textbf {\bibinfo {volume} {D32}},\ \bibinfo {pages} {389}
  (\bibinfo {year} {1985})}\BibitemShut {NoStop}%
%%CITATION = PHRVA,D32,389;%%
\bibitem [{\citenamefont {Kruczenski}\ \emph {et~al.}(2003)\citenamefont
  {Kruczenski}, \citenamefont {Mateos}, \citenamefont {Myers},\ and\
  \citenamefont {Winters}}]{Kruczenski:2003be}%
  \BibitemOpen
  \bibfield  {author} {\bibinfo {author} {\bibfnamefont {Martin}\ \bibnamefont
  {Kruczenski}}, \bibinfo {author} {\bibfnamefont {David}\ \bibnamefont
  {Mateos}}, \bibinfo {author} {\bibfnamefont {Robert~C.}\ \bibnamefont
  {Myers}}, \ and\ \bibinfo {author} {\bibfnamefont {David~J.}\ \bibnamefont
  {Winters}},\ }\bibfield  {title} {\enquote {\bibinfo {title} {{Meson
  spectroscopy in AdS / CFT with flavor}},}\ }\href {\doibase
  10.1088/1126-6708/2003/07/049} {\bibfield  {journal} {\bibinfo  {journal}
  {JHEP}\ }\textbf {\bibinfo {volume} {0307}},\ \bibinfo {pages} {049}
  (\bibinfo {year} {2003})},\ \Eprint {http://arxiv.org/abs/hep-th/0304032}
  {arXiv:hep-th/0304032 [hep-th]} \BibitemShut {NoStop}%
%%CITATION = HEP-TH/0304032;%%
\bibitem [{\citenamefont {Hong}\ \emph {et~al.}(2004)\citenamefont {Hong},
  \citenamefont {Yoon},\ and\ \citenamefont {Strassler}}]{Hong:2003jm}%
  \BibitemOpen
  \bibfield  {author} {\bibinfo {author} {\bibfnamefont {Sungho}\ \bibnamefont
  {Hong}}, \bibinfo {author} {\bibfnamefont {Sukjin}\ \bibnamefont {Yoon}}, \
  and\ \bibinfo {author} {\bibfnamefont {Matthew~J.}\ \bibnamefont
  {Strassler}},\ }\bibfield  {title} {\enquote {\bibinfo {title} {{Quarkonium
  from the fifth-dimension}},}\ }\href {\doibase 10.1088/1126-6708/2004/04/046}
  {\bibfield  {journal} {\bibinfo  {journal} {JHEP}\ }\textbf {\bibinfo
  {volume} {0404}},\ \bibinfo {pages} {046} (\bibinfo {year} {2004})},\ \Eprint
  {http://arxiv.org/abs/hep-th/0312071} {arXiv:hep-th/0312071 [hep-th]}
  \BibitemShut {NoStop}%
%%CITATION = HEP-TH/0312071;%%
\bibitem [{\citenamefont {Polchinski}(1998)}]{Polchinski:1998rr}%
  \BibitemOpen
  \bibfield  {author} {\bibinfo {author} {\bibfnamefont {J.}~\bibnamefont
  {Polchinski}},\ }\bibfield  {title} {\enquote {\bibinfo {title} {{String
  theory. Vol. 2: Superstring theory and beyond}},}\ }\href@noop {} {\
  (\bibinfo {year} {1998})}\BibitemShut {NoStop}%
%%CITATION = INSPIRE-487241;%%
\bibitem [{\citenamefont {Balasubramanian}\ \emph {et~al.}(2001)\citenamefont
  {Balasubramanian}, \citenamefont {Gimon}, \citenamefont {Minic},\ and\
  \citenamefont {Rahmfeld}}]{Balasubramanian:2000pq}%
  \BibitemOpen
  \bibfield  {author} {\bibinfo {author} {\bibfnamefont {Vijay}\ \bibnamefont
  {Balasubramanian}}, \bibinfo {author} {\bibfnamefont {Eric~G.}\ \bibnamefont
  {Gimon}}, \bibinfo {author} {\bibfnamefont {Djordje}\ \bibnamefont {Minic}},
  \ and\ \bibinfo {author} {\bibfnamefont {Joachim}\ \bibnamefont {Rahmfeld}},\
  }\bibfield  {title} {\enquote {\bibinfo {title} {{Four-dimensional conformal
  supergravity from AdS space}},}\ }\href {\doibase 10.1103/PhysRevD.63.104009}
  {\bibfield  {journal} {\bibinfo  {journal} {Phys. Rev.}\ }\textbf {\bibinfo
  {volume} {D63}},\ \bibinfo {pages} {104009} (\bibinfo {year} {2001})},\
  \Eprint {http://arxiv.org/abs/hep-th/0007211} {arXiv:hep-th/0007211 [hep-th]}
  \BibitemShut {NoStop}%
%%CITATION = HEP-TH/0007211;%%
\bibitem [{\citenamefont {Ohl}\ and\ \citenamefont
  {Uhlemann}(2011)}]{Ohl:2010au}%
  \BibitemOpen
  \bibfield  {author} {\bibinfo {author} {\bibfnamefont {Thorsten}\
  \bibnamefont {Ohl}}\ and\ \bibinfo {author} {\bibfnamefont {Christoph~F.}\
  \bibnamefont {Uhlemann}},\ }\bibfield  {title} {\enquote {\bibinfo {title}
  {{The Boundary Multiplet of N=4 SU(2)xU(1) Gauged Supergravity on
  Asymptotically-AdS$_5$}},}\ }\href {\doibase 10.1007/JHEP06(2011)086}
  {\bibfield  {journal} {\bibinfo  {journal} {JHEP}\ }\textbf {\bibinfo
  {volume} {1106}},\ \bibinfo {pages} {086} (\bibinfo {year} {2011})},\ \Eprint
  {http://arxiv.org/abs/1011.3533} {arXiv:1011.3533 [hep-th]} \BibitemShut
  {NoStop}%
%%CITATION = ARXIV:1011.3533;%%
\bibitem [{\citenamefont {Klare}\ \emph {et~al.}(2012)\citenamefont {Klare},
  \citenamefont {Tomasiello},\ and\ \citenamefont {Zaffaroni}}]{Klare:2012gn}%
  \BibitemOpen
  \bibfield  {author} {\bibinfo {author} {\bibfnamefont {Claudius}\
  \bibnamefont {Klare}}, \bibinfo {author} {\bibfnamefont {Alessandro}\
  \bibnamefont {Tomasiello}}, \ and\ \bibinfo {author} {\bibfnamefont
  {Alberto}\ \bibnamefont {Zaffaroni}},\ }\bibfield  {title} {\enquote
  {\bibinfo {title} {{Supersymmetry on Curved Spaces and Holography}},}\ }\href
  {\doibase 10.1007/JHEP08(2012)061} {\bibfield  {journal} {\bibinfo  {journal}
  {JHEP}\ }\textbf {\bibinfo {volume} {1208}},\ \bibinfo {pages} {061}
  (\bibinfo {year} {2012})},\ \Eprint {http://arxiv.org/abs/1205.1062}
  {arXiv:1205.1062 [hep-th]} \BibitemShut {NoStop}%
%%CITATION = ARXIV:1205.1062;%%
\bibitem [{\citenamefont {Fradkin}\ and\ \citenamefont
  {Tseytlin}(1985)}]{Fradkin:1985am}%
  \BibitemOpen
  \bibfield  {author} {\bibinfo {author} {\bibfnamefont {E.S.}\ \bibnamefont
  {Fradkin}}\ and\ \bibinfo {author} {\bibfnamefont {Arkady~A.}\ \bibnamefont
  {Tseytlin}},\ }\bibfield  {title} {\enquote {\bibinfo {title} {{Conformal
  Supergravity}},}\ }\href {\doibase 10.1016/0370-1573(85)90138-3} {\bibfield
  {journal} {\bibinfo  {journal} {Phys.Rept.}\ }\textbf {\bibinfo {volume}
  {119}},\ \bibinfo {pages} {233--362} (\bibinfo {year} {1985})}\BibitemShut
  {NoStop}%
%%CITATION = PRPLC,119,233;%%
\bibitem [{\citenamefont {Gaiotto}\ and\ \citenamefont
  {Witten}(2009)}]{Gaiotto:2008sa}%
  \BibitemOpen
  \bibfield  {author} {\bibinfo {author} {\bibfnamefont {Davide}\ \bibnamefont
  {Gaiotto}}\ and\ \bibinfo {author} {\bibfnamefont {Edward}\ \bibnamefont
  {Witten}},\ }\bibfield  {title} {\enquote {\bibinfo {title} {{Supersymmetric
  Boundary Conditions in N=4 Super Yang-Mills Theory}},}\ }\href {\doibase
  10.1007/s10955-009-9687-3} {\bibfield  {journal} {\bibinfo  {journal} {J.
  Statist. Phys.}\ }\textbf {\bibinfo {volume} {135}},\ \bibinfo {pages}
  {789--855} (\bibinfo {year} {2009})},\ \Eprint
  {http://arxiv.org/abs/0804.2902} {arXiv:0804.2902 [hep-th]} \BibitemShut
  {NoStop}%
%%CITATION = ARXIV:0804.2902;%%
\bibitem [{\citenamefont {Breitenlohner}\ and\ \citenamefont
  {Freedman}(1982{\natexlab{a}})}]{Breitenlohner:1982bm}%
  \BibitemOpen
  \bibfield  {author} {\bibinfo {author} {\bibfnamefont {Peter}\ \bibnamefont
  {Breitenlohner}}\ and\ \bibinfo {author} {\bibfnamefont {Daniel~Z.}\
  \bibnamefont {Freedman}},\ }\bibfield  {title} {\enquote {\bibinfo {title}
  {{Positive Energy in anti-De Sitter Backgrounds and Gauged Extended
  Supergravity}},}\ }\href {\doibase 10.1016/0370-2693(82)90643-8} {\bibfield
  {journal} {\bibinfo  {journal} {Phys. Lett.}\ }\textbf {\bibinfo {volume}
  {B115}},\ \bibinfo {pages} {197} (\bibinfo {year}
  {1982}{\natexlab{a}})}\BibitemShut {NoStop}%
%%CITATION = PHLTA,B115,197;%%
\bibitem [{\citenamefont {Breitenlohner}\ and\ \citenamefont
  {Freedman}(1982{\natexlab{b}})}]{Breitenlohner:1982jf}%
  \BibitemOpen
  \bibfield  {author} {\bibinfo {author} {\bibfnamefont {Peter}\ \bibnamefont
  {Breitenlohner}}\ and\ \bibinfo {author} {\bibfnamefont {Daniel~Z.}\
  \bibnamefont {Freedman}},\ }\bibfield  {title} {\enquote {\bibinfo {title}
  {{Stability in Gauged Extended Supergravity}},}\ }\href {\doibase
  10.1016/0003-4916(82)90116-6} {\bibfield  {journal} {\bibinfo  {journal}
  {Annals Phys.}\ }\textbf {\bibinfo {volume} {144}},\ \bibinfo {pages} {249}
  (\bibinfo {year} {1982}{\natexlab{b}})}\BibitemShut {NoStop}%
%%CITATION = APNYA,144,249;%%
\bibitem [{\citenamefont {Clark}\ \emph {et~al.}(2005)\citenamefont {Clark},
  \citenamefont {Freedman}, \citenamefont {Karch},\ and\ \citenamefont
  {Schnabl}}]{Clark:2004sb}%
  \BibitemOpen
  \bibfield  {author} {\bibinfo {author} {\bibfnamefont {A.B.}\ \bibnamefont
  {Clark}}, \bibinfo {author} {\bibfnamefont {D.Z.}\ \bibnamefont {Freedman}},
  \bibinfo {author} {\bibfnamefont {A.}~\bibnamefont {Karch}}, \ and\ \bibinfo
  {author} {\bibfnamefont {M.}~\bibnamefont {Schnabl}},\ }\bibfield  {title}
  {\enquote {\bibinfo {title} {{Dual of the Janus solution: An interface
  conformal field theory}},}\ }\href {\doibase 10.1103/PhysRevD.71.066003}
  {\bibfield  {journal} {\bibinfo  {journal} {Phys.Rev.}\ }\textbf {\bibinfo
  {volume} {D71}},\ \bibinfo {pages} {066003} (\bibinfo {year} {2005})},\
  \Eprint {http://arxiv.org/abs/hep-th/0407073} {arXiv:hep-th/0407073 [hep-th]}
  \BibitemShut {NoStop}%
%%CITATION = HEP-TH/0407073;%%
\bibitem [{\citenamefont {D'Hoker}\ \emph
  {et~al.}(2006{\natexlab{a}})\citenamefont {D'Hoker}, \citenamefont {Estes},\
  and\ \citenamefont {Gutperle}}]{D'Hoker:2006uv}%
  \BibitemOpen
  \bibfield  {author} {\bibinfo {author} {\bibfnamefont {Eric}\ \bibnamefont
  {D'Hoker}}, \bibinfo {author} {\bibfnamefont {John}\ \bibnamefont {Estes}}, \
  and\ \bibinfo {author} {\bibfnamefont {Michael}\ \bibnamefont {Gutperle}},\
  }\bibfield  {title} {\enquote {\bibinfo {title} {{Interface Yang-Mills,
  supersymmetry, and Janus}},}\ }\href {\doibase
  10.1016/j.nuclphysb.2006.07.001} {\bibfield  {journal} {\bibinfo  {journal}
  {Nucl. Phys.}\ }\textbf {\bibinfo {volume} {B753}},\ \bibinfo {pages}
  {16--41} (\bibinfo {year} {2006}{\natexlab{a}})},\ \Eprint
  {http://arxiv.org/abs/hep-th/0603013} {arXiv:hep-th/0603013 [hep-th]}
  \BibitemShut {NoStop}%
%%CITATION = HEP-TH/0603013;%%
\bibitem [{\citenamefont {de~Haro}\ \emph {et~al.}(2001)\citenamefont
  {de~Haro}, \citenamefont {Solodukhin},\ and\ \citenamefont
  {Skenderis}}]{deHaro:2000xn}%
  \BibitemOpen
  \bibfield  {author} {\bibinfo {author} {\bibfnamefont {Sebastian}\
  \bibnamefont {de~Haro}}, \bibinfo {author} {\bibfnamefont {Sergey~N.}\
  \bibnamefont {Solodukhin}}, \ and\ \bibinfo {author} {\bibfnamefont {Kostas}\
  \bibnamefont {Skenderis}},\ }\bibfield  {title} {\enquote {\bibinfo {title}
  {{Holographic reconstruction of space-time and renormalization in the AdS /
  CFT correspondence}},}\ }\href {\doibase 10.1007/s002200100381} {\bibfield
  {journal} {\bibinfo  {journal} {Commun. Math. Phys.}\ }\textbf {\bibinfo
  {volume} {217}},\ \bibinfo {pages} {595--622} (\bibinfo {year} {2001})},\
  \Eprint {http://arxiv.org/abs/hep-th/0002230} {arXiv:hep-th/0002230 [hep-th]}
  \BibitemShut {NoStop}%
%%CITATION = HEP-TH/0002230;%%
\bibitem [{\citenamefont {Bianchi}\ \emph {et~al.}(2002)\citenamefont
  {Bianchi}, \citenamefont {Freedman},\ and\ \citenamefont
  {Skenderis}}]{Bianchi:2001kw}%
  \BibitemOpen
  \bibfield  {author} {\bibinfo {author} {\bibfnamefont {Massimo}\ \bibnamefont
  {Bianchi}}, \bibinfo {author} {\bibfnamefont {Daniel~Z.}\ \bibnamefont
  {Freedman}}, \ and\ \bibinfo {author} {\bibfnamefont {Kostas}\ \bibnamefont
  {Skenderis}},\ }\bibfield  {title} {\enquote {\bibinfo {title} {{Holographic
  renormalization}},}\ }\href {\doibase 10.1016/S0550-3213(02)00179-7}
  {\bibfield  {journal} {\bibinfo  {journal} {Nucl.Phys.}\ }\textbf {\bibinfo
  {volume} {B631}},\ \bibinfo {pages} {159--194} (\bibinfo {year} {2002})},\
  \Eprint {http://arxiv.org/abs/hep-th/0112119} {arXiv:hep-th/0112119 [hep-th]}
  \BibitemShut {NoStop}%
%%CITATION = HEP-TH/0112119;%%
\bibitem [{\citenamefont {Karch}\ \emph {et~al.}(2006)\citenamefont {Karch},
  \citenamefont {O'Bannon},\ and\ \citenamefont {Skenderis}}]{Karch:2005ms}%
  \BibitemOpen
  \bibfield  {author} {\bibinfo {author} {\bibfnamefont {Andreas}\ \bibnamefont
  {Karch}}, \bibinfo {author} {\bibfnamefont {Andy}\ \bibnamefont {O'Bannon}},
  \ and\ \bibinfo {author} {\bibfnamefont {Kostas}\ \bibnamefont {Skenderis}},\
  }\bibfield  {title} {\enquote {\bibinfo {title} {{Holographic renormalization
  of probe D-branes in AdS/CFT}},}\ }\href {\doibase
  10.1088/1126-6708/2006/04/015} {\bibfield  {journal} {\bibinfo  {journal}
  {JHEP}\ }\textbf {\bibinfo {volume} {0604}},\ \bibinfo {pages} {015}
  (\bibinfo {year} {2006})},\ \Eprint {http://arxiv.org/abs/hep-th/0512125}
  {arXiv:hep-th/0512125 [hep-th]} \BibitemShut {NoStop}%
%%CITATION = HEP-TH/0512125;%%
\bibitem [{\citenamefont {Shapere}\ \emph {et~al.}(2012)\citenamefont
  {Shapere}, \citenamefont {Wilczek},\ and\ \citenamefont
  {Xiong}}]{Shapere:2012wn}%
  \BibitemOpen
  \bibfield  {author} {\bibinfo {author} {\bibfnamefont {Alfred~D.}\
  \bibnamefont {Shapere}}, \bibinfo {author} {\bibfnamefont {Frank}\
  \bibnamefont {Wilczek}}, \ and\ \bibinfo {author} {\bibfnamefont {Zhaoxi}\
  \bibnamefont {Xiong}},\ }\bibfield  {title} {\enquote {\bibinfo {title}
  {{Models of Topology Change}},}\ }\href@noop {} {\  (\bibinfo {year}
  {2012})},\ \Eprint {http://arxiv.org/abs/1210.3545} {arXiv:1210.3545
  [hep-th]} \BibitemShut {NoStop}%
%%CITATION = ARXIV:1210.3545;%%
\bibitem [{\citenamefont {Witten}(1988)}]{Witten:1988ze}%
  \BibitemOpen
  \bibfield  {author} {\bibinfo {author} {\bibfnamefont {Edward}\ \bibnamefont
  {Witten}},\ }\bibfield  {title} {\enquote {\bibinfo {title} {{Topological
  Quantum Field Theory}},}\ }\href {\doibase 10.1007/BF01223371} {\bibfield
  {journal} {\bibinfo  {journal} {Commun. Math. Phys.}\ }\textbf {\bibinfo
  {volume} {117}},\ \bibinfo {pages} {353} (\bibinfo {year}
  {1988})}\BibitemShut {NoStop}%
%%CITATION = CMPHA,117,353;%%
\bibitem [{\citenamefont {Vafa}\ and\ \citenamefont
  {Witten}(1994)}]{Vafa:1994tf}%
  \BibitemOpen
  \bibfield  {author} {\bibinfo {author} {\bibfnamefont {Cumrun}\ \bibnamefont
  {Vafa}}\ and\ \bibinfo {author} {\bibfnamefont {Edward}\ \bibnamefont
  {Witten}},\ }\bibfield  {title} {\enquote {\bibinfo {title} {{A Strong
  coupling test of S duality}},}\ }\href {\doibase
  10.1016/0550-3213(94)90097-3} {\bibfield  {journal} {\bibinfo  {journal}
  {Nucl. Phys.}\ }\textbf {\bibinfo {volume} {B431}},\ \bibinfo {pages} {3--77}
  (\bibinfo {year} {1994})},\ \Eprint {http://arxiv.org/abs/hep-th/9408074}
  {arXiv:hep-th/9408074 [hep-th]} \BibitemShut {NoStop}%
%%CITATION = HEP-TH/9408074;%%
\bibitem [{\citenamefont {Bobev}\ \emph {et~al.}(2014)\citenamefont {Bobev},
  \citenamefont {Elvang}, \citenamefont {Freedman},\ and\ \citenamefont
  {Pufu}}]{Bobev:2013cja}%
  \BibitemOpen
  \bibfield  {author} {\bibinfo {author} {\bibfnamefont {Nikolay}\ \bibnamefont
  {Bobev}}, \bibinfo {author} {\bibfnamefont {Henriette}\ \bibnamefont
  {Elvang}}, \bibinfo {author} {\bibfnamefont {Daniel~Z.}\ \bibnamefont
  {Freedman}}, \ and\ \bibinfo {author} {\bibfnamefont {Silviu~S.}\
  \bibnamefont {Pufu}},\ }\bibfield  {title} {\enquote {\bibinfo {title}
  {{Holography for $N = 2^*$ on $S^4$}},}\ }\href {\doibase
  10.1007/JHEP07(2014)001} {\bibfield  {journal} {\bibinfo  {journal} {JHEP}\
  }\textbf {\bibinfo {volume} {1407}},\ \bibinfo {pages} {001} (\bibinfo {year}
  {2014})},\ \Eprint {http://arxiv.org/abs/1311.1508} {arXiv:1311.1508
  [hep-th]} \BibitemShut {NoStop}%
%%CITATION = ARXIV:1311.1508;%%
\bibitem [{\citenamefont {Clark}\ and\ \citenamefont
  {Karch}(2005)}]{Clark:2005te}%
  \BibitemOpen
  \bibfield  {author} {\bibinfo {author} {\bibfnamefont {A.}~\bibnamefont
  {Clark}}\ and\ \bibinfo {author} {\bibfnamefont {A.}~\bibnamefont {Karch}},\
  }\bibfield  {title} {\enquote {\bibinfo {title} {{Super Janus}},}\ }\href
  {\doibase 10.1088/1126-6708/2005/10/094} {\bibfield  {journal} {\bibinfo
  {journal} {JHEP}\ }\textbf {\bibinfo {volume} {10}},\ \bibinfo {pages} {094}
  (\bibinfo {year} {2005})},\ \Eprint {http://arxiv.org/abs/hep-th/0506265}
  {arXiv:hep-th/0506265 [hep-th]} \BibitemShut {NoStop}%
%%CITATION = HEP-TH/0506265;%%
\bibitem [{\citenamefont {D'Hoker}\ \emph
  {et~al.}(2006{\natexlab{b}})\citenamefont {D'Hoker}, \citenamefont {Estes},\
  and\ \citenamefont {Gutperle}}]{D'Hoker:2006uu}%
  \BibitemOpen
  \bibfield  {author} {\bibinfo {author} {\bibfnamefont {Eric}\ \bibnamefont
  {D'Hoker}}, \bibinfo {author} {\bibfnamefont {John}\ \bibnamefont {Estes}}, \
  and\ \bibinfo {author} {\bibfnamefont {Michael}\ \bibnamefont {Gutperle}},\
  }\bibfield  {title} {\enquote {\bibinfo {title} {{Ten-dimensional
  supersymmetric Janus solutions}},}\ }\href {\doibase
  10.1016/j.nuclphysb.2006.08.017} {\bibfield  {journal} {\bibinfo  {journal}
  {Nucl. Phys.}\ }\textbf {\bibinfo {volume} {B757}},\ \bibinfo {pages}
  {79--116} (\bibinfo {year} {2006}{\natexlab{b}})},\ \Eprint
  {http://arxiv.org/abs/hep-th/0603012} {arXiv:hep-th/0603012 [hep-th]}
  \BibitemShut {NoStop}%
%%CITATION = HEP-TH/0603012;%%
\bibitem [{\citenamefont {Wetterich}(2011)}]{Wetterich:2010ni}%
  \BibitemOpen
  \bibfield  {author} {\bibinfo {author} {\bibfnamefont {C.}~\bibnamefont
  {Wetterich}},\ }\bibfield  {title} {\enquote {\bibinfo {title} {{Spinors in
  euclidean field theory, complex structures and discrete symmetries}},}\
  }\href {\doibase 10.1016/j.nuclphysb.2011.06.013} {\bibfield  {journal}
  {\bibinfo  {journal} {Nucl. Phys.}\ }\textbf {\bibinfo {volume} {B852}},\
  \bibinfo {pages} {174--234} (\bibinfo {year} {2011})},\ \Eprint
  {http://arxiv.org/abs/1002.3556} {arXiv:1002.3556 [hep-th]} \BibitemShut
  {NoStop}%
%%CITATION = ARXIV:1002.3556;%%
\bibitem [{\citenamefont {Nicolai}(1978)}]{Nicolai:1978vc}%
  \BibitemOpen
  \bibfield  {author} {\bibinfo {author} {\bibfnamefont {H.}~\bibnamefont
  {Nicolai}},\ }\bibfield  {title} {\enquote {\bibinfo {title} {{A Possible
  constructive approach to (SUPER phi**3) in four-dimensions. 1. Euclidean
  formulation of the model}},}\ }\href {\doibase 10.1016/0550-3213(78)90537-0}
  {\bibfield  {journal} {\bibinfo  {journal} {Nucl. Phys.}\ }\textbf {\bibinfo
  {volume} {B140}},\ \bibinfo {pages} {294} (\bibinfo {year}
  {1978})}\BibitemShut {NoStop}%
%%CITATION = NUPHA,B140,294;%%
\end{thebibliography}%
\end{document}